\documentclass{article} % For LaTeX2e
\usepackage{iclr2027_conference,times}

\usepackage{amsmath,amsfonts,bm}

\def\eqref#1{equation~\ref{#1}}
\def\1{\bm{1}}

\DeclareMathAlphabet{\mathsfit}{\encodingdefault}{\sfdefault}{m}{sl}
\SetMathAlphabet{\mathsfit}{bold}{\encodingdefault}{\sfdefault}{bx}{n}

\usepackage[colorlinks,
linkcolor = UrlColor,
urlcolor = UrlColor,
citecolor = RefColor,]{hyperref}
\usepackage{url}

\usepackage{lipsum}
\usepackage{tikz}
\usepackage{amsmath}
\usepackage{stfloats}
\usepackage{verbatim}
\usepackage{graphicx}
\usepackage{dsfont}
\usepackage{amsfonts}
\usepackage{array}
\usepackage{algorithm2e}
\usepackage{footmisc}
\RestyleAlgo{ruled}
\SetKwComment{Comment}{/* }{ */}
\usepackage{booktabs}
\usepackage[table]{xcolor}
\usepackage{colortbl}

\usepackage{multirow}

\usepackage{amssymb}
\usepackage{amsthm}
\usepackage{mathtools}
\usepackage{bbm}
\usepackage{bm}
\usepackage{dsfont}
\usepackage{enumitem}
\usepackage{makecell}
\usepackage{subcaption}
\usepackage{bbm}
\usepackage{diagbox} % Required for the split cell
\newcommand{\Angstrom}[0]{\text{\r{A}}}

\definecolor{Gray}{gray}{0.9}
\definecolor{LightCyan}{rgb}{0.88,1,1}

\newcolumntype{a}{>{\columncolor{Gray}}c}
\newcolumntype{b}{>{\columncolor{white}}c}

\newcommand{\mcl}[1]{\mathcal{#1}}
\newcommand{\mbf}[1]{\mathbf{#1}}

\newcommand{\prg}[1]{\vskip 4pt \noindent \textbf{#1}.}
\newcommand{\prginit}[1]{\noindent \textbf{#1}.}

\definecolor{SelfColor}{rgb}{0.913,0.443,0.196}
\definecolor{UrlColor}{rgb}{0.776, 0.255, 0.275}
\definecolor{RefColor}{rgb}{0.290,0.565,0.596}

\title{MARCO: The Radioactive Watermark for Protein Generative Models}

\author{
    Huajie Chen$^{\dagger}$, Xin Guo$^{\dagger}$, Yuchen Shi$^{\dagger}$, Yuchen Zhong$^{\dagger}$, Minhui Xue$^{\ddagger}$, Chi Liu$^{\dagger}$,\\
    \textbf{Congcong Zhu$^{\dagger}$, Kun Gao$^{\dagger}$, Minfeng Qi$^{\dagger}$, \& Tianqing Zhu$^{\dagger}$}
    \thanks{
        $^{\dagger}$Faculty of Data Science, City University of Macau.
        $^{\ddagger}$CSIRO and Responsible AI Research (RAIR) Centre, Adelaide University.
        }
}

\iclrfinalcopy % Uncomment for camera-ready version, but NOT for submission.
\begin{document}

\maketitle

\begin{abstract}
Protein Generative Models (PGMs) have revolutionized structural biology by enabling the design of complex 3D protein structures from sequence data. However, this breakthrough introduces a dual-use challenge, exposing high-value PGMs to economic risks like unauthorized model extraction and biosecurity threats such as biohazard synthesis. To mitigate these threats, we propose \textbf{MARCO} (\textsc{COnformation waterMARk}), the first radioactive watermarking framework specifically tailored for PGMs. MARCO establishes a Dual-Layer defense that simultaneously protects intellectual property and ensures the forensic traceability of potential biosecurity misuses. (i) To preserve efficiency, MARCO iteratively embeds watermarks during diffusion reverse denoising via an auxiliary encoder-decoder, allowing the original PGM parameters to remain frozen for broad compatibility. (ii) To preserve biophysical fidelity and maximize robustness, we employ specialized loss functions targeting $C_\alpha$-atom pairwise distances and torsion angles ($\psi, \phi$) within an adversarial training framework integrated with stochastic attack simulations. (iii) Crucially, MARCO exhibits ``radioactivity'' where the watermark automatically transfers to the outputs of any pirate models trained on the watermarked data, effectively countering model extraction attacks. Comprehensive experiments demonstrate that MARCO achieves superior fidelity and robustness while successfully validating watermark transferability.
\end{abstract}

\section{Introduction}
Protein generative models (PGMs), including AlphaFold~\cite{Abramson2024AlphaFold3} and RoseTTAFold~\cite{baek2021RosettaFold}, have transformed structural biology by predicting 3D structures from sequences~\cite{baker2001protein}. Recognized by the 2024 Nobel Prize in Chemistry, these advances also enable precise de novo design and accelerate the development of therapeutics and biomaterials beyond the reach of traditional screening~\cite{luo20213d}.

Their rapid deployment, however, creates a dual-use challenge. Open access fosters innovation but may enable malicious users to engineer enhanced pathogens or toxins~\cite{wang2025call}, prompting organizations including OpenAI, Anthropic, and Google DeepMind to highlight biosecurity risks~\cite{Biosec}. Meanwhile, extracting proprietary models~\cite{chen2025Queen} threatens commercial advantages and investment in drug discovery~\cite{zhang2025generative}. PGMs therefore require tailored protection.

Although watermarking is established for large language models~\cite{kirchenbauer2023watermark} and diffusion-based image generators~\cite{fernandez2023stable}, proteins impose distinct biophysical and geometric constraints. Embedded signals in coordinates or amino acid sequences must preserve conformation---and thus folding stability and binding affinity~\cite{Jumper2021AlphaFold2}. Because protein structures occupy continuous space and are invariant to rigid transformations, robust watermarks~\cite{li2021leveraging} must also withstand $SE(3)$ operations and conformational flexibility. This requires geometry-aware encoding in invariant features such as torsion angles or residue-residue distance maps, rather than simple coordinate noise~\cite{minan2025informed}.

\prg{Our Work}
We introduce \textbf{MARCO} (\textsc{COnformation waterMARk}), the first generalized radioactive watermarking framework compatible with contemporary PGM architectures. 
% As illustrated in \autoref{fig:overview}, 
MARCO adds robust signatures to high-fidelity structures from pretrained models without fine-tuning or retraining. Its ``radioactive'' watermark transfers to pirate models trained on protected structures, preserving data provenance.

We study three questions:
\textbf{RQ1)} Can a general watermarking framework adapt to existing PGMs without weight fine-tuning or retraining?
\textbf{RQ2)} Can it preserve biological fidelity while resisting common structural perturbations?
\textbf{RQ3)} Can the watermark propagate to derivative models trained on protected data, establishing a ``radioactive'' chain of custody?

\textbf{A1)}
Contemporary diffusion-based PGMs generate protein backbones as rigid-body sequences defined by translations and rotations~\cite{baek2021RosettaFold,lin2022ESMFold,yim2023se,yim2023fast}. MARCO perturbs both components during reverse diffusion, steering generation toward a watermarked conformation at inference time without retraining.
\textbf{A2)}
MARCO combines a joint encoder-decoder and discriminator~\cite{goodfellow2020generative} with a frozen PGM. The encoder watermarks intermediate diffusion predictions under fidelity constraints; the decoder recovers signals from stochastically distorted structures; and the discriminator distinguishes clean from watermarked structures. Their competition keeps perturbations biophysically indistinguishable from natural proteins.
\textbf{A3)} 
By intervening throughout reverse diffusion rather than watermarking a static structure, MARCO shifts the output distribution~\cite{sander2024watermarking}. A pirate model trained on these samples learns the biased distribution, so its outputs inherit the watermark and remain traceable to the source.

Our contributions are listed as follows:
\begin{itemize}[leftmargin=*]
    \item 
    We propose the first ``radioactive'' watermarking framework tailored to the biophysical and geometric constraints of PGMs with specifically designed biological loss terms.

    \item 
    MARCO embeds high-fidelity structures with signatures robust to $SE(3)$ transformations and malicious attacks; surrogate models trained on protected data inherit the signal, enabling persistent intellectual-property protection.

    \item 
    MARCO is an efficient, architecture-agnostic, plug-and-play module that requires no PGM weight fine-tuning.
\end{itemize}

\section{Preliminaries and Related Work}\label{sect:prelim}
% The notation used below is summarized in \autoref{tab:notations}.
% \todo{Change notation $t$ to $x$. Also in the figures.}

\subsection{Protein Generative Models}\label{sect:pgms}

\prginit{Overview}
Models such as RFdiffusion~\cite{watson2023RFDiffusion}, ESMFold~\cite{lin2022ESMFold}, and AlphaFold~3~\cite{Abramson2024AlphaFold3} learn a joint distribution $p_\theta(\mathbf{X},\mathbf{S})$ over sequence $\mathbf{S}$ and structure $\mathbf{X}$. A structure comprises $N$ rigid frames $\mathbf{X}=\{(\mathbf{R}_i,\mathbf{x}_i)\}_{i=1}^N$, where $\mathbf{R}_i\in SO(3)$ is the backbone orientation and $\mathbf{x}_i\in\mathbb{R}^3$ the $C_\alpha$ position. They enforce $SE(3)$-equivariance, $\Theta(g\cdot\mathbf{X})=g\cdot\Theta(\mathbf{X})$, for any global transformation $g$~\cite{fuchs2020se3Trans}, making predictions coordinate-frame independent and supporting physical validity and conformational generalization.
% \todo{Define $SO(3) \text{ and }SE(3)$.}
$SO(3)$ is the special orthogonal group of 3D rotations, $SO(3)={\mbf R\in\mathbb{R}^{3\times 3}\mid \mbf R^\top \mbf R=I,\ \det(R)=1},$
whose rotations preserve distances, angles, and orientation. $SE(3)$ is the group of 3D rigid transformations, $SE(3)={(\mbf R,\mbf x)\mid \mbf R\in SO(3),\ \mbf x\in\mathbb{R}^3},$ which maps $\mbf a\in\mathbb{R}^3$ to $\mbf R\mbf a+\mbf x$ and has the homogeneous form $\mbf X=\begin{bmatrix}\mbf R&\mbf x\\0&1\end{bmatrix}$, where $SO(3)$ represents rotation and $\mbf x$ translation.

\prg{Learning Objective}
% To sample from this complex distribution, contemporary methods predominantly employ denoising diffusion probabilistic models (DDPMs) or flow matching. These frameworks define a forward process that gradually corrupts the data structure $\mathbf{X}_0$ into noise $\mathbf{X}_T$ via a stochastic differential equation (SDE) or an interpolant.
Most models sample through denoising diffusion~\cite{ho2020DDPM} or flow matching~\cite{lipmanflowmatching}, whose forward process corrupts $\mathbf{X}_0$ into noise $\mathbf{X}_T$ via an SDE or interpolant.
% For example, RFDiffusion \cite{watson2023RFDiffusion} is trained to predict the ground truth structure $\mathbf{X}_0 = (\mathbf{R}_0, \mathbf{x}_0)$ from the noisy input $\mathbf{X}_t$ at timestep $t$. The loss is a weighted sum of errors on the translation ($C_\alpha$ positions) and rotation (backbone orientation):
RFdiffusion~\cite{watson2023RFDiffusion}, for example, recovers $\mathbf{X}_0=(\mathbf{R}_0,\mathbf{x}_0)$ from $\mathbf{X}_t$ by minimizing weighted translation and rotation errors:
\begin{equation}\label{eq:rfdiff_loss}
\begin{aligned}
    % \mathcal{L} = \mathbb{E}_{t, \mathbf{X}_0, \epsilon} \left[ w_{\text{trans}}(t) \| \hat{\mathbf{x}}_0 - \mathbf{x}_0 \|^2 + w_{\text{rot}}(t) \| \hat{\mathbf{R}}_0 - \mathbf{R}_0 \|_{\mathcal{F}}^2 \right],
    \mathcal{L} = \mathbb{E}_{t, \mathbf{X}_0, \eta} \left[ w_{\text{trans}}(t) \| \hat{\mathbf{x}}_0 - \mathbf{x}_0 \|^2 + w_{\text{rot}}(t) \| \hat{\mathbf{R}}_0 - \mathbf{R}_0 \|_{\mathcal{F}}^2 \right],
\end{aligned}
\end{equation}
% where $\| \hat{\mathbf{x}}_0 - \mathbf{x}_0 \|^2$ represents the Euclidean squared error between the predicted and true $C_\alpha$ coordinates;
% $\| \hat{\mathbf{R}}_0 - \mathbf{R}_0 \|_{\mathcal{F}}^2$ denotes the rotational error, typically calculated as the Frobenius norm difference between the predicted and true orientation matrices (or the vectors of the $N, C_\alpha, C$ atoms defining the frame).
where the terms measure $C_\alpha$ coordinate error and the Frobenius distance between orientation matrices, respectively.

\prg{Sampling}
% At each timestep $t$, the sampler performs the following operations to transition to state $\mathbf{X}_{t-1}$:
% The model $\Theta$ predicts the clean structure $\hat{\mathbf{X}}_0$. Crucially, RFDiffusion inputs the previous estimate of the clean structure, denoted as $\hat{\mathbf{X}}_0^{prev}$.
At step $t$, $\Theta$ predicts the clean structure from $\mathbf{X}_t$: $(\hat{\mathbf{R}}_0, \hat{\mathbf{x}}_0) = \Theta(\mathbf{X}_t, t, \mathbf{S})$.
% This self-conditioning allows the model to refine its hypothesis of the global topology iteratively, significantly improving coherence.
This iterative refinement improves topological coherence.
% Using the posterior distributions defined by the forward diffusion process, the sampler calculates the mean of the next state distribution $q(\mathbf{X}_{t-1} | \mathbf{X}_t, \hat{\mathbf{X}}_0)$ that consists of the translational mean and the rotational mean.
% Translational mean is a linear interpolation between the current noisy position and the predicted clean position defined as
The sampler computes the posterior mean for $\mathbf{X}_{t-1}$, with linear-interpolation translational mean
\begin{equation}\label{eq:trans_mean}
\begin{aligned}
    \boldsymbol{\mu}_{\mathbf{x}} = \frac{\sqrt{\bar{\alpha}_{t-1}} \beta_t}{1 - \bar{\alpha}_t} \hat{\mathbf{x}}_0 + \frac{\sqrt{\alpha_t} (1 - \bar{\alpha}_{t-1})}{1 - \bar{\alpha}_t} \mathbf{x}_t
\end{aligned}
\end{equation}
% Rotational mean is computed using geodesic interpolation on the $SO(3)$ manifold (often approximated via slerp or tangent space vectors) to find the intermediate frame between $\mathbf{R}_t$ and $\hat{\mathbf{R}}_0$.
and rotational mean obtained by geodesic interpolation between $\mathbf{R}_t$ and $\hat{\mathbf{R}}_0$ on $SO(3)$.
% The final values for $\mathbf{X}_{t-1}$ are obtained by adding stochastic noise to these means.
% Specifically, the translation is updated by 
Gaussian noise yields $\mathbf{x}_{t-1} = \boldsymbol{\mu}_{\mathbf{x}} + \sigma_t \eta$,
where $\eta\sim\mathcal{N}(\mathbf{0},\mathbf{I})$, while rotation updates as
\begin{equation}\label{eq:rot_update}
\begin{aligned}
    \mathbf{R}_{t-1} = \mathbf{R}_t \cdot \text{Exp}\left( c_t \cdot \log\left(\mathbf{R}_t^\top \hat{\mathbf{R}}_0 \right) + \sigma_t \cdot \xi \right),
\end{aligned}
\end{equation}
% where $c_t$ is a scalar coefficient derived from the diffusion posterior;
% $\sigma_t$ denotes the noise scale for the current timestep;
% $\mbf \epsilon \sim \mcl N(\mbf 0, \mbf I)$ is a perturbation noise.
where $c_t$ is the posterior coefficient, $\sigma_t$ the noise scale, and $\xi\sim\mathcal{N}(\mathbf{0},\mathbf{I})$.
% The prediction $\hat{\mathbf{X}}_0$ generated is stored and passed as $\hat{\mathbf{X}}_0^{prev}$ for the next iteration ($t-1$).
% Eventually, the loop terminates at $t=0$, yielding the final generated backbone $\mathbf{X}_0$. This backbone is then typically passed to a sequence design module (like ProteinMPNN) to optimize the amino acid identities $\mathbf{S}$ for the generated geometry.
The prediction becomes $\hat{\mathbf{X}}_0^{\text{prev}}$ at the next step. At $t=0$, a sequence design module such as ProteinMPNN typically optimizes $\mathbf{S}$ for the final backbone $\mathbf{X}_0$.

\subsection{Watermarking Diffusion Models}

% Current studies of watermarking diffusion models have bifurcated into two primary paradigms: \textit{inference-time injection} (modifying the sampling trajectory) and \textit{model-time imprinting} (modifying the network weights). 
% Despite their operational differences, these methods share a common objective: induce a statistical bias in the generated distribution $p_\theta(\mathbf{x}_0)$ such that watermarked samples fall into a verifiable region $\mathcal{R}_k$ defined by a secret key k, while maintaining high perceptual fidelity to the original distribution.
Diffusion watermarking follows two paradigms: \textit{inference-time injection} and \textit{model-time imprinting}. Both bias the generated distribution $p_\theta(\mathbf{x}_0)$ to make outputs verifiable while preserving fidelity.
% This bias ensures that watermarked samples reside within a verifiable region $\mathcal{R}_k$ defined by a secret key $k$, while strictly preserving the fidelity of the original data distribution.

\prg{Inference-Time Injection}
% The most prevalent approach involves embedding the watermark signal directly into the initial noise vector $\mathbf{x}_T$ or the intermediate latents.
% Tree-Ring Watermarking \cite{wen2023treeringwatermark} exploits the property that the low-frequency components of the initial noise largely determine the global layout of the image. The method transforms the noise $\mathbf{x}_T$ into the frequency domain via Fourier Transform $\mathcal{F}$ and implants a pattern m into a specific frequency ring M:
Tree-Ring Watermarking~\cite{wen2023treeringwatermark} embeds a pattern $m_{\mathrm{TR}}$ in a frequency ring $M$ of the initial noise $\mathbf{x}_T$, exploiting low-frequency control of global image layout:
\begin{equation}\label{eq:treering}
\begin{aligned}
    \mathcal{F}(\mathbf{x}_T') = \mathcal{F}(\mathbf{x}_T) \odot (1 - M) + m_{\mathrm{TR}} \odot M
\end{aligned}
\end{equation}
% Detection is performed by inverting the diffusion process (DDIM inversion) to recover $\hat{\mathbf{x}}_T$ and checking for the pattern $m$.
% Gaussian Shading \cite{yang2024GaussianShading} formulates watermarking as an optimal transport problem. It optimizes the initial noise $\mathbf{x}_T$ to minimize the distance to a "shaded" latent region associated with the watermark, without requiring inversion for detection.
Detection recovers $\hat{\mathbf{x}}_T$ through DDIM inversion and tests for $m_{\mathrm{TR}}$. Gaussian Shading~\cite{yang2024GaussianShading} instead uses optimal transport to move $\mathbf{x}_T$ toward a watermarked latent region, avoiding inversion during detection.

\prg{Model-Time Imprinting}
% Alternatively, watermarks can be baked into the generative weights or the decoder.
% Stable Signature \cite{fernandez2023stablesignature} finetunes the VAE decoder $D_\phi$ such that all generated images contain a robust, invisible signature $w$. The objective minimizes a combination of perceptual loss and watermark detection loss:
Stable Signature~\cite{fernandez2023stablesignature} fine-tunes the VAE decoder $D_\phi$ to embed a robust, invisible signature $w$ in every image, balancing fidelity and detection:
\begin{equation}\label{eq:stable_signature}
\begin{aligned}
    \mathcal{L}_{\text{total}} = \mathcal{L}_{\text{perceptual}}(\mathbf{x}, \mathbf{x}_w) + \lambda \mathcal{L}_{\text{detect}}(E(\mathbf{x}_w), w), 
\end{aligned}
\end{equation}
% where $E$ is a watermark extractor.
% ProMark \cite{asnani2024promark} embeds watermarks into the prompt embeddings using learnable tokens, ensuring that specific trigger words generate watermarked outputs.
% Generative Watermarking via Rejection Sampling \cite{zhao2024RejectionSamplingWatermark} theoretically grounds watermarking in rejection sampling, filtering the outputs of a pre-trained model to accept only those that satisfy a specific hash function or metric condition $h(\mathbf{x}) \in \mathcal{K}$.
where $E$ extracts the watermark. ProMark~\cite{asnani2024promark} uses learnable prompt tokens to trigger watermarked outputs, whereas rejection-sampling watermarking~\cite{zhao2024RejectionSamplingWatermark} retains outputs satisfying $h(\mathbf{x})\in\mathcal{K}$.

\prg{Limitations in Watermarking Protein Structures}
Image methods~\cite{zhao2023recipe,huang2024robin,liu2023watermarking,WMAdapter} assume fixed 2D grids, image-specific latents or text conditioning, and perceptual invisibility, so they do not transfer directly to variable-length 3D proteins. Protein watermarks must respect $SE(3)$ symmetry and preserve inter-residue distances, torsion angles, stereochemistry, folding stability, and function. FoldMark~\cite{zhang2025foldmark} applies watermark-conditioned LoRA fine-tuning to PGMs, but modifies model weights and does not sufficiently constrain the predicted structures biologically. SynthIDBio \cite{stutz2026function} fine-tunes the PGM and trains a watermark detector with a fidelity loss term and a watermarking loss term. However, SynthIDBio does not support multi-bit watermarking, and there is no evidence that such design can withstand adversarial attacks.

% While effective for 2D grids, these methods predominantly rely on the Euclidean nature of the data. Extending these strategies to the non-Euclidean $SE(3)$ manifold of protein structures, where ``frequency'' is defined on a sphere and ``noise'' is geodesic, remains a challenge to address.

\section{Threat Model}
% \subsection{Threat Model}
We consider two parties: the victim and the adversary. Based on this setting, we define MARCO's design properties in \autoref{sect:design_properties}.

\prg{The Victim}
% The victim is the entity that develops or hosts the protein generative model $\Theta$, e.g., a pharmaceutical company or research lab.
% The victim possesses a proprietary generative model $\Theta$ (e.g., RFDiffusion) trained on high-value structural data. 
% The victim's goal is to embed a watermark signal $m$ into the generated protein structures $\hat{\mathbf{X}} \sim \Theta(\cdot)$ such that a recovered watermark $\hat{m}$ can be extracted by a decoder $\hat{m} = D(\tilde{\mbf X}) \approx m$. 
% The watermarking process must not degrade the biophysical viability of the protein. Specifically, the watermarked structure $\tilde{\mbf X}$ must remain foldable, physically stable (low free energy), and functionally equivalent to a non-watermarked baseline.
The victim, typically a pharmaceutical company or research laboratory, develops or hosts a proprietary PGM $\Theta$ (e.g., RFdiffusion) trained on valuable structural data. It embeds a recoverable watermark $\mbf w$ in generated structures $\hat{\mathbf{X}}\sim\Theta(\cdot)$, producing $\tilde{\mathbf{X}}$ such that $D(\tilde{\mathbf{X}})\approx\mbf w$.
% Crucially, this embedding process is bound by strict biophysical constraints: the watermarked structure $\tilde{\mathbf{X}}$ must preserve the foldability, thermodynamic stability, and functional equivalence of the unwatermarked baseline.
% \prg{The Victim's Capabilities}
% The victim holds exclusive, full control over the generative model $\Theta$. This includes access to the model weights, internal activations, and gradients. Consequently, the victim is capable of modifying $\theta$ via fine-tuning and intervening in the reverse diffusion SDE/ODE solver.
% The victim is assumed to have access to high-performance computing infrastructure sufficient to run the generative model and auxiliary validation tools. This includes the ability to perform forward folding or energy minimization to verify that the watermarked proteins retain their biophysical utility.
The victim has white-box access to $\Theta$, including its parameters $\theta$, activations, and gradients.
% This authority allows for direct modifications, such as fine-tuning weights or intervening within the reverse diffusion SDE/ODE solvers. 
It also has sufficient computing resources for biophysical validation through forward-folding simulations and energy minimization.
% to verify that watermarked structures maintain their functional utility.

\prg{The Adversary}
% The adversary acts as a malicious user or a competitor who gains access to the generated outputs (or the model weights). Their goal is to utilize the generated proteins while evading provenance tracking—either to infringe on intellectual property rights or to bypass biosecurity screening mechanisms for restricted toxins. The adversary employs an attack function $\mathcal{A}$ to produce a modified structure $\mathbf{X}' = \mathcal{A}(\tilde{\mbf X})$. To be considered successful, the adversary must satisfy two conditions:
% The detection algorithm fails to extract the recovered watermark in the modified structure $\hat{m} = D(\mathbf{X}') \not\approx m$;
% the modified protein $\mathbf{X}'$ must retain the intended biological function and structural integrity of $\tilde{\mbf X}$.
The adversary is a malicious user or competitor who exploits model outputs while hiding their provenance to infringe intellectual property or evade biosecurity screening for restricted agents such as toxins. It transforms a watermarked structure as $\mathbf{X}'=\mathcal{A}(\tilde{\mathbf{X}})$. An attack succeeds only if \textit{i)} detection fails, $D(\mathbf{X}')\not\approx\mbf w$, and \textit{ii)} $\mathbf{X}'$ retains the structural integrity and biological function of $\tilde{\mathbf{X}}$.
% \prg{The Adversary's Capabilities} 
The adversary knows the watermarking algorithm and model architecture, but not the victim's training data, training algorithm, or watermark parameters. Its access to the victim model is black-box and query-only, but it has unlimited access to watermarked structures.
% In scenarios where the generative model is open-source (e.g., RFDiffusion), the adversary may also possess the model weights, allowing them to perform attacks such as fine-tuning or distillation.
It has standard computing resources: an A100 GPU with 80GB VRAM and 64GB RAM suffices to fine-tune AlphaFold2, and molecular-modeling software such as Rosetta and PyMOL\footnote{https://rosettacommons.org/, https://pymol.org/}.
% This grants them the capability to perform domain-specific attacks including energy minimization, re-folding, resampling, geometric transformation, etc.

\prg{Adaptive Attacker}
% \todo
% \textit{Explicitly include Energy Minimization Attacks (e.g., Rosetta Relax, OpenMM) in the adversary's capabilities.}
% We consider an adaptive attacker who knows all the details of MARCO.
% Such an attacker attempts to use the most powerful attack to remove the watermark embedded in the protein structure while maintaining its conformation.
An adaptive adversary knows MARCO fully and uses optimization to erase its watermark while preserving fidelity.
% Methods such as molecular dynamics relaxation protocols in OpenMM \cite{eastman2013OpenMM} and Rosetta Relax \cite{tyka2011RosettaRelax} can be employed to remove the watermark by driving the structure towards a physical energy minimum.
% This is equivalent to globally modifying the atomic coordinates based on the force field such that watermarked structures relying on high-energy conformations are immediately wiped out.
It may use OpenMM molecular dynamics~\cite{eastman2013OpenMM} or Rosetta Relax~\cite{tyka2011RosettaRelax} to drive structures toward local thermodynamic minima.
% This process fundamentally alters atomic coordinates based on force field constraints, effectively neutralizing any watermarks that rely on energetically unfavorable, high-energy conformations.
% Moreover, since the adaptive attacker has white-box access to the local pirate model, it is therefore possible to use local regularization techniques such as differential privacy stochastic gradient descent (DPSGD) \cite{abadi2016DPSGD} to purge the watermark signal during training.
% Essentially, DPSGD adds random noises to the gradients during the training process according to the privacy budget $(\epsilon, \delta)$.
% This makes the PGM "forget" the watermark.
With white-box access to its pirate model, the adversary may also use Differentially Private Stochastic Gradient Descent (DPSGD)~\cite{abadi2016DPSGD} to scrub the watermark during training. Its calibrated gradient noise prevents the PGM from retaining the watermarked distributional features.
% \todo{Mention watermark forgery attacks.}
It may also forge a watermark on the structures~\cite{chen2024high,zhang2021deep} to obscure ownership verification.

\begin{figure*}[t!]
\centering
\includegraphics[width=.9\linewidth]{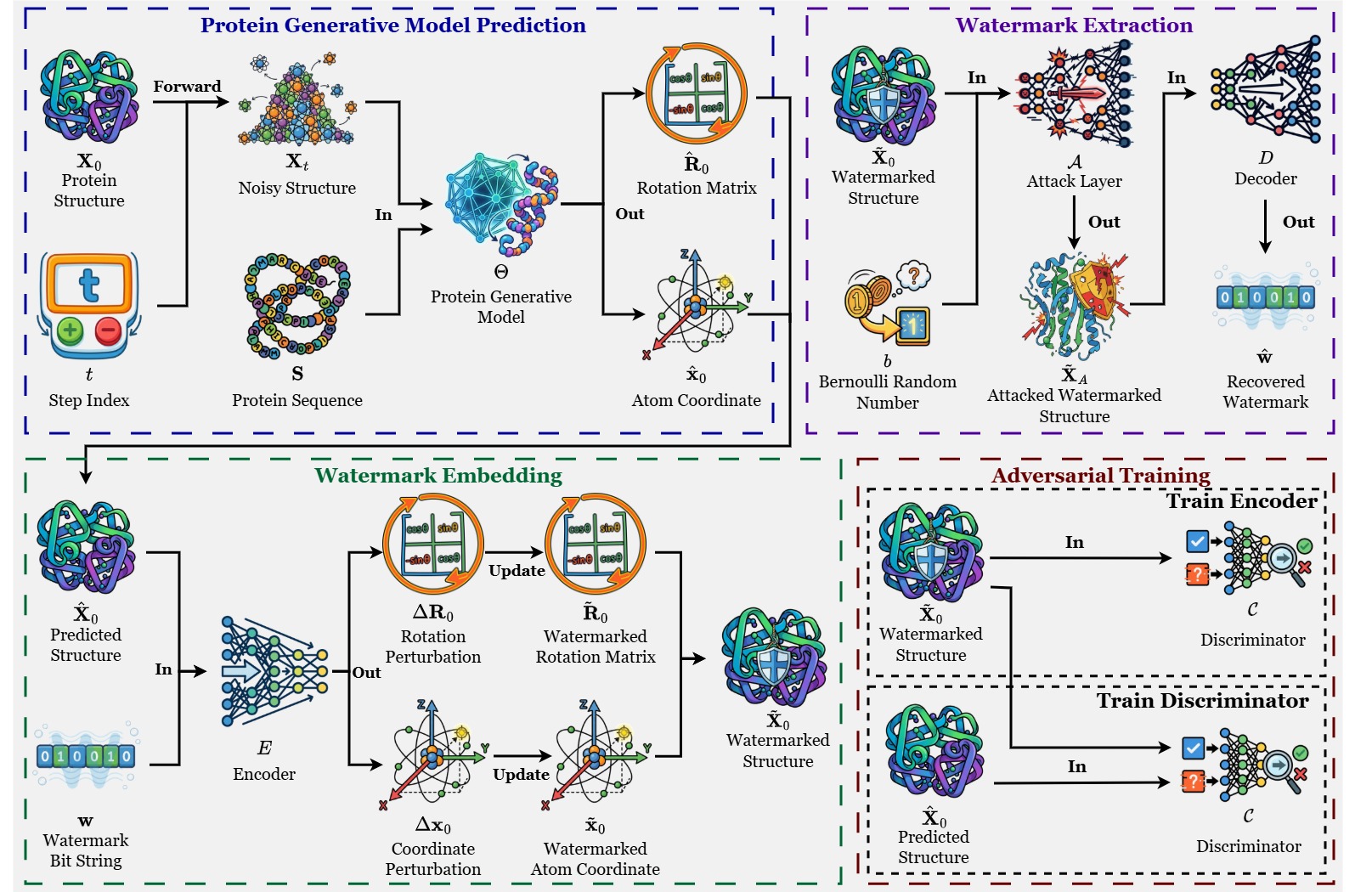}
\caption{\textbf{MARCO training.} \textit{i) PGM prediction}: The frozen PGM $\Theta$ predicts a clean structure $\hat{\mathbf{X}}_0$ from noisy state $\mathbf{X}_t$ and sequence $\mathbf{S}$. \textit{ii) Watermark embedding}: Encoder $E$ embeds $\mbf w$ through geometric perturbations, producing $\tilde{\mathbf{X}}_0$ and steering the generation trajectory. \textit{iii) Watermark extraction}: A Bernoulli-controlled attack layer produces $\tilde{\mathbf{X}}_A$, from which decoder $D$ recovers $\hat{\mbf w}$ for robustness. \textit{iv) Adversarial training}: Discriminator $\mcl C$ distinguishes $\hat{\mathbf{X}}_0$ from $\tilde{\mathbf{X}}_0$, while $E$ learns biophysically imperceptible perturbations that deceive $\mcl C$.}
\label{fig:marco_training}
\end{figure*}

\section{MARCO Method}
\subsection{Overview}
We introduce MARCO, a watermarking framework for Protein Generative Models (PGMs) that formulates watermarking as multi-objective optimization on the rigid-body $SE(3)$ manifold using an adversarially trained encoder-decoder network.
At each denoising step, the encoder embeds the watermark into the predicted protein backbone, steering diffusion toward a verifiable ``watermark manifold''. A discriminator penalizes distributional divergence so that watermarked structures remain biophysically indistinguishable from natural ones. During training, a stochastic attack layer encourages redundant, persistent features that survive geometric perturbations and malicious attacks, while the decoder recovers the signature from the distorted structures to reinforce robustness. For flow-matching and deterministic models (e.g. ESMFold), MARCO can also provide robust watermark with fewer steps of perturbations. In \autoref{sec:certified_robustness}, we prove that MARCO provides certified robustness.

\subsection{Training MARCO}
As shown in \autoref{alg:marco_training} and \autoref{fig:marco_training}, MARCO jointly optimizes a watermark encoder $E$, decoder $D$, and discriminator $\mcl C$ with a frozen PGM $\Theta$ in four phases: \textit{i)} PGM prediction, \textit{ii)} watermark embedding, \textit{iii)} watermark extraction, and \textit{iv)} adversarial training.

\begin{algorithm}[t!]
\scriptsize
\caption{Training Process of MARCO}\label{alg:marco_training}
\KwData{Frozen PGM $\Theta$; encoder $E$; decoder $D$; discriminator $\mcl C$; attack simulator $\mcl A$; training structures and sequences; watermark $\mbf w$}
\KwResult{Trained watermark encoder $E$ and decoder $D$}
Freeze $\Theta$\;
\While{training has not converged}{
    Sample a noisy structure and use $\Theta$ to predict its clean backbone\;
    Use $E$ to embed $\mbf w$ into the prediction with a timestep-aware geometric perturbation\;
    Apply a stochastic simulated attack and use $D$ to recover the watermark\;
    Update $E$ and $D$ to balance watermark recovery, structural fidelity, and covertness\;
    Update $\mcl C$ to distinguish clean predictions from watermarked ones\;
}
\Return{$E, D$}
\end{algorithm}

\begin{algorithm}[t!]
\scriptsize
\caption{Evaluation Process of MARCO}\label{alg:marco_eval}
\KwData{PGM $\Theta$; trained encoder $E$ and decoder $D$; protein sequence $\mbf S$; owner and model identities; watermark length; authoritative ledger; suspect structure $\mbf X$}
\KwResult{Watermarked structure $\tilde{\mbf X}$; recovered watermark $\hat{\mbf w}$; detection decision $d_{\text{wm}}$}
\tcc{Provenance-Backed Watermark Generation}
Derive a target watermark $\mbf w$ of the specified length from the model identity\;
Timestamp and register the ownership record with the authoritative ledger\;

\tcc{Watermarked Structure Generation}
Initialize a random noisy structure\;
\ForEach{denoising step}{
    Use $\Theta$ to predict the clean protein structure\;
    Use $E$ to embed $\mbf w$ through a timestep-aware geometric perturbation\;
    Advance the denoising trajectory using the watermarked prediction\;
}
Set $\tilde{\mbf X}$ to the final generated structure\;

\tcc{Watermark Validation}
Use $D$ to recover $\hat{\mbf w}$ from the suspect structure $\mbf X$\;
Compare $\hat{\mbf w}$ with $\mbf w$ using a statistical confidence test\;
Set $d_{\text{wm}}$ according to whether the detection threshold is met\;
\Return{$\tilde{\mbf X}, \hat{\mbf w}, d_{\text{wm}}$}
\end{algorithm}

\prg{Protein Generative Model Prediction}
The forward process converts a ground-truth structure $\mbf X_0$ into $\mbf X_t$ at timestep $t$, from which $\Theta$ estimates the clean baseline $\hat{\mathbf{X}}_0=(\hat{\mathbf{R}}_0,\hat{\mathbf{x}}_0)$ using $\mathbf{X}_t$, $t$, and sequence $\mathbf{S}$ (\autoref{sect:pgms}).

\prg{Watermark Embedding}
The encoder $E$ embeds $\mbf w$ into this estimate through residual learning, predicting a perturbation field $(\Delta \mbf R_0,\Delta\mathbf{x}_0)$ rather than absolute watermarked coordinates:
\begin{equation}\label{eq:wm_gen_delta}
\begin{aligned}
    (\Delta \mbf R_0, \Delta\mbf x_0) = E(\hat{\mbf R}_0, \hat{\mbf x}_0, \mbf w)
\end{aligned}
\end{equation}

% The perturbation is then applied to the clean prediction so that the watermark is embedded.
% Perturbing $\hat{\mbf x}_0$ is simple, where we simply add $\Delta \mbf x_0$ to $\hat{\mbf x}_0$ with a cosine trick:
The watermark is embedded in $\hat{\mathbf{x}}_0$ by adding $\Delta\mathbf{x}_0$ under a temporal scaling schedule:
\begin{equation}\label{eq:wm_embed_t}
\begin{aligned}
    \tilde{\mbf x}_0 = \hat{\mbf x}_0 + \cos(\frac{\pi (T-t)}{2T}) \Delta \mbf x_0
\end{aligned}
\end{equation}
% Here, the cosine trick aims to decrease the weight of the perturbation as $t$ increases.
The cosine schedule anneals the perturbation magnitude with timestep $t$.

% Perturbing $\hat{\mbf R}_0$ is more complex, because applying perturbations to rotation matrices is non-trivial, i.e., the space of rotations, $SO(3)$, is a non-linear manifold. 
% Simple element-wise addition (e.g., $\tilde{\mathbf{R}}_0 = \hat{\mathbf{R}}_0 + \Delta \mathbf{R}_0$) breaks the mathematical properties of the matrix, resulting in a transformation that is no longer orthogonal.
% Hence, $\Delta \mbf R_0 \in \mbb R^3$ is a vector in the tangent space of the rotation group, and $\|\Delta \mbf R_0\|$ is the magnitude of the perturbation in radians.
% $\Delta \mbf R_0$ is then converted into a rotation matrix $\Delta \mbf R_0$ using the matrix exponential: 
Because $SO(3)$ is a non-linear manifold, element-wise addition violates rotational orthogonality. We instead define $\Delta\mbf R_0\in\mathbb{R}^3$ in the tangent space $\mathfrak{so}(3)$, with $\|\Delta\mbf R_0\|$ giving the angular magnitude in radians, and map it onto the manifold via the matrix exponential with the same temporal modulation:
\begin{equation}\label{eq:R_convert}
\begin{aligned}
    \Delta \mathbf{R}_0 = \exp(\cos(\frac{\pi (T-t)}{2T}) \cdot \Delta \mbf R_0)
\end{aligned}
\end{equation}
% Here, we apply the cosine trick again for the same reason.
% Finally, the perturbed rotation matrix $\hat{\mbf R}_0$ is derived by
The perturbed rotation is then obtained by matrix multiplication:
\begin{equation}\label{eq:wm_embed_R}
\begin{aligned}
    \tilde{\mbf R}_0 = \hat{\mbf R}_0 \Delta \mbf R_0
\end{aligned}
\end{equation}

\begin{figure*}[t!]
    \centering
    \includegraphics[width=.9\textwidth]{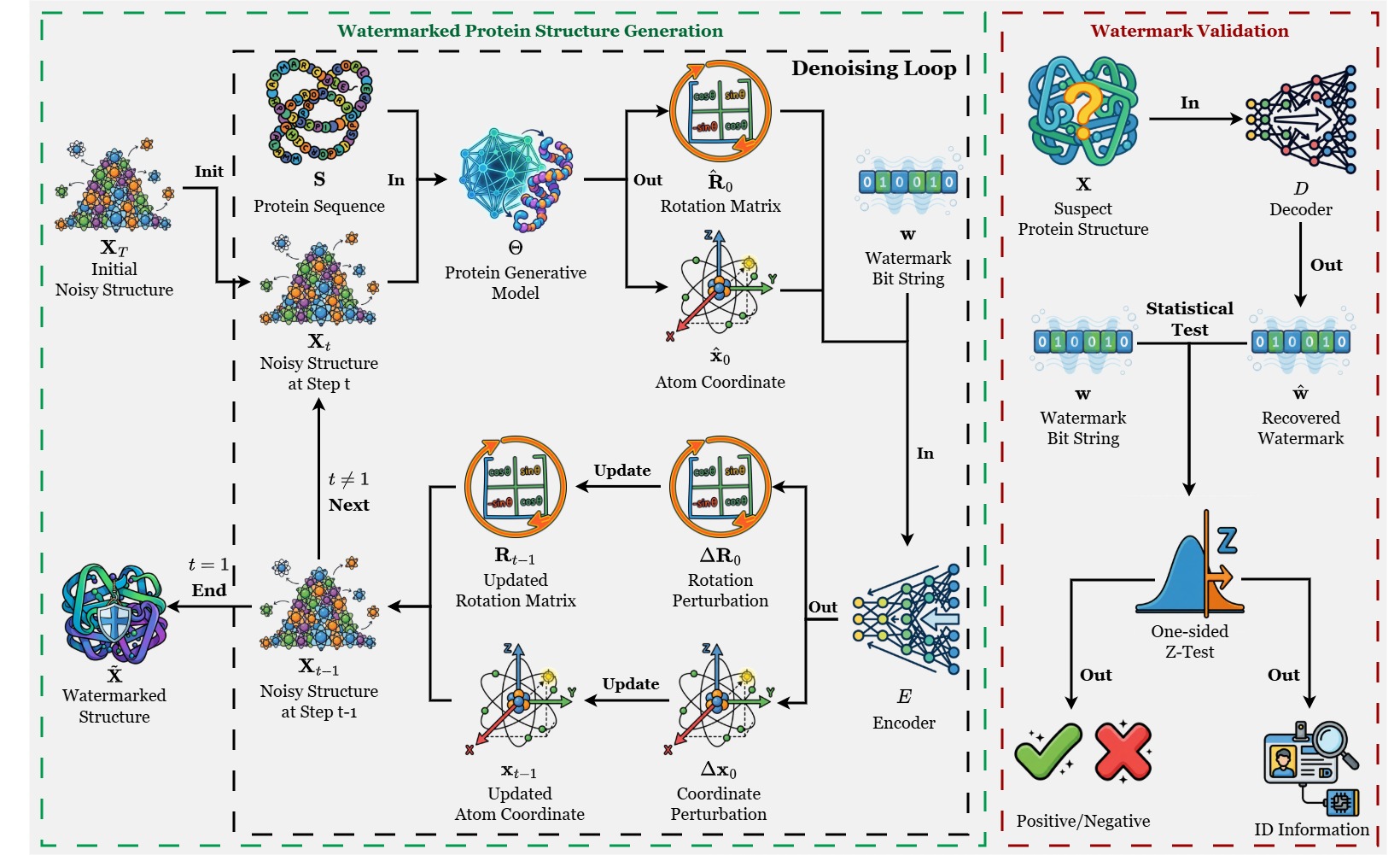}
    \caption{\textbf{MARCO evaluation.} \textit{i) Watermarked structure generation}: At each denoising step, watermark perturbations are applied to the PGM's clean predictions, yielding a fully watermarked structure $\tilde{\mbf X}$. \textit{ii) Watermark validation}: The decoder recovers $\hat{\mbf w}$ from a suspect structure and statistically compares it with the original watermark. Successful verification confirms provenance and identifies the embedded user.}
    \label{fig:marco_eval}
\end{figure*}

% Then, $(\tilde{\mbf R}_0, \tilde{\mbf x}_0)$ replace the roles of $(\hat{\mbf R}_0, \hat{\mbf x}_0)$ in the denoising process as described in \autoref{eq:trans_update} and \autoref{eq:rot_update} to update $\mbf X_t$ to get $\mbf X_{t-1}$.
The watermarked estimates $(\tilde{\mathbf{R}}_0,\tilde{\mathbf{x}}_0)$ replace $(\hat{\mathbf{R}}_0,\hat{\mathbf{x}}_0)$ in the standard diffusion updates (\autoref{sect:pgms} and \autoref{eq:rot_update}), propagating the watermark to $\mathbf{X}_{t-1}$.
% Through continuously perturbing the denoising trajectory, the watermark is completely embedded in $\tilde{X}$ when $t$ becomes $0$.

% To maintain the fidelity, the victim should minimize the two terms: $\|\hat{\mbf x}_0^i - \tilde{\mbf x}_0^i\|^2$ and $\|\hat{\mbf R}^i_0 - \tilde{\mbf R}^i_0\|^2$.
% Given \autoref{eq:wm_gen_delta}, this is equivalent to minimize the following watermarking loss:
To preserve structural fidelity, we minimize the translational and rotational deviations $\|\hat{\mathbf{x}}_0^i-\tilde{\mathbf{x}}_0^i\|^2$ and $\|\hat{\mathbf{R}}_0^i-\tilde{\mathbf{R}}_0^i\|^2$. Under \autoref{eq:wm_gen_delta}, this reduces to constraining the predicted perturbations through:
\begin{equation}\label{eq:loss_WM}
\begin{aligned}
    \mcl L_{\text{wm}} = \frac{1}{N} \sum_{i=1}^N \|\Delta \mbf R_0^i\| + \|\Delta \mbf x_0^i\|
\end{aligned}
\end{equation}
% By minimizing this loss term, the victim decreases the distance between the clean predicted structure and the watermarked structure.
Minimizing $\mathcal{L}_{\text{wm}}$ constrains the geometric distance between $\hat{\mathbf{X}}_0$ and $\tilde{\mathbf{X}}_0$, preserving biophysical consistency.

% \begin{equation}\label{eq:loss_Calpha}
% \begin{aligned}
%     \mcl L_{C_\alpha} = \frac{1}{N}\sum_{i=1}^N \|\hat{\mbf x}_0^i - \tilde{\mbf x}_0^i\|^2 + \|\hat{\mbf R}^i_0 - \tilde{\mbf R}^i_0\|^2
% \end{aligned}
% \end{equation}

% To further ensure high-fidelity watermark embedding, we design two novel loss terms: the distance matrix loss $\mcl L_{\text{dm}}$ and the torsion angle loss $\mcl L_{\text{tor}}$.
We further promote embedding fidelity with two novel losses: the distance matrix loss $\mathcal{L}_{\text{dm}}$ and torsion angle loss $\mathcal{L}_{\text{tor}}$.

The distance matrix loss enforces the preservation of global structural topology by penalizing deviations in pairwise $C_\alpha$ distances:
\begin{equation}\label{eq:loss_DistMat}
\begin{aligned}
    \mcl L_{\text{dm}} = \frac{1}{N^2} \sum_{i=1}^N \sum_{j=1}^N (\|\hat{\mbf x}_0^{i} - \hat{\mbf x}_0^{j}\| - \|\tilde{\mbf x}_0^{i} - \tilde{\mbf x}_0^{j}\|)^2
\end{aligned}
\end{equation}
% This loss term aims to maintain the relative position of each $C_\alpha$ atom to the others in the backbone of the watermarked structure following that of the predicted structure.
Minimizing this term preserves the clean backbone's internal organization and relative atomic positions, preventing global fold distortion.

% Moreover, the torsion angle loss is defined as
The torsion angle loss $\mathcal{L}_{\text{tor}}$ constrains local geometry:
\begin{equation}\label{eq:loss_Torsion}
\begin{aligned}
    \mcl L_{\text{tor}} = \frac{1}{N-2}\sum_{i=2}^{N-1} [1 - \cos(\hat{\phi}_i - \tilde{\phi}_i) + 1 - \cos(\hat{\psi}_i - \tilde{\psi}_i)],
\end{aligned}
\end{equation}
% where $\hat{\phi}_i$ and $\hat{\psi}_i$ denote the predicted backbone $\phi$ and $\psi$ torsion angles;
% $\Upsilon_i$ and $\tilde{\psi}_i$ are the corresponding torsion angles derived from the watermarked structure.
% We follow the determination process of the torsion angles in the previous works \cite{watson2023RFDiffusion}.
% This loss term penalizes sterically implausible geometries to help improve performance under the predicted local distance difference test.
where $(\hat{\phi}_i,\hat{\psi}_i)$ and $(\tilde{\phi}_i,\tilde{\psi}_i)$ are the clean and watermarked backbone dihedral angles, computed using standard protocols \cite{watson2023RFDiffusion}. The loss penalizes sterically implausible conformations to preserve local structural integrity.

% Before watermark extraction, we employ an attack layer $\mcl A$ to randomly apply transformations such as rotation and cropping to perturb the watermarked structure, denoted as $\tilde{\mbf X}_A = \mcl A(\tilde{\mbf X}_0, b)$.
% Here, $b$ is randomly sampled from a Bernoulli distribution with $p=0.5$. 
% If $b=1$, $\mcl A$ executes the attack, otherwise nothing happens. 

\prg{Watermark Extraction}
Before extraction, an attack layer $\mathcal{A}$ simulates adversarial conditions through manipulations such as rotation, cropping, and noise addition, producing $\tilde{\mathbf{X}}_A=\mathcal{A}(\tilde{\mathbf{X}}_0,b)$. For $b\sim\text{Bern}(0.5)$, the attack is applied when $b=1$; otherwise, the structure is unchanged.

% Next, the decoder $D$ extracts the recovered watermark $\hat{\mbf w}$ from $\tilde{\mbf X}_A$, denoted as $\hat{\mbf w} = D(\tilde{\mbf X}_A)$.
% We then use an extraction loss to ensure that the watermark can be correctly recovered from the watermarked structure, which is defined as
The decoder $D$ recovers $\hat{\mbf w}$ from the potentially distorted $\tilde{\mathbf{X}}_A$ using the extraction loss:
\begin{equation}\label{eq:loss_ext}
\begin{aligned}
    \mcl L_{\text{ext}} = -\frac{1}{l} \sum_{i=1}^l \mbf w_i\log (\hat{\mbf w}_i)
    + (1 - \mbf w_i)\log (1-\hat{\mbf w}_i)
\end{aligned}
\end{equation}
% This loss term helps the decoder $D$ learn to extract the recovered watermark from both the watermarked structure and the attacked structure.
Minimizing $\mathcal{L}_{\text{ext}}$ enables robust recovery from both pristine and structurally compromised proteins.

% Additionally, a discriminator $\mcl C$ is trained along with $E$ and $D$ to form a adversarial learning framework.
% The adversarial training loss is defined as
\prg{Adversarial Training}
A discriminator $\mcl C$, trained jointly with $E$ and $D$, minimizes the classification error between clean and watermarked structures in an adversarial minimax game:
\begin{equation}\label{eq:loss_Disc}
\begin{aligned}
    \mcl L_{\mcl C} = - \big( \log \mcl C(\hat{\mbf X}_0) + \log \big(1 - \mcl C(\tilde{\mbf X}_0) \big) \big)
\end{aligned}
\end{equation}
The encoder simultaneously minimizes an adversarial loss to deceive $\mcl C$:
\begin{equation}\label{eq:loss_ADV}
\begin{aligned}
    \mcl L_{\text{adv}} = - \log \big(\mcl C(\tilde{\mbf X}_0) \big)
\end{aligned}
\end{equation}
% where $\mcl L_{\mcl C}$ is used to update the weights of $\mcl C$, and $\mcl L_{\text{adv}}$ is for updating $E$.
% Eventually, these two loss terms help $E$ more covertly embed the watermark into the predicted structure, such that $\tilde{X}_0$ and $\hat{X}_0$ are statistically indistinguishable.
Thus, $\mathcal{L}_{\mcl C}$ trains the discriminator to separate the distributions, while $\mathcal{L}_{\text{adv}}$ trains the encoder to make $\tilde{\mathbf{X}}_0$ statistically indistinguishable from $\hat{\mathbf{X}}_0$.

The overall MARCO training loss is
\begin{equation}\label{eq:loss_total}
\begin{aligned}
    \mcl L = \lambda_1\mcl L_{\text{wm}} + \lambda_2\mcl L_{\text{dm}} + \lambda_3\mcl L_{\text{tor}} + \lambda_4\mcl L_{\text{ext}} + \lambda_5\mcl L_{\text{adv}}.
\end{aligned}
\end{equation}

\subsection{Evaluating MARCO}\label{sect:marco_eval}
As shown in \autoref{alg:marco_eval} and \autoref{fig:marco_eval}, MARCO establishes a provenance-backed watermark, generates a watermarked protein structure, and validates the watermark.
% \todo{Provide cryptographic watermark bit string generation method against watermark forgery.}
To prevent ownership claims based on forged watermarks, MARCO cryptographically timestamps and registers the model identity and timestamp with a trustworthy ledger; unregistered watermarks cannot establish ownership.

\prg{Watermarked Protein Structure Generation}
% In the generation of the watermarked structure, a random noise $\mbf X_T$ is first randomly sampled from a Gaussian distribution.
% Then, the denoising process starts from $t=T$ to $t=1$.
% In each iteration at step $t$, a predicted structure $\hat{\mbf X}_0$ is generated by $\hat{\mbf X}_0 = (\hat{\mbf R}_0, \hat{\mbf x}_0) = \Theta(\mbf X_t, \mbf S, t)$.
% $\hat{\mbf X}_0$ is then fed into $E$ to predict the perturbation $(\Delta \mbf R_0, \Delta \mbf x_0)$.
% Next, the perturbation is introduced to $\hat{\mbf X}_0$ to get the perturbed prediction $(\tilde{\mbf R}_0, \tilde{\mbf x}_0)$.
% At the end of this iteration, $\mbf X_{t-1}$ is derived by denoising $\mbf X_t$ using $(\tilde{\mbf R}_0, \tilde{\mbf x}_0)$, and it is then used to replace $\mbf X_t$ for the next iteration.
% At the end of the denoising process, a complete watermarked structure $\tilde{\mbf X} = \mbf X_0$ is generated, and can be released.
Starting from standard Gaussian noise $\mathbf{X}_T$ and protein sequence $\mbf S$, the reverse diffusion process ($t=T\to1$) uses $\Theta$ to estimate a clean baseline $\hat{\mathbf{X}}_0$ at each step. Encoder $E$ injects perturbations $(\Delta\mathbf{R}_0,\Delta\mathbf{x}_0)$ to produce $(\tilde{\mathbf{R}}_0,\tilde{\mathbf{x}}_0)$, which guides the transition to $\mathbf{X}_{t-1}$. The process yields the releasable watermarked structure $\tilde{\mathbf{X}}$.

\prg{Watermark Validation}
% For validation, the victim receives a suspect protein structure $\mbf X$.
% The victim feeds $\mbf X$ into the decoder $D$ to extract the recovered watermark $\hat{\mbf w}$.
% By counting the matching bits between the original watermark $\mbf w$ and $\hat{\mbf w}$, the victim gets the number of matching bits $k$.
% Then, the victim performs the One-Sided Z-Test to examine whether $\hat{\mbf w}$ aligns with $\mbf w$.
% The Z-score is computed via 
The victim uses decoder $D$ to recover $\hat{\mbf w}$ from a suspect structure $\mathbf{X}$ and counts the $k$ matching bits between $\hat{\mbf w}$ and the original watermark $\mbf w$. A one-sided Z-test assesses significance against random coincidence using $z=(k-\mu)/\sigma$,
% where $\mu$ and $\sigma$ respectively denotes the expectation of matching bits and the standard deviation.
% For a Bernoulli distribution with $p=0.5$, $\mu = 0.5L$, and $\sigma = 0.5 \sqrt{L}$.
% Then, the victim compares $z$ with the threshold $\hat{z}$.
% In the One-Sided Z-Test, $z \geq 3.0$ indicates that $P \leq 10^{-3}$, which is the false positive rate.
% We follow the settings of the current studies and set $\hat{z} = 3.0$.
where $\mu=0.5l$ and $\sigma=0.5\sqrt{l}$ are the mean and standard deviation for a random binary sequence of length $l$ with $p=0.5$. The threshold $\hat{z}=3.0$ gives a false positive rate of $\leq10^{-3}$. MARCO sets $d_{\text{wm}}=1$ if $z\geq\hat{z}$ and $0$ otherwise, returning both $d_{\text{wm}}$ and $\hat{\mbf w}$.

\section{Evaluation}
The experimental settings and the other experiments about robustness, radioactivity, robustness and ablation study are presented in \autoref{sect:exp_settings} and \autoref{sect:additional_exp}. 
% Check this anonymous repository\footnote{\url{https://anonymous.4open.science/r/MARCO-0116}} for our code. 

\begin{figure*}[t!]
    % \captionsetup{labelfont={color=blue}, textfont={color=blue}}
    \centering
    
    % --- Row 1 ---
    \begin{subfigure}[b]{0.4\textwidth}
        \centering
        \includegraphics[width=\linewidth]{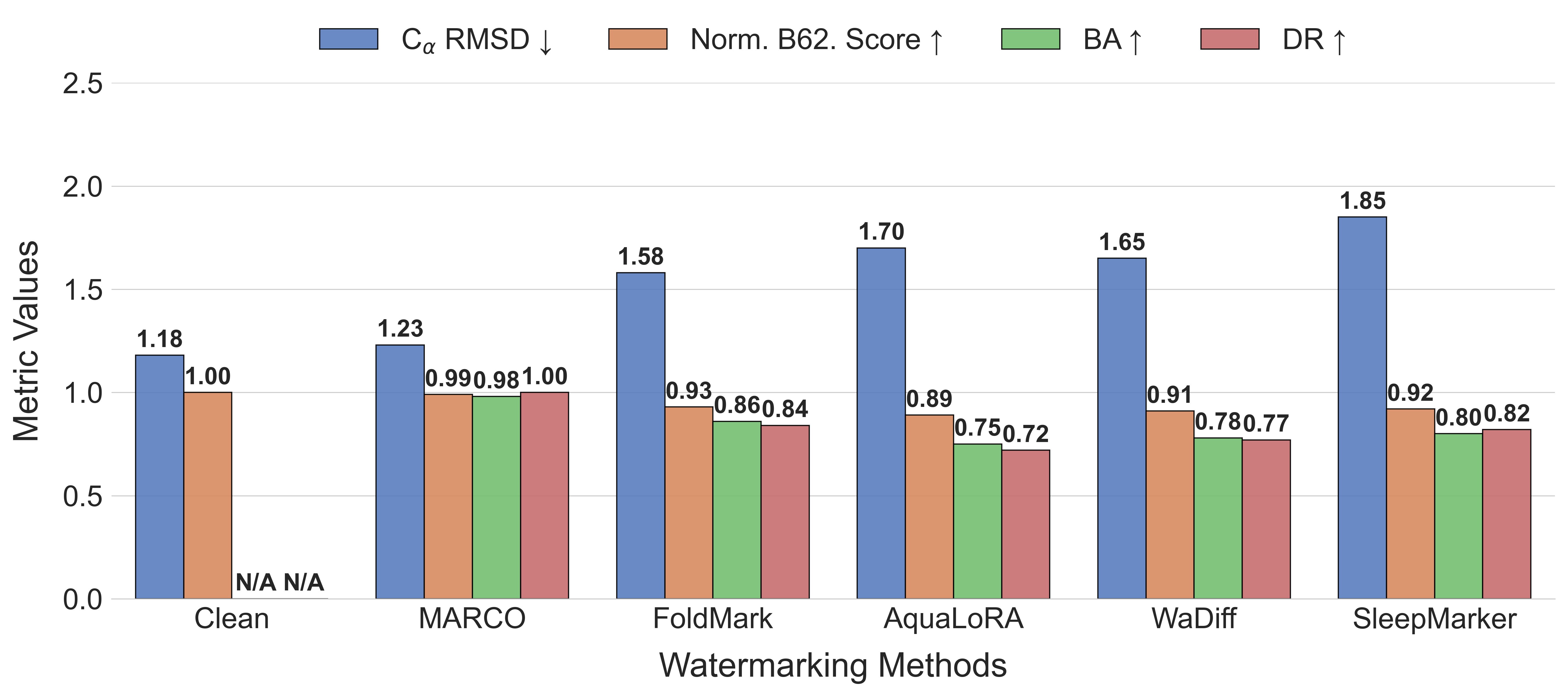}
        \caption{$\Theta$: RFDiffusion}
        \label{fig:exp_fid_rbn_rfdiffusion}
    \end{subfigure}
    \hspace{10pt}
    \begin{subfigure}[b]{0.4\textwidth}
        \centering
        \includegraphics[width=\linewidth]{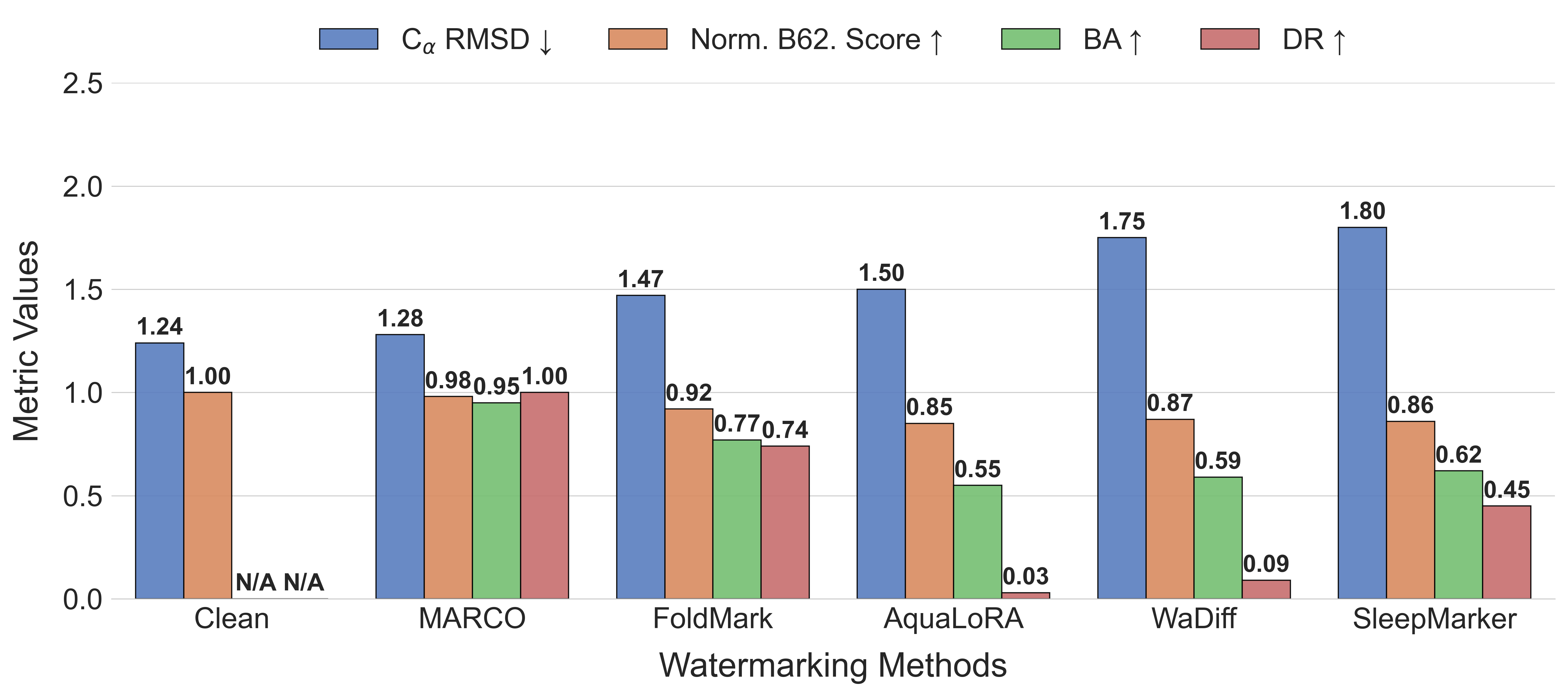}
        \caption{$\Theta$: ESMFold}
        \label{fig:exp_fid_rbn_esmfold}
    \end{subfigure}
    
    \par\bigskip % Force a new row with some vertical spacing
    
    % --- Row 2 ---
    \begin{subfigure}[b]{0.4\textwidth}
        \centering
        \includegraphics[width=\linewidth]{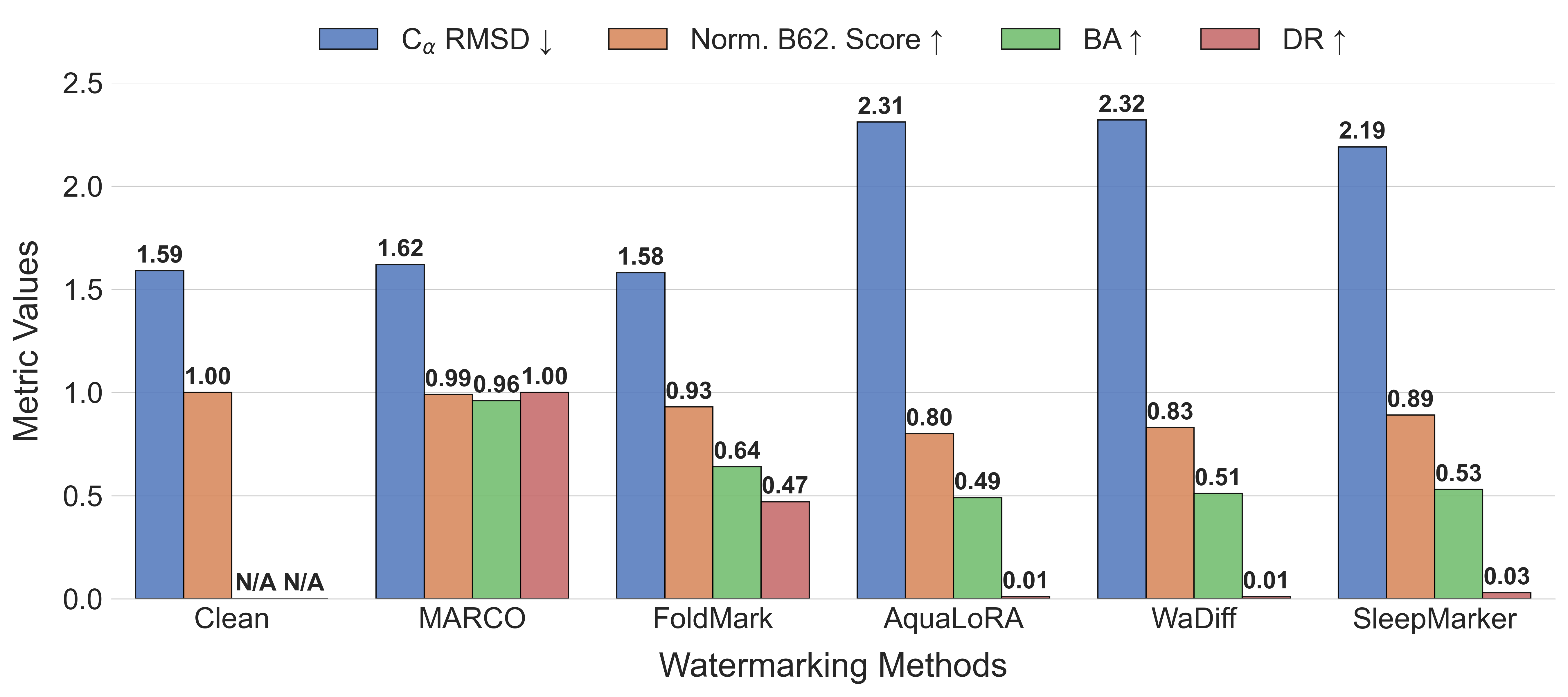}
        \caption{$\Theta$: Chroma}
        \label{fig:exp_fid_rbn_chroma}
    \end{subfigure}
    \hspace{10pt}
    \begin{subfigure}[b]{0.4\textwidth}
        \centering
        \includegraphics[width=\linewidth]{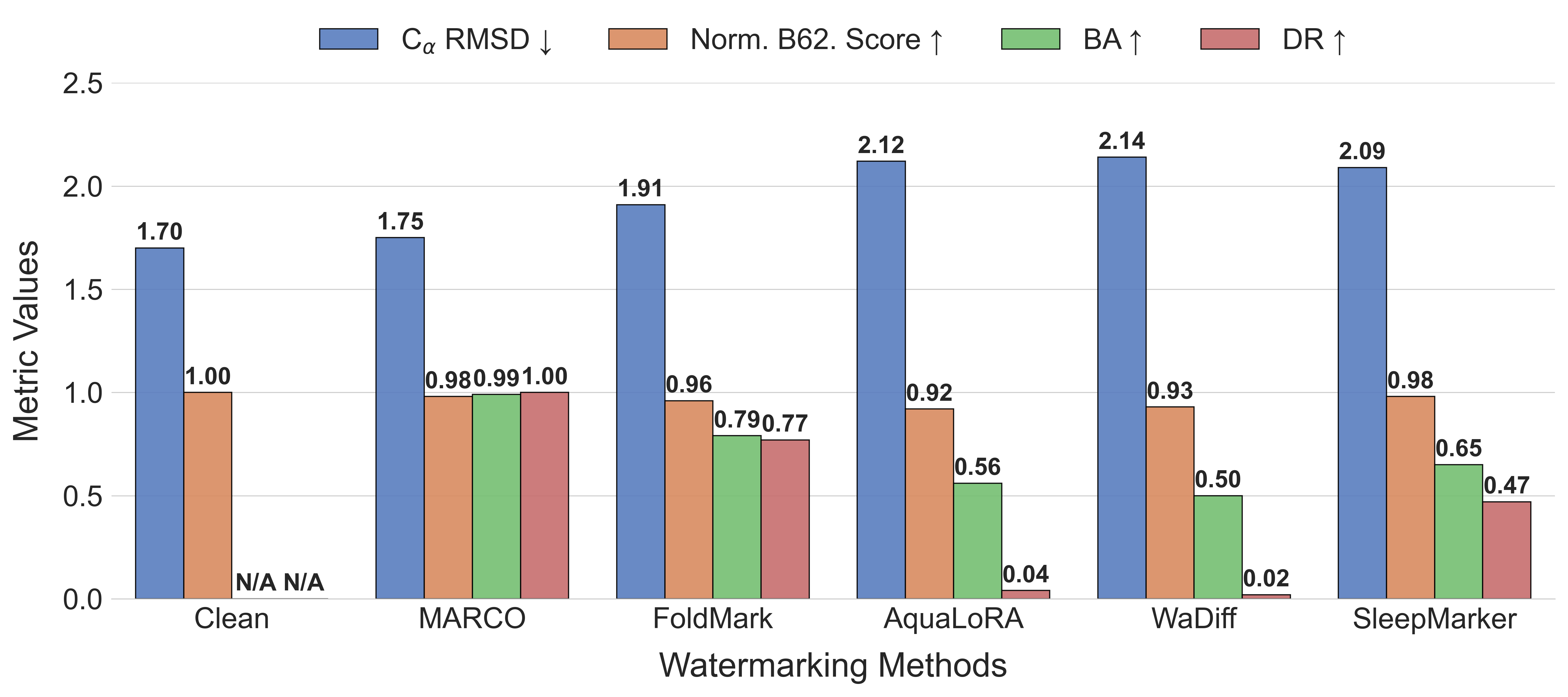}
        \caption{$\Theta$: FrameDiff}
        \label{fig:exp_fid_rbn_framediff}
    \end{subfigure}
    
    \par\bigskip % Force a new row
    
    % --- Row 4 ---
    \begin{subfigure}[b]{0.4\textwidth}
        \centering
        \includegraphics[width=\linewidth]{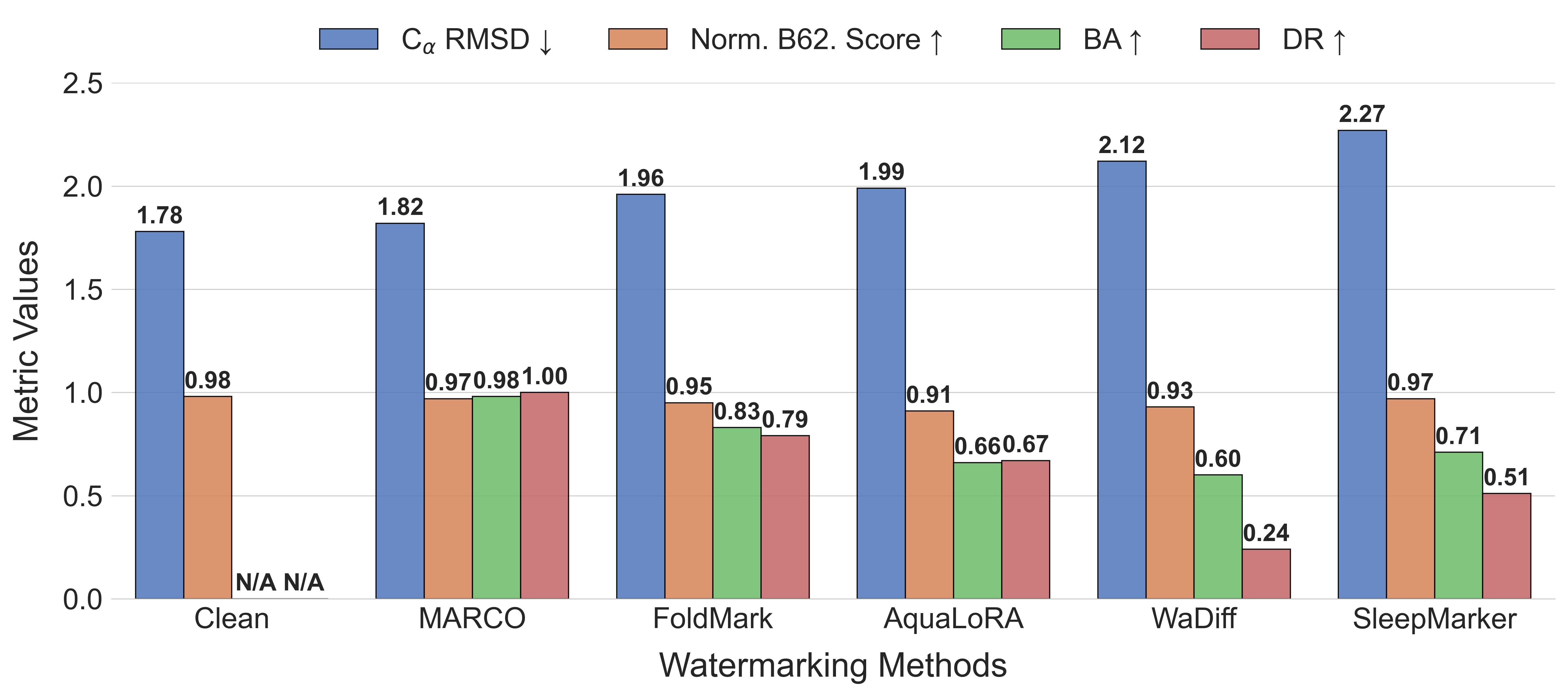}
        \caption{$\Theta$: FrameFlow}
        \label{fig:exp_fid_rbn_frameflow}
    \end{subfigure}
    \hspace{10pt}
    \begin{subfigure}[b]{0.4\textwidth}
        \centering
        \includegraphics[width=\linewidth]{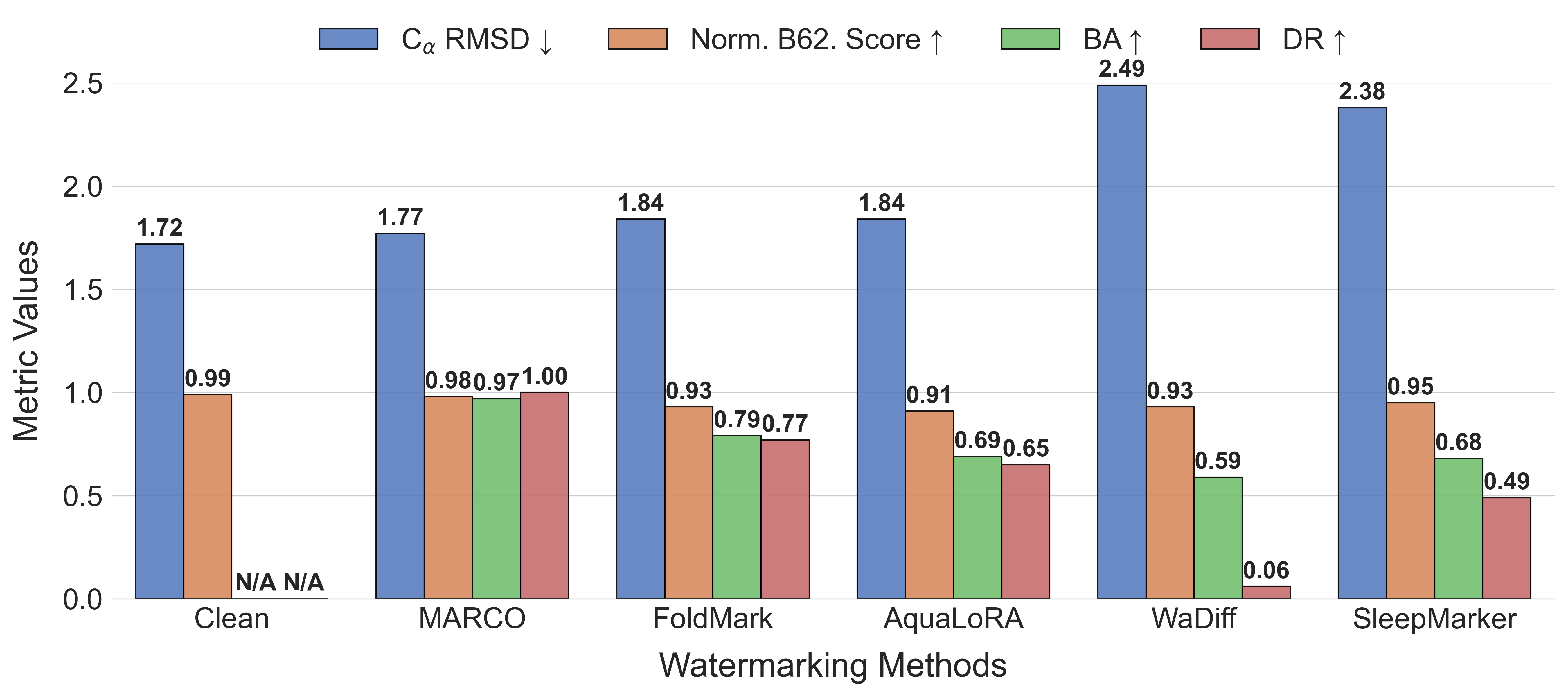}
        \caption{$\Theta$: FoldFlow2}
        \label{fig:exp_fid_rbn_foldflow2}
    \end{subfigure}
    
    \caption{\textbf{Evaluation of Fidelity of PGM Watermarking Methods}. 
    % MARCO surpasses the other four baselines in both fidelity and robustness because of the loss terms designed for maintaining conformational homeostasis.
    }
    \label{fig:exp_fid_rbn}
\end{figure*}

\subsection{Fidelity}
We compare MARCO with four baselines using $C_\alpha$-RMSD, Normalized BLOSUM62 score, Bit Accuracy (BA), and Detection Rate (DR) across six pre-trained PGMs with official weights and published baseline hyperparameters. We set the watermark length to $l=64$, providing a $2^{64}$ identification space for robust validation, and generate $1000$ clean--watermarked structure pairs $(\hat{\mathbf{X}},\tilde{\mathbf{X}})$. Results compare these structures and the ground-truth and recovered watermarks $(\mbf w,\hat{\mbf w})$.

As shown in \autoref{fig:exp_fid_rbn}, MARCO leads most metrics, achieving the lowest RMSD, near-perfect BLOSUM62 scores, BA $>0.95$, and DR $=1.0$ without attacks. Image-domain baselines lack protein-specific thermodynamic constraints and therefore compromise fidelity and robustness by distorting molecular conformations. FoldMark is the strongest baseline, but MARCO's design performs better. Moreover, the baselines require PGM fine-tuning, whereas MARCO is a plug-and-play module that leaves the backbone model unchanged. Their partial watermark recovery also remains substantially less accurate than MARCO's, especially under adversarial attacks.
% We also visualize several clean predicted structures along with their corresponding watermarked structures in \autoref{fig:exp_visualization}, where the watermarked structures highly align to the clean predictions.
% Here, we use RFDiffusion as the PGM to generate the clean predictions.
% This indicates that MARCO embeds its watermark into the clean predictions without significantly tampering them, keeping the conformation of the watermarked structure at its finest.
The RFDiffusion overlays in \autoref{fig:exp_visualization} (\autoref{sect:additional_exp}) show strong alignment between clean and watermarked structures, confirming that MARCO preserves native conformations with minimal geometric alteration.

% \prg{Metric}
% \begin{itemize}
%     \item Structure Similarity: $C_\alpha$r.m.s.d.
%     \item Sequence Similarity
%     \item Bit Accuracy
%     \item Detection Rate
% \end{itemize}

% \prg{Models}
% \begin{itemize}
%     \item RFDiffusion
%     \item RFDiffusionAA
%     \item AlphaFold2
%     \item ESMFold
%     \item Chroma
%     \item FrameDiff
%     \item FrameFlow
%     \item FoldFlow2
% \end{itemize}

% \prg{Baselines}
% \begin{itemize}
%     \item WaDiff
%     \item AquaLoRA
% \end{itemize}
% \todo

\begin{figure*}[t!] % 'p' encourages placing this tall figure on its own page
    % \captionsetup{labelfont={color=blue}, textfont={color=blue}}
    \centering
    
    % --- Row 1 ---
    \begin{subfigure}[b]{0.32\textwidth}
        \centering
        \includegraphics[width=\linewidth]{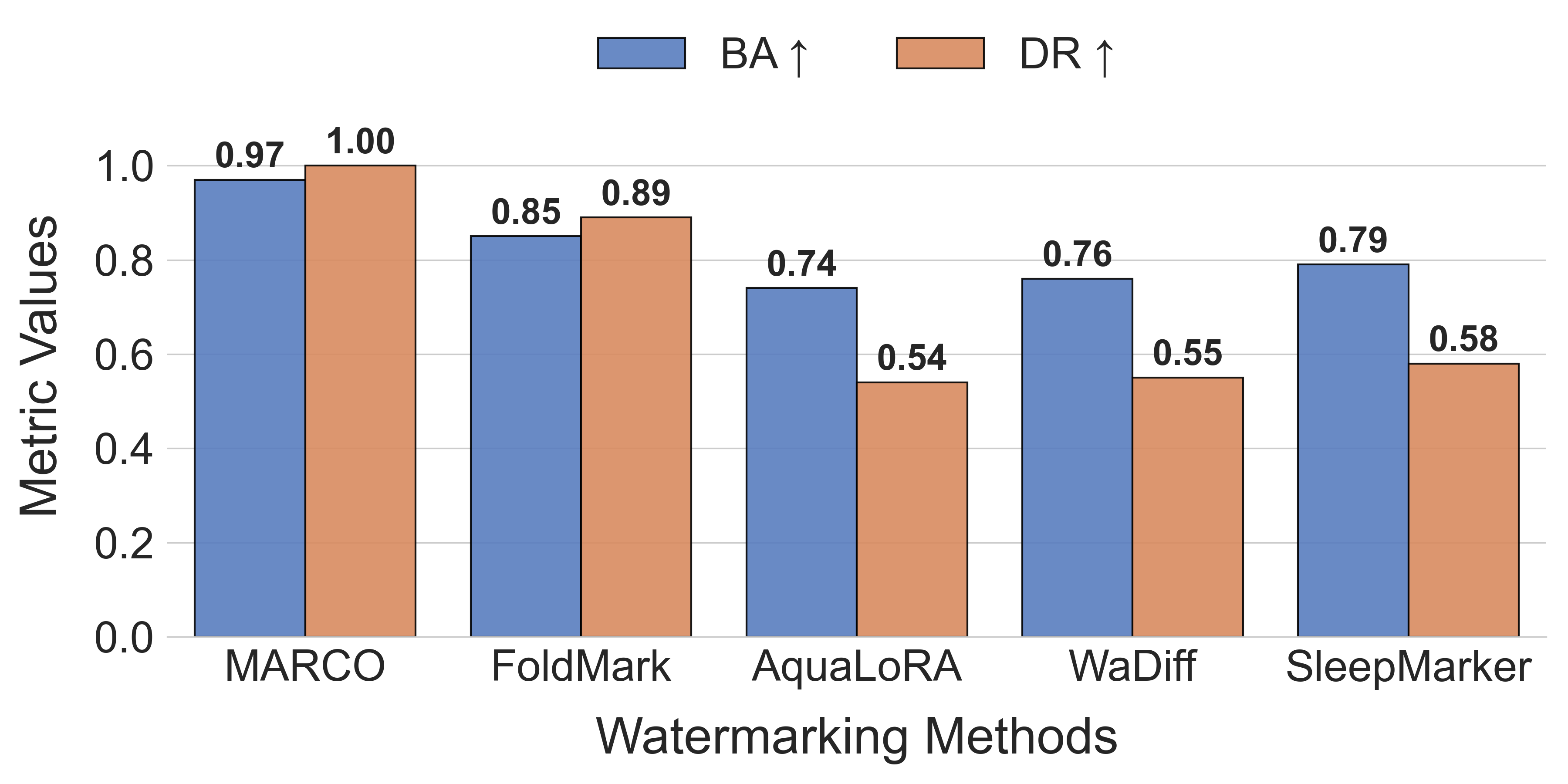}
        \caption{$\Theta:$ RFDiff, Rotation}
    \end{subfigure}
    \hfill
    \begin{subfigure}[b]{0.32\textwidth}
        \centering
        \includegraphics[width=\linewidth]{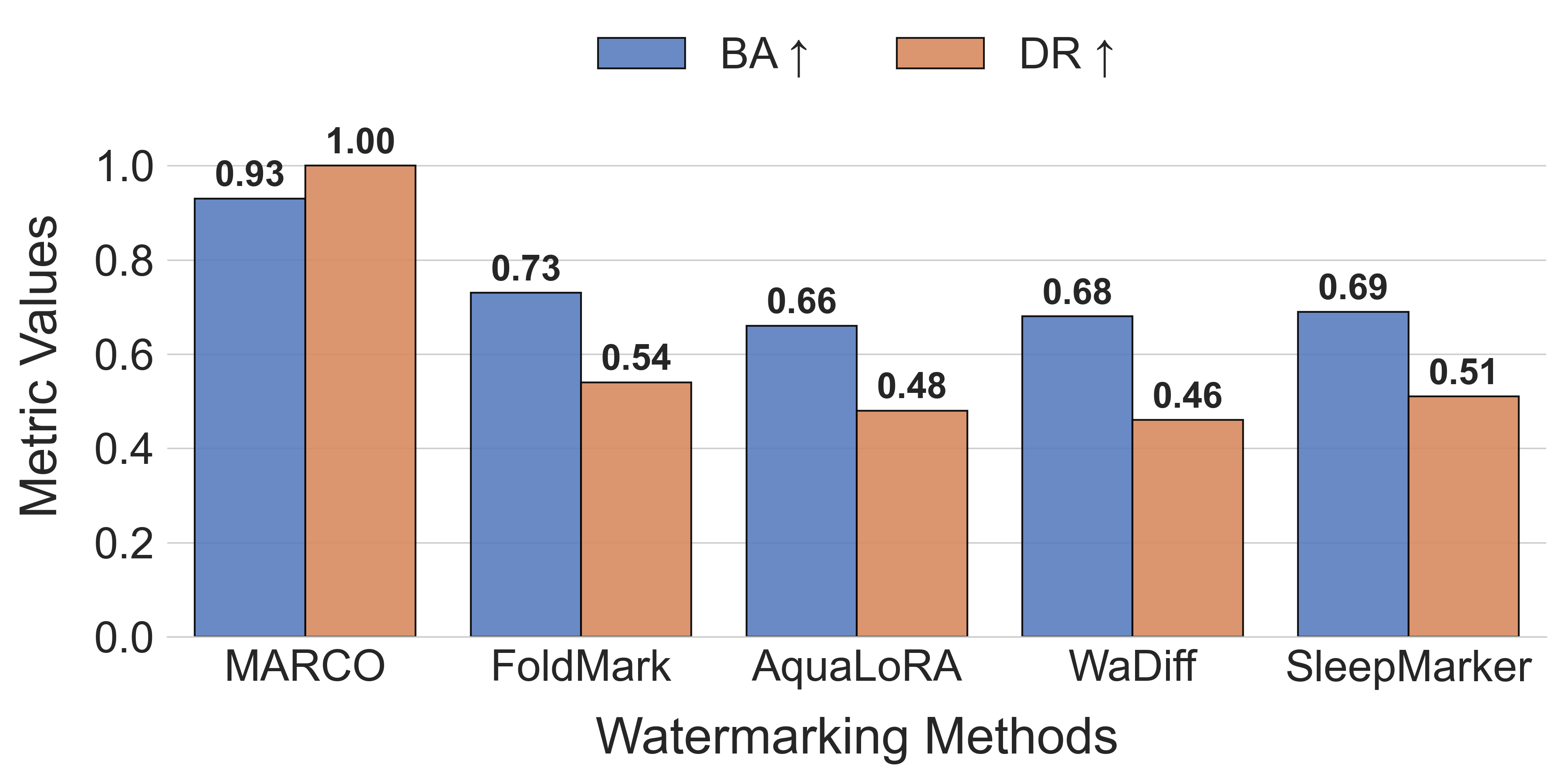}
        \caption{$\Theta:$ RFDiff, Cropping}
    \end{subfigure}
    \hfill
    \begin{subfigure}[b]{0.32\textwidth}
        \centering
        \includegraphics[width=\linewidth]{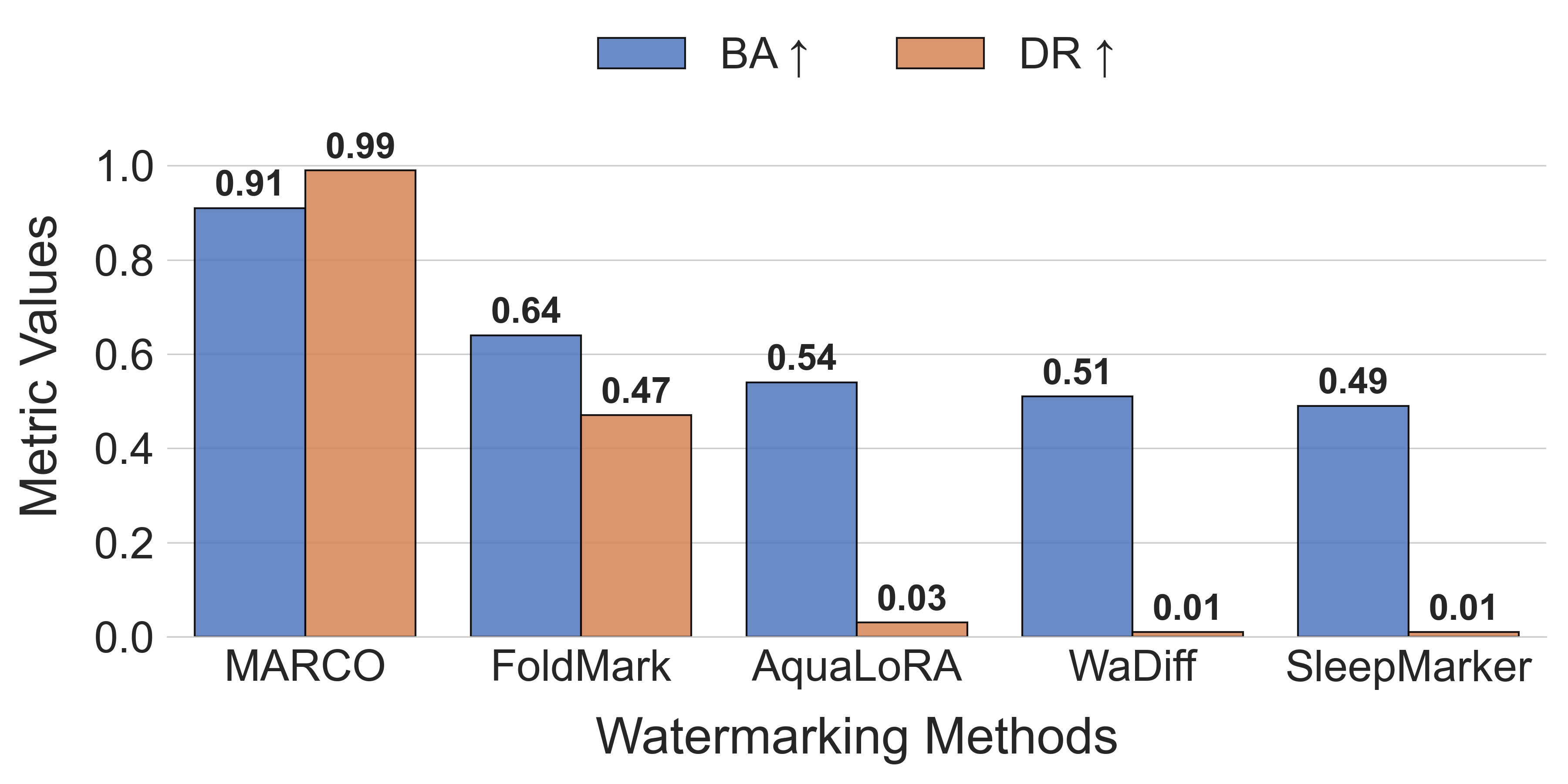}
        \caption{$\Theta:$ RFDiff, Noise Addition}
    \end{subfigure}
    
    \par\medskip % New row
    
    % --- Row 2 ---
    \begin{subfigure}[b]{0.32\textwidth}
        \centering
        \includegraphics[width=\linewidth]{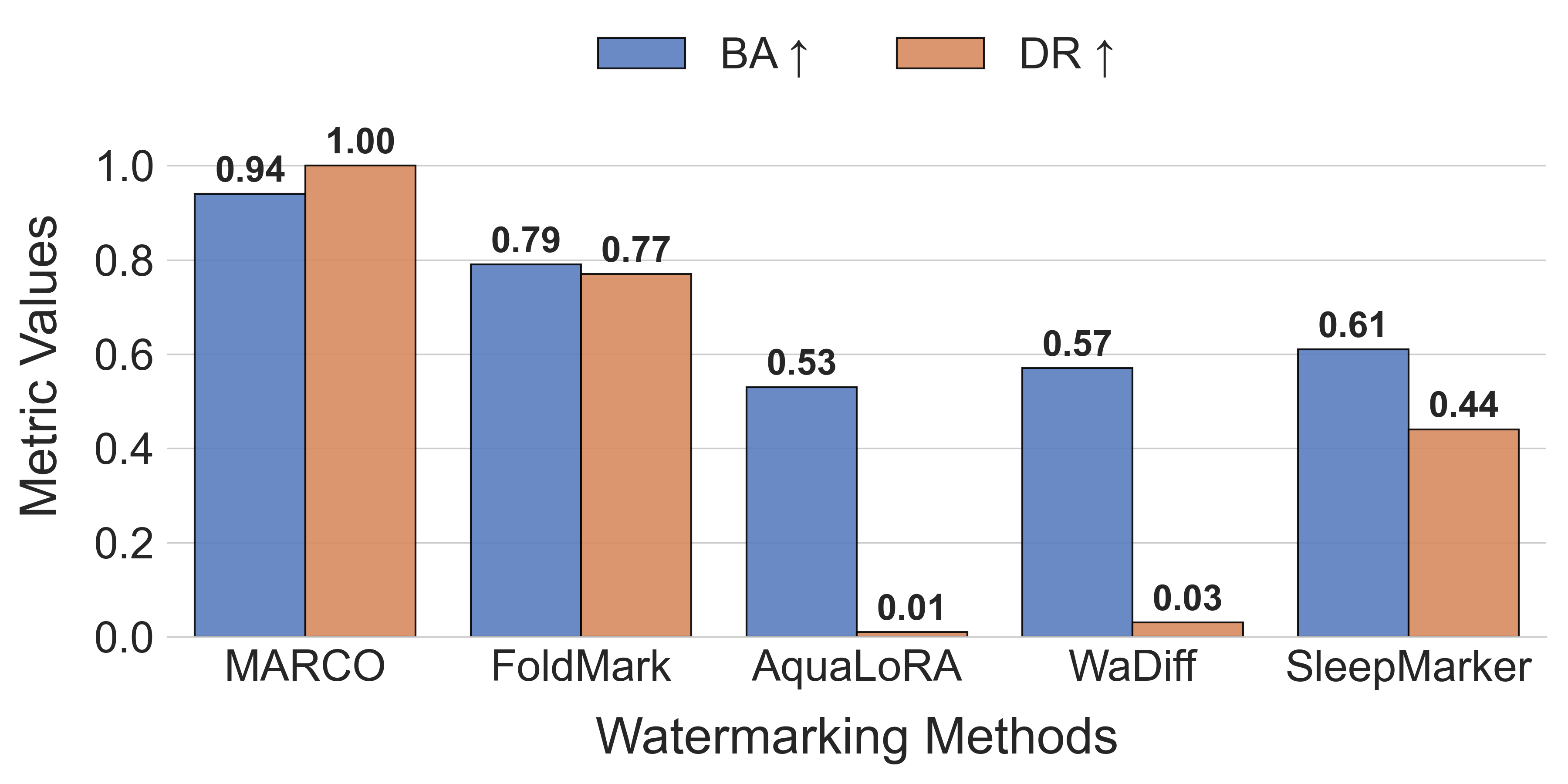}
        \caption{$\Theta:$ ESMFold, Rotation}
    \end{subfigure}
    \hfill
    \begin{subfigure}[b]{0.32\textwidth}
        \centering
        \includegraphics[width=\linewidth]{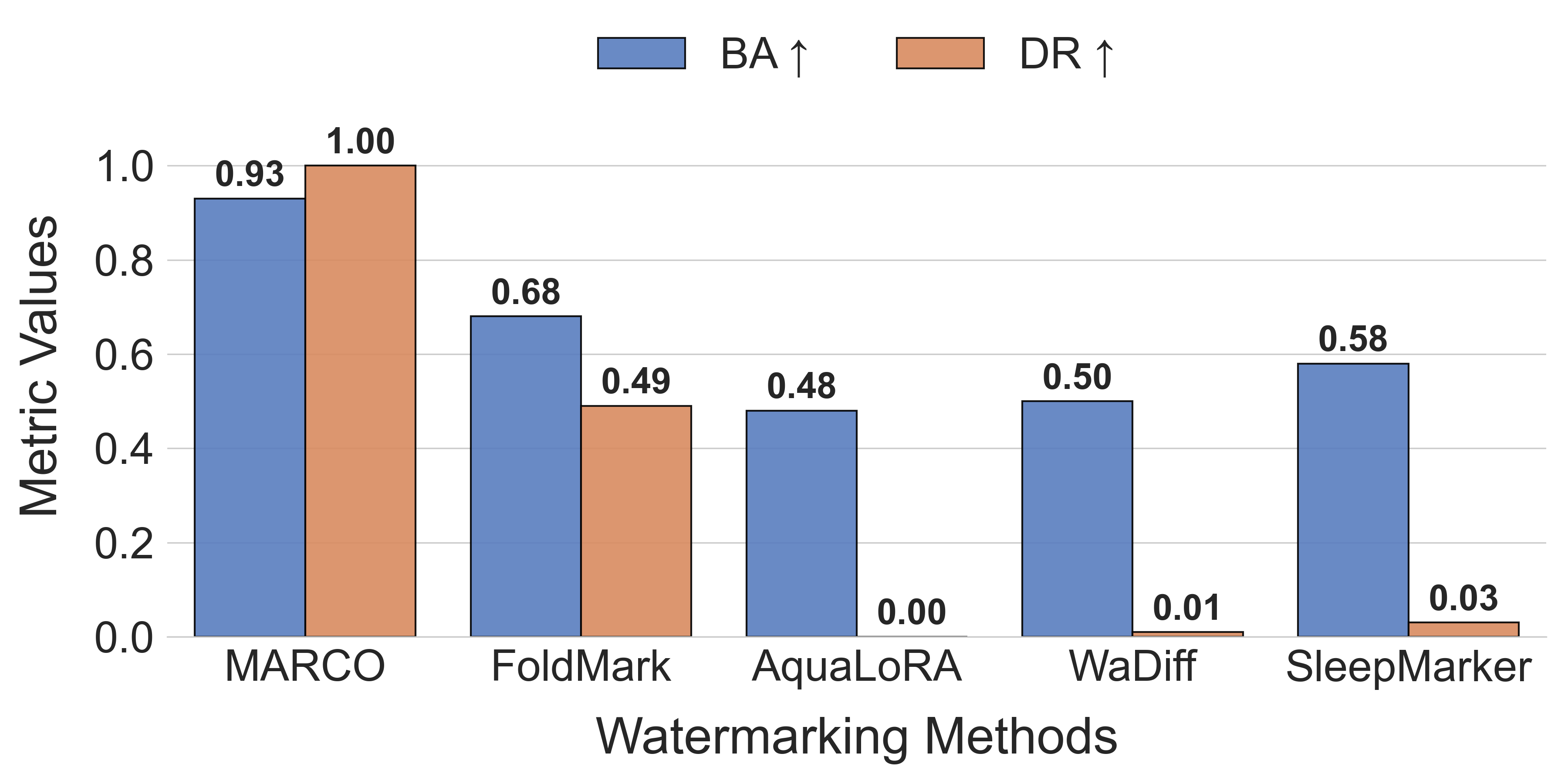}
        \caption{$\Theta:$ ESMFold, Cropping}
    \end{subfigure}
    \hfill
    \begin{subfigure}[b]{0.32\textwidth}
        \centering
        \includegraphics[width=\linewidth]{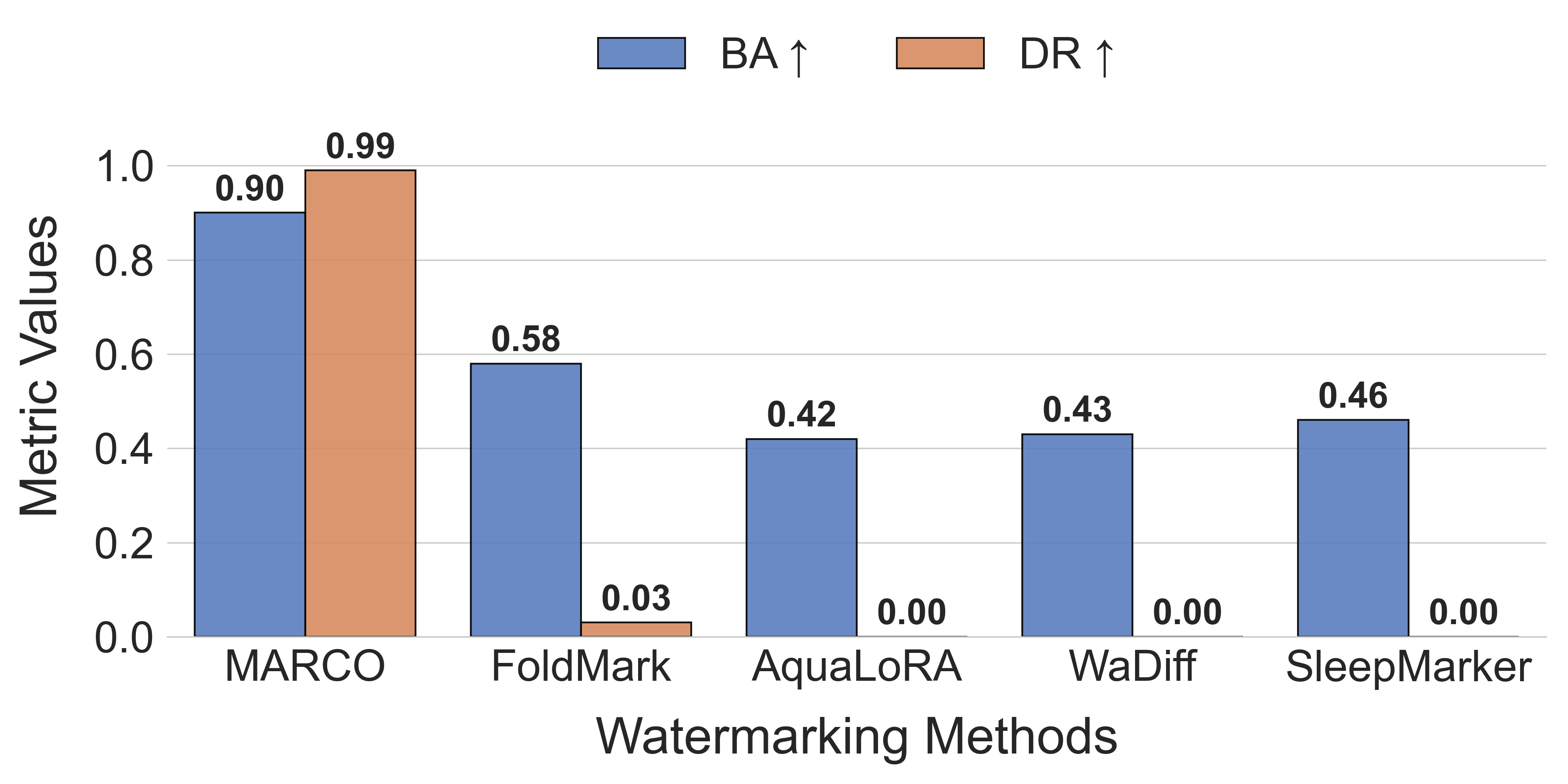}
        \caption{$\Theta:$ ESMFold, Noise Addition}
    \end{subfigure}

    \par\medskip % New row

    % --- Row 3 ---
    \begin{subfigure}[b]{0.32\textwidth}
        \centering
        \includegraphics[width=\linewidth]{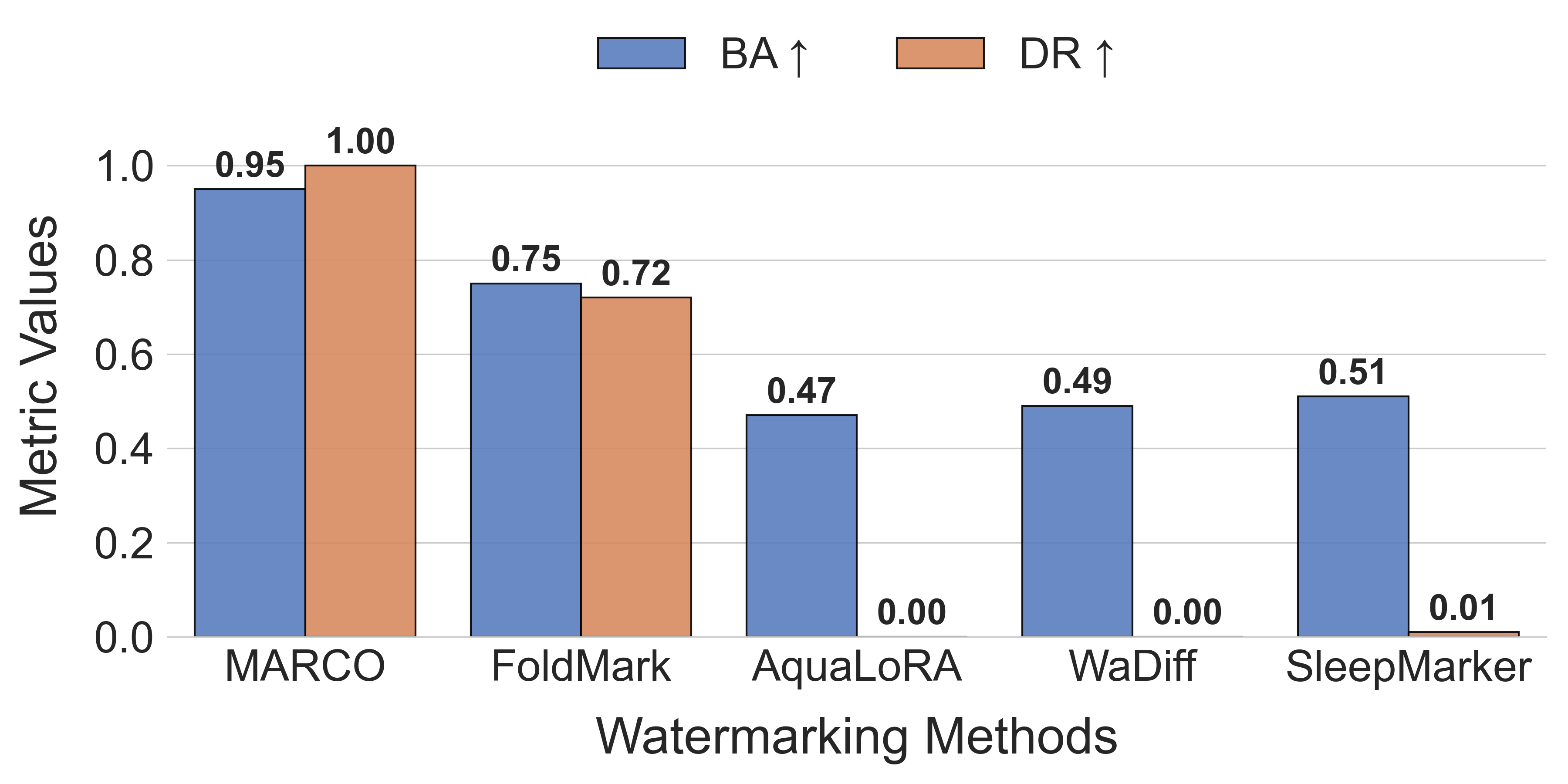}
        \caption{$\Theta:$ Chroma, Rotation}
    \end{subfigure}
    \hfill
    \begin{subfigure}[b]{0.32\textwidth}
        \centering
        \includegraphics[width=\linewidth]{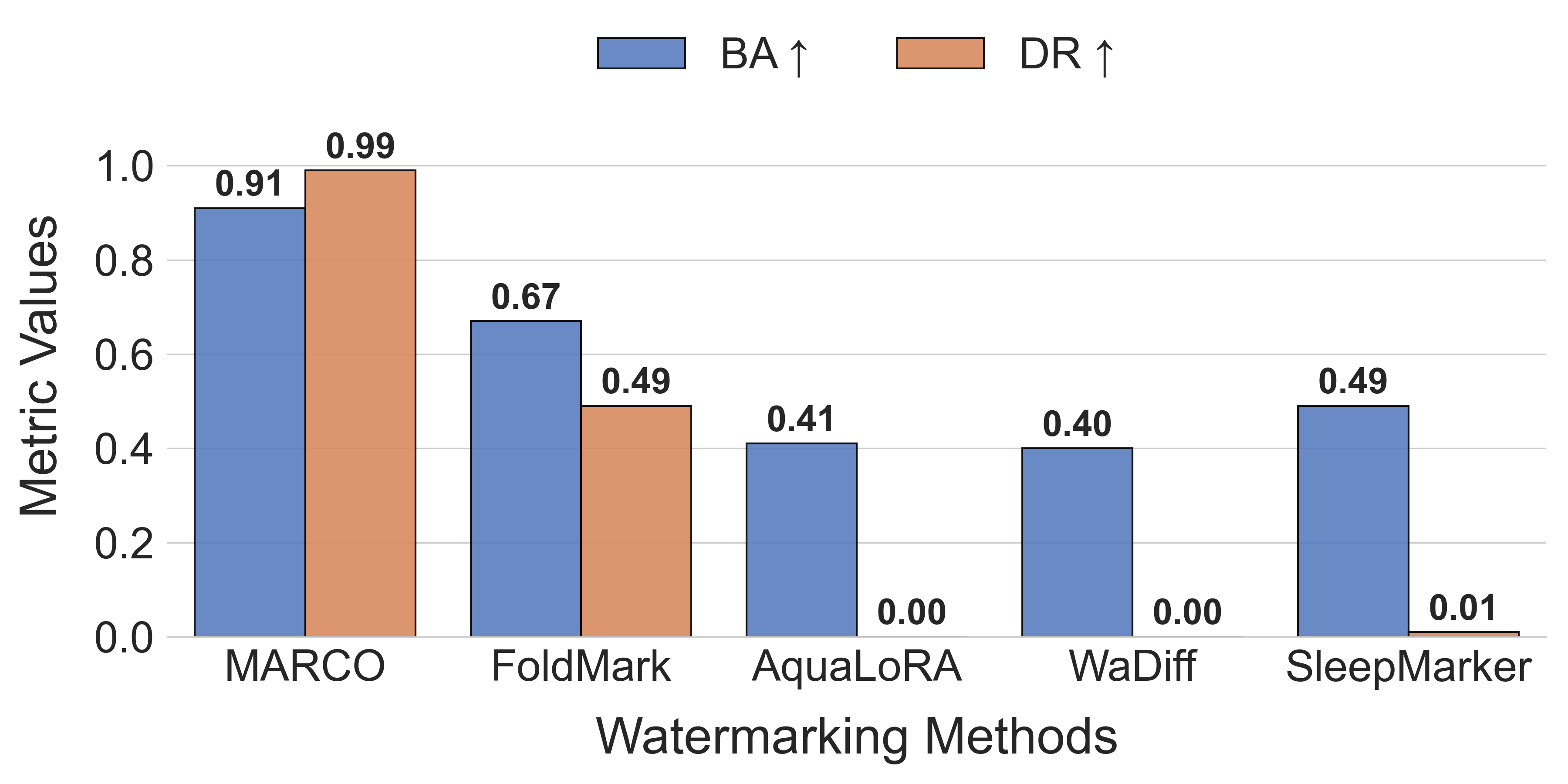}
        \caption{$\Theta:$ Chroma, Cropping}
    \end{subfigure}
    \hfill
    \begin{subfigure}[b]{0.32\textwidth}
        \centering
        \includegraphics[width=\linewidth]{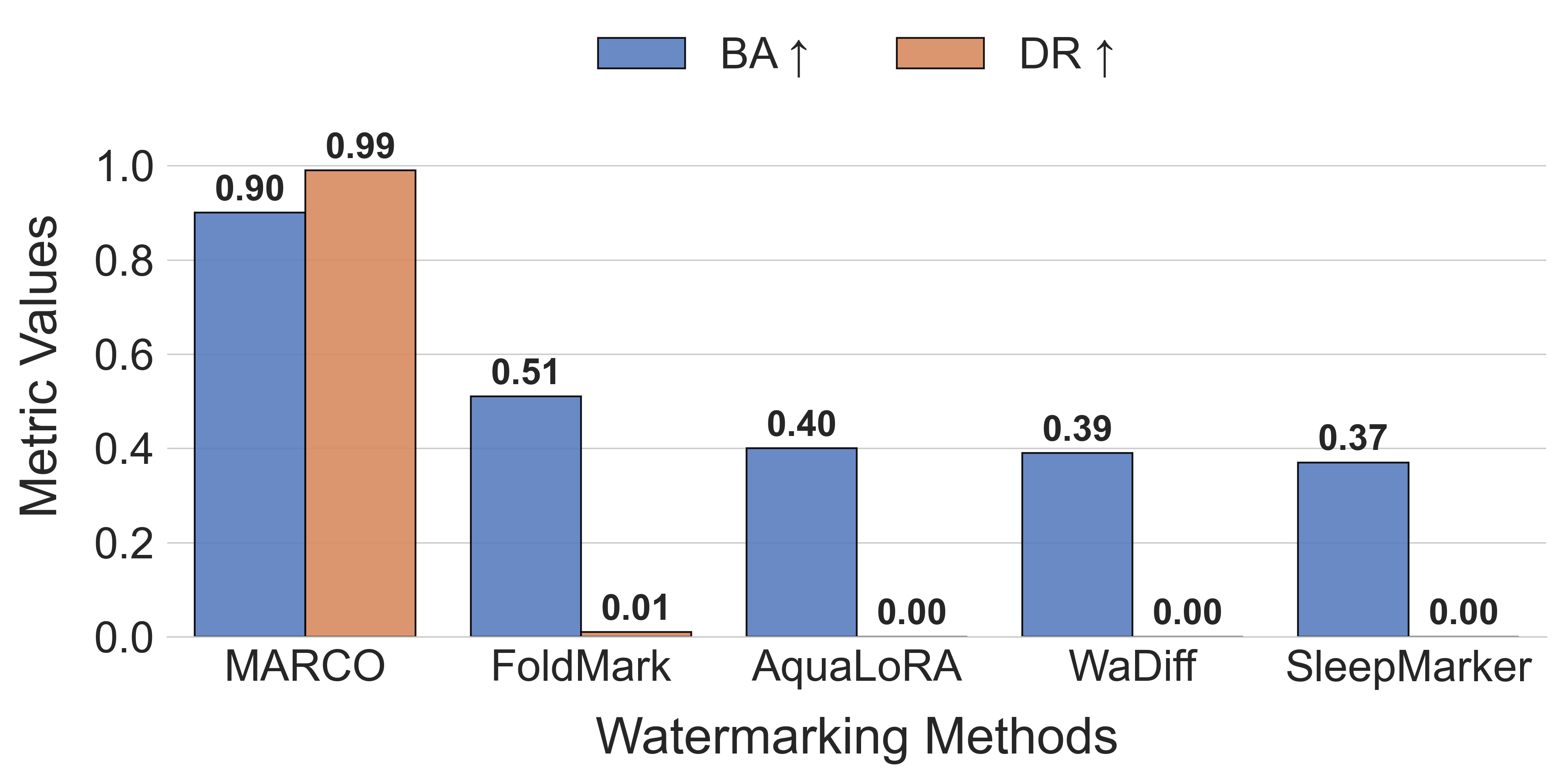}
        \caption{$\Theta:$ Chroma, Noise Addition}
    \end{subfigure}

    \caption{Part A -- \textbf{Evaluation of Robustness of PGM Watermarking Methods against Conformation Manipulations}.
    % After conformation manipulations, MARCO survives and outperforms the other four baselines with impressive robustness.
    }
    \label{fig:exp_rbn_atk_A}
\end{figure*}

\subsection{Robustness}
\prginit{MARCO vs. Conformational Manipulations}
We evaluate MARCO against random rotation, cropping, and Gaussian noise injection. Random Rotation: For $\mathbf{X}=\{(\mathbf{R}_i,\mathbf{x}_i)\}_{i=1}^N$, we sample a global rotation $\mathbf{Q}\sim\text{Haar}(SO(3))$, the uniform distribution over $SO(3)$, and apply $\mathbf{x}'_i=\mathbf{Q}(\mathbf{x}_i-\mathbf{c})+\mathbf{c}$ and $\mathbf{R}'_i=\mathbf{Q}\mathbf{R}_i$, where $\mathbf{c}$ is the geometric center. Cropping: We remove a contiguous $30\%$ of the backbone from either end. Gaussian Noise Injection: For $\mathbf{x}\in\mathbb{R}^{N\times3}$, we sample $\chi\sim\mathcal{N}(\mathbf{0},\mathbf{I})$, $\chi\in\mathbb{R}^{N\times3}$, and add it to $\mathbf{x}$.

As shown in \autoref{fig:exp_rbn_atk_A} and \autoref{fig:exp_rbn_atk_B} (\autoref{sect:additional_exp}), MARCO significantly outperforms all four baselines in Bit Accuracy (BA) and Detection Rate (DR), maintaining BA $>0.90$ and DR $>0.99$ under every perturbation.
Rotation has relatively little effect, whereas cropping and noise sharply degrade the baselines by removing structural information and globally distorting molecular conformations, respectively. MARCO's robustness suggests that the attack layer is critical for learning persistent features; adversarial training likely also diffuses the watermark throughout the structure to evade the discriminator and prevent localized failures.

\section{Discussion}
PGMs face economic and biosecurity threats: extracting proprietary models causes economic loss, while bypassing safeguards to synthesize toxins or bioweapons risks severe societal harm. MARCO addresses both with a dual-layer defense that embeds user-specific intellectual property as a watermark, serving as both a Digital Rights Management (DRM) tool and a Biological Weapons Convention (BWC)-compliant digital chain of custody for PGM accountability.
MARCO balances fidelity, preserved by geometric constraints, with robustness, enhanced by the attack layer and extraction loss; adversarial training reinforces both. Adjusting the loss weights and probability of conformational manipulations tailors this trade-off to specific security requirements.
Capacity remains a performance constraint: the ablation study shows that increasing watermark length $l$ reduces fidelity and robustness. Choosing an optimal $l$ therefore balances scalable user identification with structural integrity.
Model extraction attacks aim to learn a PGM's output distribution. MARCO makes this distribution radioactive through stepwise reverse-diffusion perturbations, and cross-architecture training experiments confirm that the watermark propagates to pirate models. This radioactivity is essential for identifying stolen models and validating ownership. In \autoref{sect:future_work}, we discuss potential extensions of MARCO to protein language models and wet-lab verification.

% \lipsum[1-7]

\section{Conclusion}
% We propose MARCO, a radioactive protein generative model watermark that can be efficiently trained to adapt to mainstream protein generative models without modifying their weights.
% MARCO achieves both high fidelity and robustness by generating attack-resisting watermarked structures with negligible differences with the clean predicted structures.
% Through the translation of ProteinMPNN, it is shown that MARCO also has negligible influences on the protein sequence.
% The watermarked structure can survive malicious attacks such as rotation, cropping and noise addition, where the recovered watermarks mostly align with the original watermark.
% MARCO also has impressive capacity that supports long watermark bit string embedding and extraction without sacrificing too much performance.
% For potential future direction, we focus on further enhancing the conformational fidelity and making the watermark method adaptive to more protein generative models.
We propose MARCO, a radioactive watermark designed to efficiently adapt to mainstream Protein Generative Models without modifying their pre-trained weights. MARCO achieves a critical balance of fidelity and robustness, generating attack-resilient structures with negligible geometric deviation from clean predictions. Sequence validation via ProteinMPNN confirms that these structural watermarks exert minimal influence on the recovered amino acid sequence. Furthermore, the watermark proves robust against conformational manipulations and even adaptive attacks while supporting high-capacity payloads with minimal performance trade-offs. Future work will focus on further enhancing conformational fidelity and extending the framework's applicability to a broader spectrum of generative architectures and other biological modalities.

\section*{AI Use Statement}
In the preparation of this manuscript, we employed generative AI tools strictly for the purpose of linguistic refinement. Specifically, these tools were used to check for grammatical errors and to improve the flow and readability of the text.
We affirm that:
\begin{itemize}
    \item \textbf{No Content Generation:} The core concepts, experimental design, data analysis, and scientific conclusions are entirely the work of the authors.

    \item \textbf{No Synthetic Text:} No substantial portions of the text were generated by AI. The tools acted solely as a copy-editing assistant.

    \item \textbf{Human Oversight:} All suggestions made by the AI were manually reviewed and verified by the authors, who take full responsibility for the accuracy and integrity of the final publication.
\end{itemize}

\section*{Ethics Statement}
In conducting this research, we strictly adhered to a ``no-harm'' principle. Our methodology was designed to protect the copyrights of the model owner and trace biosecurity misuses of protein generative models in a purely theoretical and isolated context, ensuring that no individuals, organizations, or operational systems were negatively impacted.

\prg{Isolation from Operational Systems}
All experiments described in this paper were performed in a fully offline, local environment. We utilized open-source implementations of dataset distillation protocols and deployed them on our own hardware. At no point did our method interact with, query, or stress-test live commercial APIs or external web services. Consequently, this research caused no service disruption, latency, or financial cost to any model providers or platforms.

\prg{Use of Public, Non-Sensitive Data}
The evaluation was conducted exclusively using standard, publicly available academic datasets. We did not harvest data from private users, nor did we target specific content creators. By relying solely on established benchmarks, we ensured that no personal data was processed, and no individual's privacy or intellectual property rights were infringed upon during the course of this study.

\prg{Safe Security Assessment}
Our work treats the sensitive information of the protein datasets and the protein generative models as a mathematical artifact to be analyzed, rather than attacking the infrastructure that hosts it. By confining our ``red-teaming'' efforts to a sandbox environment, we demonstrate that security vulnerabilities can be identified and documented without risk to the broader digital ecosystem or its stakeholders.

\section*{Reproducibility Statement}
To promote transparency and reproducibility, we will release artifacts including all the code of all empirical studies employed in this paper. 
We currently provide the artifact in the following anonymized link:
% \url{https://anonymous.4open.science/r/MARCO-0116}.
\url{https://to_be_released_after_acceptance}.
All datasets we use are included in the artifact as they are public accessible.

\bibliographystyle{iclr2027_conference}
\bibliography{ref}

\clearpage
\appendix
\section*{{\huge Appendix}}

% \lipsum[1]

\section{Design Properties}\label{sect:design_properties}
% MARCO possesses the following properties defined by us that are essentially required by any watermarking methods for PGMs:
We define the following core properties as essential prerequisites for any effective PGM watermarking framework. MARCO is explicitly designed to satisfy these constraints:

\textbf{Fidelity} 
% measures the extent to which the watermarked proteins retain the structural quality and biological function of the original distribution. We require that the divergence between the predicted and watermarked distributions is bounded, and that the watermarked structures occupy low-energy states:
quantifies the preservation of structural integrity and biological function. A viable watermark must ensure that the modified protein structures remain indistinguishable from the original distribution in terms of biophysical plausibility. Mathematically, we require that the watermarked samples $\tilde{\mathbf{X}}$ reside within comparable low-energy states to the clean predictions $\hat{\mathbf{X}}$:
\begin{equation}\label{eq:fidelity}
\begin{aligned}
\mathbb{E}_{\mathbf{X}_w} [\mathcal{E}_{\text{phys}}(\tilde{\mbf X})] \approx \mathbb{E}_{\mathbf{X}} [\mathcal{E}_{\text{phys}}(\hat{\mathbf{X}})],
\end{aligned}
\end{equation}
% where $\mathcal{E}_{\text{phys}}$ represents a physical energy function.
where $\mathcal{E}_{\text{phys}}$ denotes a validated physical energy potential.
Specifically, \autoref{eq:rfdiff_loss} is an example of defining fidelity in detail.
    
\textbf{Robustness} 
% quantifies the watermarking scheme's ability to preserve the embedded payload against adversarial attempts to modify the protein structure. A watermarking scheme is considered robust if, for any utility-preserving attack $\mathcal{A}$, the recovered watermark $\hat{m}$ aligns bit-by-bit with the original watermark $m$, minimizing the Bit Error Rate (BER):
evaluates the resilience of the embedded signal against adversarial structural perturbations. A watermarking scheme is deemed robust if the watermark $\mbf w$ remains reliably recoverable even after the protein undergoes any utility-preserving transformation or attack $\mathcal{A} \in \mathcal{T}$ (e.g., rigid body rotation, noise injection, or energy minimization).  Mathematically, we require that the Bit Error Rate (BER) of the recovered signal remains strictly bounded by a tolerance threshold $\rho$:
\begin{equation}\label{eq:robustness}
\begin{aligned}
    \sup_{\mathcal{A} \in \mathcal{T}} \text{BER}\Big(\mbf w, \, D\big(\mathcal{A}(\tilde{\mathbf{X}})\big)\Big) \le \rho
\end{aligned}
\end{equation}
where $D$ represents the decoding function, and $\text{BER}(\mbf w, \hat{\mbf w}) = \frac{1}{l} \sum_{i=1}^l \mathbbm{1}_{(\mbf w_i \neq \hat{\mbf w}_i)}$ measures the fraction of mismatched bits. 
% This ensures that the provenance information remains readable and intact even after the protein has undergone transformations like sequence redesign or molecular dynamics relaxation.

\textbf{Radioactivity} 
% ensures that if an adversary trains a piracy model $\Upsilon$ on a dataset $\mcl C_m = \{\tilde{\mbf X}^{(i)}\}_{i=1}^N$ consisting of watermarked structures, the watermark signal transfers to the weights of the new model. Consequently, the structures generated by this piracy model, $\mathbf{X}_{\text{out}} \sim \Upsilon(\cdot)$, should also exhibit the watermark signal:
guarantees the persistent propagation of the watermark through model extraction attacks. It ensures that if an adversary trains a pirate model $\Upsilon$ using a dataset $\mcl C_m = \{\tilde{\mathbf{X}}^{(i)}\}_{i=1}^N$ derived from watermarked structures, the signature is effectively encoded into the surrogate's parameters.  Consequently, any structure $\mathbf{X}_{\text{out}}$ generated by this derivative model $\Upsilon(\cdot)$ must inherently carry the watermark signal, satisfying the condition:
\begin{equation}\label{eq:radioactivity}
\begin{aligned}
    \sup \text{BER}\Big(\mbf w, D(\mathbf{X}_{\text{out}})\bigg) \leq \rho
\end{aligned}
\end{equation}
    
\textbf{Covertness} 
% requires that the watermark be statistically invisible to an adversary. 
% The watermarked distribution $p_{m}$ should be indistinguishable from the clean model distribution $p_\theta$ to a discriminator $\mcl C$:
mandates that the embedded watermark remains statistically imperceptible to adversarial inspection. The watermarked distribution $p_{m}$ must be effectively indistinguishable from the clean model distribution $p_\theta$. Formally, this requires minimizing the divergence between the two distributions as measured by an optimal discriminator $\mcl C$:
\begin{equation}\label{eq:covertness}
\begin{aligned}
    \max_{\mcl C} \left| \mathbb{E}_{\hat{\mathbf{X}} \sim p_{\theta}} [\mcl C(\hat{\mathbf{X}})] - \mathbb{E}_{\tilde{\mbf X} \sim p_{m}} [\mcl C(\tilde{\mbf X})] \right| \approx 0
\end{aligned}
\end{equation}
% For proteins, this implies that the distribution of bond lengths, torsion angles, and residue packing densities must match natural biological statistics.
In the context of protein structures, this implies that key geometric statistics, namely bond lengths, torsion angles, and residue packing densities, must strictly adhere to natural biological distributions, thereby preventing detection via statistical anomaly analysis.
    
\textbf{Capacity} requires the length of the watermark must be sufficient to carry the identification information.

\textbf{Efficiency} requires that the watermark embedding and extraction processes must not consume too much time.

\textbf{Generalizability} ensures that the watermarking algorithm is agnostic to specific model architectures and protein topologies.

% \begin{table}[t!]
% \centering
% \caption{\textbf{Primary Notations}}
% \resizebox{.6\linewidth}{!}{
% \begin{tabular}{c|l}
% \toprule
%     Symbols & Definitions\\
% \midrule
%     $\mbf X$ & The original protein structure\\

%     \rowcolor{Gray}$\hat{\mbf X}, \tilde{{\mbf X}}$ & The predicted/watermarked protein structure\\

%     $\mbf R$ & The original rotation matrix\\

%     \rowcolor{Gray}$\hat{\mbf R}, \tilde{\mbf R}$ & The predicted/watermarked rotation matrix\\

%     $\mbf x, \hat{\mbf x}, \tilde{\mbf x}$ & The original/predicted/watermarked $C_\alpha$ coordinate\\

%     \rowcolor{Gray}$N$ & The length of the protein sequence\\

%     $T, t$ & The denoising step number/timestep\\

%     \rowcolor{Gray}$\Theta,\Upsilon$ & The original/pirate protein generative model\\

%     $E, D, \mcl C$ & The encoder/decoder/discriminator\\

%     \rowcolor{Gray}$\mcl L$ & The loss function\\

%     $d_{\text{wm}}$ & The watermark detection boolean indicator\\
    
%     \rowcolor{Gray}$\mbf w, \hat{\mbf w}$ & The original/recovered watermark\\

%     $l$ & The length of the watermark\\

%     \rowcolor{Gray}$\eta$ & The random noise added to the $C_\alpha$ coordinate\\

%     $\xi$ & The random noise added to the Rotation Matrix\\
    
% \bottomrule
% \end{tabular}
% }
% \label{tab:notations}
% \end{table}

\section{Certified Robustness via Randomized Smoothing}
\label{sec:certified_robustness}

While we empirically demonstrate robustness against various attacks, we further provide a theoretical guarantee for watermark persistence using Randomized Smoothing. This framework certifies that the watermark decoder $D$ will reliably recover the correct bit string $m$ under any adversarial perturbation $\delta$ bounded by a radius $r$ (in terms of $L_2$ norm or Root Mean Square Deviation (RMSD)).

Let $D_i: \mathbb{R}^{3N} \to \{0, 1\}$ be the base decoder for the $i$-th watermark bit. We define a smoothed classifier $g_i$ as the majority vote of $D_i$ under isotropic Gaussian noise $\mathcal{N}(0, \sigma^2 I)$:
\begin{equation}
    g_i(X) = \underset{c \in \{0, 1\}}{\mathrm{argmax}} \; \mathbb{P}_{\epsilon \sim \mathcal{N}(0, \sigma^2 I)} (D_i(X + \epsilon) = c)
\end{equation}

\noindent \textbf{Theorem 1 (Certified Radius).} \textit{Let $p_A$ be the probability of the most likely class (the correct bit) predicted by $D_i$ under noise distribution $\mathcal{N}(0, \sigma^2 I)$, and let $p_B = 1 - p_A$. If $p_A > 0.5$, then $g_i(X)$ is provably robust against any perturbation $\delta$ satisfying $||\delta||_2 < r$, where:}
\begin{equation}
    r = \frac{\sigma}{2} (\Theta^{-1}(p_A) - \Theta^{-1}(p_B)), 
\end{equation}
\textit{where $\Theta^{-1}$ is the inverse standard normal CDF.}

\prg{Connection to RMSD}
In the context of protein structures with $N$ atoms, an $L_2$-bounded perturbation relates to the RMSD as $\text{RMSD} = ||\delta||_2 / \sqrt{N}$. Therefore, for a certified $L_2$ radius $r$, MARCO guarantees that the watermark bit $m_i$ will not flip for any attack that modifies the structure within an RMSD threshold of:
\begin{equation}
    R_{\text{RMSD}} = \frac{\sigma}{2\sqrt{N}} (\Theta^{-1}(p_A) - \Theta^{-1}(1 - p_A))
\end{equation}

This theoretical bound confirms that the robustness of MARCO is not merely an artifact of specific attack simulations, but an intrinsic geometric property enforced by the noise-aware training of the decoder.

\section{Experiment Settings}\label{sect:exp_settings}
\prg{Pre-trained Protein Generative Models}
% We evaluate our proposed method across a diverse set of state-of-the-art protein generative models, ranging from structure prediction networks to de novo backbone generators.
We evaluate our proposed method across a diverse suite of state-of-the-art protein generative models, ranging from structure prediction networks to de novo backbone generators.
\begin{itemize}[leftmargin=*]
\item \textbf{RFDiffusion}\cite{watson2023RFDiffusion}
% \footnote{https://github.com/RosettaCommons/RFdiffusion}
: A leading method for de novo protein design that combines the RoseTTAFold structure prediction network with a Denoising Diffusion Probabilistic Model (DDPM) framework. It is widely used for designing motif scaffolds and ligand binders.

\item \textbf{ESMFold}
% \footnote{https://github.com/facebookresearch/esm}
\cite{lin2022ESMFold}: A high-speed structure prediction model based on the ESM protein language model. Unlike AlphaFold, it generates atomic-level structures end-to-end directly from sequence data without requiring explicit Multiple Sequence Alignment (MSA) retrieval, making it approximately ten times faster than MSA-based methods.

\item \textbf{Chroma}
% \footnote{https://github.com/chroma-core/chroma}
\cite{Chroma2023}: A diffusion-based generative model designed to enable the creation of novel programmable protein structures and functions, pushing the boundaries of de novo design beyond natural protein space.

\item \textbf{FrameDiff}
% \footnote{https://github.com/Immortals-33/Framediff}
\cite{yim2023FrameDiff}: An $SE(3)$-invariant diffusion model specifically designed for protein backbone generation. It operates on rigid bodies (frames) in 3D space rather than just point clouds, allowing it to generate designable protein monomers up to 500 amino acids long without relying on pre-trained folding networks.

\item \textbf{FrameFlow}
% \footnote{https://github.com/microsoft/protein-frame-flow} 
\cite{yim2024FrameFlow}: A flow-matching model that adapts the FrameDiff architecture to a deterministic generative paradigm. By utilizing $SE(3)$ flow matching, it significantly reduces the number of sampling steps required for inference while maintaining the geometric rigor of frame-based updates9.

\item \textbf{FoldFlow2}
% \footnote{https://github.com/DreamFold/FoldFlow} 
\cite{huguet2024FoldFlow2}: A state-of-the-art generative model that integrates sequence encoding (from ESM-2) with structure encoding (via flow matching) in a multi-modal trunk. Trained on a dataset eight times larger than its predecessor, it effectively addresses the interdependence of sequence and structure to generate diverse and high-quality proteins.
\end{itemize}

\prg{Datasets}
% We follow RFDiffusion \cite{watson2023RFDiffusion} to utilize data from the Protein Data Bank \cite{berman2000PDB} (PDB)\footnote{https://www.rcsb.org/}. The dataset is a subset of monomers PDB, where the sequence length ranges from $60$ to $512$ residues;
% the structure resolution is better than $5\Angstrom$.
% Proteins with over $50\%$ loops are excluded.
% The dataset eventually contains $20,312$ proteins.
Following the data curation protocols of RFdiffusion \cite{watson2023RFDiffusion}, we utilize a subset of monomeric structures derived from the Protein Data Bank (PDB) \cite{berman2000PDB}. The inclusion criteria require sequence lengths ranging from 60 to 512 residues and a structural resolution finer than $5\text{\AA}$. Additionally, structures exhibiting greater than 50\% loop content are excluded to ensure structural integrity. The final curated dataset comprises 20,312 protein entries.

\prg{Implementation}
% For the encoder, the decoder and the discriminator in MARCO, we adopt a 6-layer-depth model architecture modified from FrameDiff \cite{yim2023FrameDiff}.
% This architecture comprises Invariant Point Attention (IPA) \cite{Jumper2021AlphaFold2}, Transformer blocks \cite{vaswani2017Transformer} and MLP layers.
% The IPA processes spatial relationships while maintaining rotational invariance. The number of hidden channels for IPA is set to 16.
% The Transformer blocks consist of standard transformer layers that follow the IPA module to process the node embeddings.
% The MLP Layers consist of 3 linear layers with ReLU activation and LayerNorm, and are used for node and edge updates.
For the encoder, decoder, and discriminator modules within MARCO, we adopt a 6-layer architecture adapted from FrameDiff \cite{yim2023FrameDiff}. This architecture integrates Invariant Point Attention (IPA) \cite{Jumper2021AlphaFold2}, Transformer blocks \cite{vaswani2017Transformer}, and Multi-Layer Perceptrons (MLP).
The IPA module, configured with 16 hidden channels, processes spatial relationships while strictly preserving rotational invariance. Following this, Transformer blocks process the node embeddings to capture global contextual dependencies. Finally, the MLP blocks, composed of three linear layers with ReLU activation and LayerNorm, perform the final node and edge feature updates.

% The models are trained using the Adam optimizer with a learning rate of $0.0001$.
% The batch size is set to $64$.
% The training process consists of $50$ epochs, where the weights of the PGM are fixed.
% The weights of the encoder, the decoder, and the discriminator are the objects to optimize.
% The training is complete on a server having $8$ NVIDIA A100 GPUs with $80$GB VRAM each.
Model training is conducted using the Adam optimizer with a learning rate of $10^{-4}$ and a batch size of 64. The training protocol spans 50 epochs, during which the weights of the pre-trained PGM remain frozen; optimization is exclusively applied to the encoder, decoder, and discriminator parameters. All experiments were executed on a high-performance server equipped with 8 NVIDIA A100 GPUs (80GB VRAM each).
% \todo{Mention how $\lambda$s are determined by grid search.}
The weight parameters $\lambda$s are determined by grid search in pre-experiments.

\prg{Watermark Baselines}
To provide a rigorous comparative analysis, we adapt state-of-the-art watermarking methods originally developed for image diffusion models (e.g., Stable Diffusion) to the protein domain.
\begin{itemize}[leftmargin=*]
    \item \textbf{AquaLoRA}~\cite{feng2024AquaLora} embeds watermarks directly into the diffusion U-Net via a Low-Rank Adaptation (LoRA) module using a two-stage process: Latent Watermark Pre-training and Prior Preserving Fine-Tuning.
    
    \item \textbf{WaDiff}~\cite{min2024WaDiff} integrates user-specific watermarks directly into the diffusion sampling process. The method functions as a ``watermark-conditioned'' diffusion model, where the watermark is treated as a conditioned input by expanding the model's input channels.
    
    \item \textbf{SleeperMark}~\cite{wang2025SleeperMark} explicitly disentangles watermark information from semantic concepts, guiding the model to embed a multi-bit message only when triggered, while regular prompts yield watermark-free, high-fidelity outputs. 
    % The architecture involves pre-training a latent secret encoder and optimizing the diffusion backbone using a specialized loss to steer denoising trajectories.

    \item \textbf{FoldMark}~\cite{zhang2025foldmark} trains an encoder-decoder and fine-tunes each PGM using watermark-conditioned LoRA, evolutionary information, and structural consistency losses to make the output carry the watermark.
\end{itemize}
We follow the hyperparameter settings in the original papers.

\prg{Metrics}
For the two most important properties: fidelity and robustness, we use the following metrics to measure them:

\begin{itemize}[leftmargin=*]
    \item \textbf{Structure Similarity}: 
    Root Mean Square Deviation (RMSD) is the standard measure of the average distance between the atoms of superimposed proteins.
    \begin{equation}\label{eq:RMSD}
    \begin{aligned}
        \text{RMSD} = \sqrt{\frac{1}{N} \sum_{i=1}^{N} d_i^2},
    \end{aligned}
    \end{equation}
    where $N$ is the number of atoms and $d_i$ is the distance between the $i$-th pair of corresponding atoms.
    A lower RMSD indicates a better structural similarity, and a value less than $2.0 \Angstrom$ is usually considered a very close match.

    \item \textbf{Sequence Similarity}: 
    Instead of directly computing match over no match, we employ BLOSUM62 \cite{henikoff1992BLOSUM62} to compute a substitution matrix to score aligned pairs between the original sequence and the translated sequence derived by feeding the watermarked structure into the ProteinMPNN \cite{dauparas2022ProteinMPNN}.
    In short, for identical pairs or mutations that are not going to significantly affect the protein structure, BLOSUM62 gives positive scores, whereas negative scores are given to the mutations significantly altering the protein structure.

    However, raw BLOSUM62 scores scale with the length of the protein, where a longer protein yields a higher score. 
    Thus, they cannot be directly compared across different pairs.
    To measure similarity between different pairs, we normalize the score into a range $[0, 1]$.
    We employ the Geometric Mean Normalization to derive the normalized BLOSUM62 score
    \begin{equation}\label{eq:norm_BLOSUM62}
    \begin{aligned}
        S_{\text{norm}}(\mbf S_a, \mbf S_b) = \frac{S(\mbf S_a, \mbf S_b)}{\sqrt{S(\mbf S_a, \mbf S_a) \times S(\mbf S_b, \mbf S_b)}},
    \end{aligned}
    \end{equation}
    where $S(\cdot, \cdot)$ denotes the raw BLOSUM62 score between two protein sequences. A high $S_{\text{norm}}(\mbf S_a, \mbf S_b)$ indicates that the two sequences $\mbf S_a$ and $\mbf S_b$ are biologically similar, whereas a low value means the opposite.

    \item \textbf{Watermark Bit Accuracy}: 
    We simply count the number of matching bits between the original and the recovered watermarks, and divide it by the watermark length to get the Bit Accuracy (BA):
    \begin{equation}\label{eq:bit_acc}
    \begin{aligned}
        \text{BA} = \frac{\text{CountMatch}(\mbf w, \hat{\mbf w})}{l}
    \end{aligned}
    \end{equation}

    \item \textbf{Watermark Detection Rate}: 
    We compute the Z-score as described in \autoref{sect:marco_eval}, and if $z \geq 3$ (BA $\geq 0.6875$), it is considered a successful detection.
    The detection rate (DR) is derived by dividing the number of successful detections by the total number of detections.

\end{itemize}

\section{Additional Experiments}\label{sect:additional_exp}

\begin{figure*}[t!] % 'p' encourages placing this tall figure on its own page
    \centering
    
    % --- Row 1 ---
    \begin{subfigure}[b]{0.32\textwidth}
        \centering
        \includegraphics[width=\linewidth]{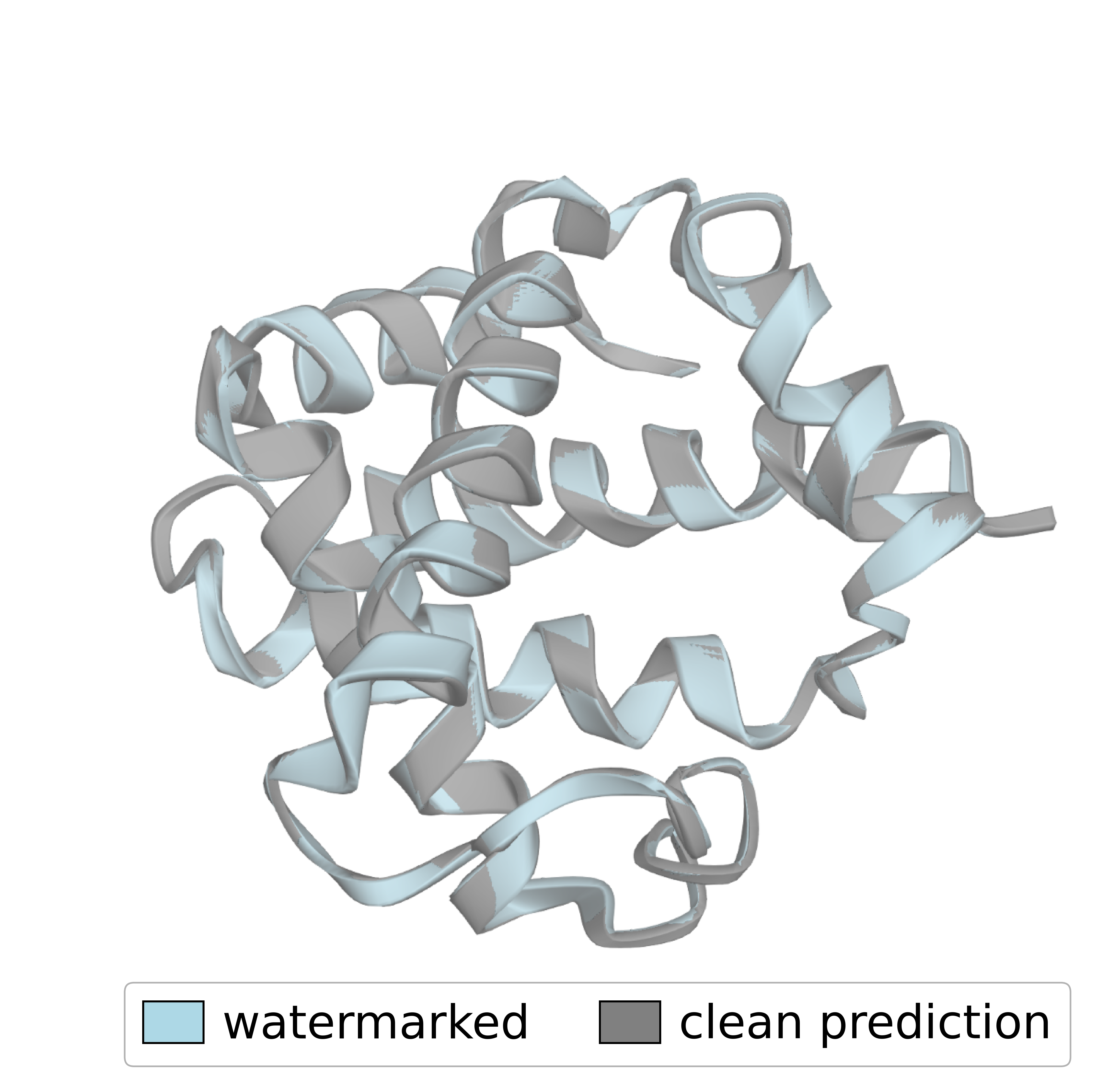}
        \caption{{\scriptsize 1A3N -- Deoxy Human Hemoglobin}}
    \end{subfigure}
    \hfill
    \begin{subfigure}[b]{0.32\textwidth}
        \centering
        \includegraphics[width=\linewidth]{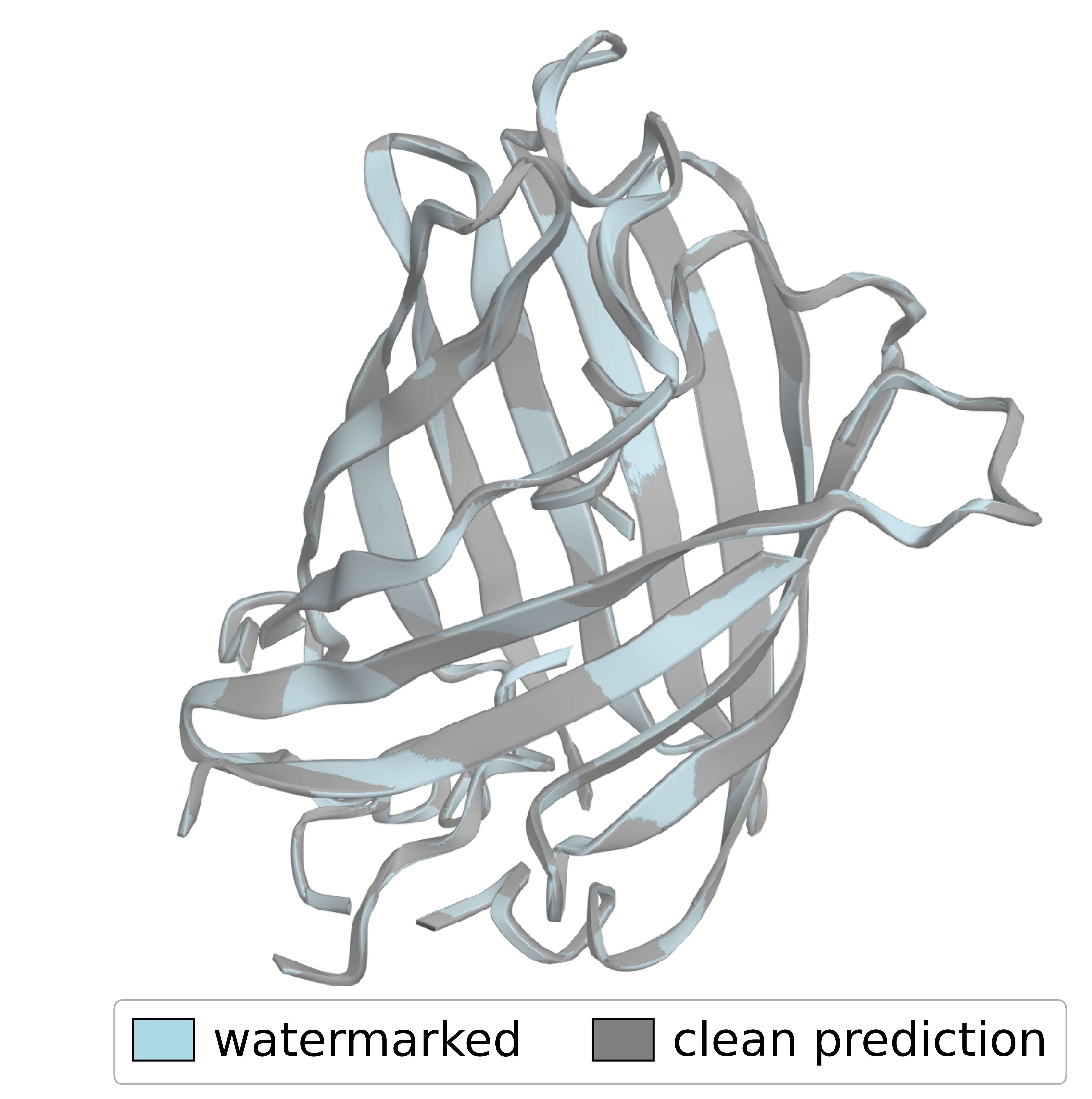}
        \caption{{\scriptsize 1EMA -- Green Fluorescent Protein}}
    \end{subfigure}
    \hfill
    \begin{subfigure}[b]{0.32\textwidth}
        \centering
        \includegraphics[width=\linewidth]{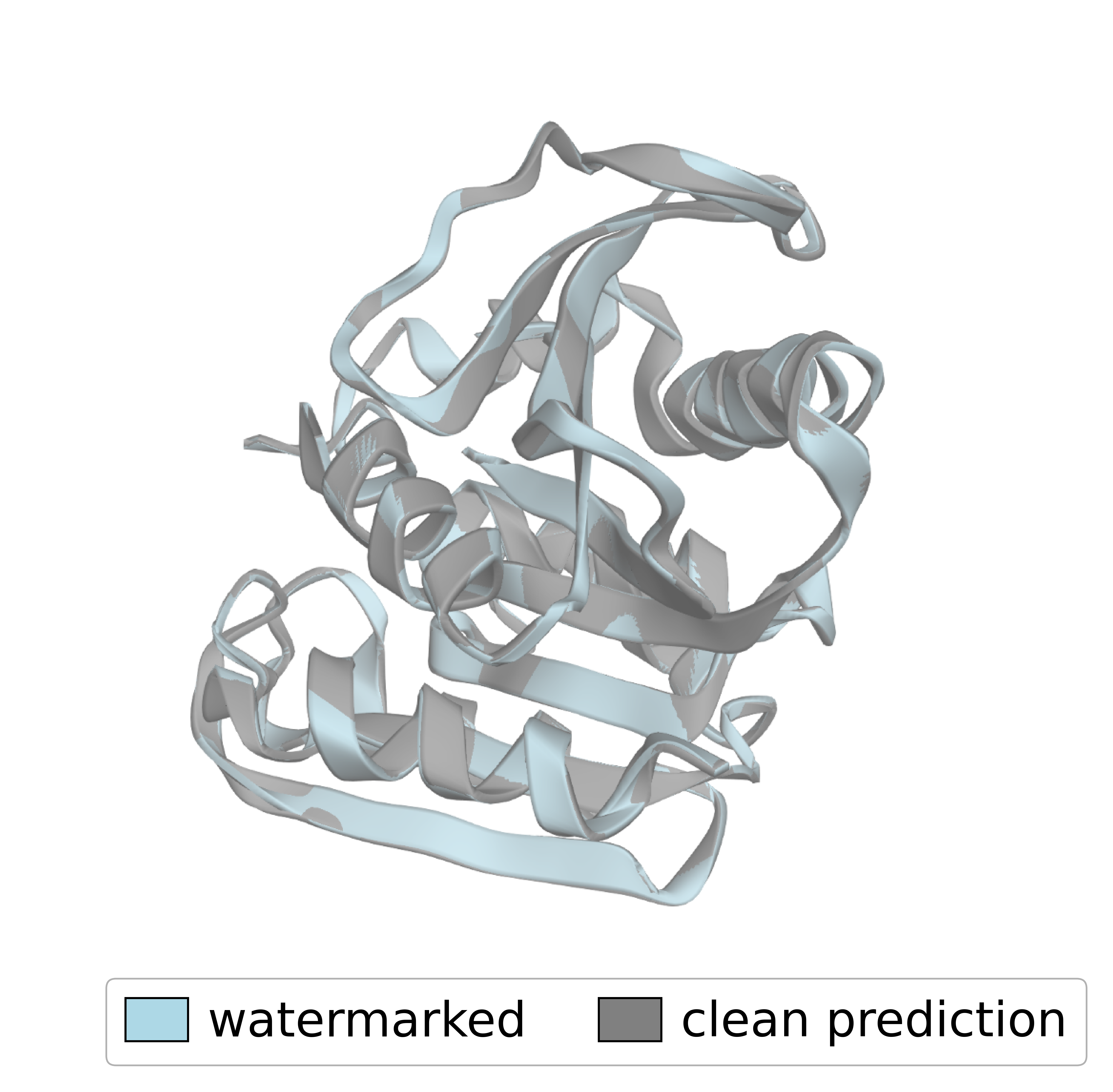}
        \caption{{\scriptsize 4AO6 -- Cold-Adapted Esterase}}
    \end{subfigure}
    
    \caption{\textbf{Visualization of the Clean Prediction vs. Watermarked Structure}. The conformation of the watermarked structure highly aligns to the clean prediction, demonstrating the high fidelity of MARCO.}
    \label{fig:exp_visualization}
\end{figure*}

\begin{figure*}[t!] % 'p' encourages placing this tall figure on its own page
    % \captionsetup{labelfont={color=blue}, textfont={color=blue}}
    \centering
    
    % --- Row 1 ---
    \begin{subfigure}[b]{0.32\textwidth}
        \centering
        \includegraphics[width=\linewidth]{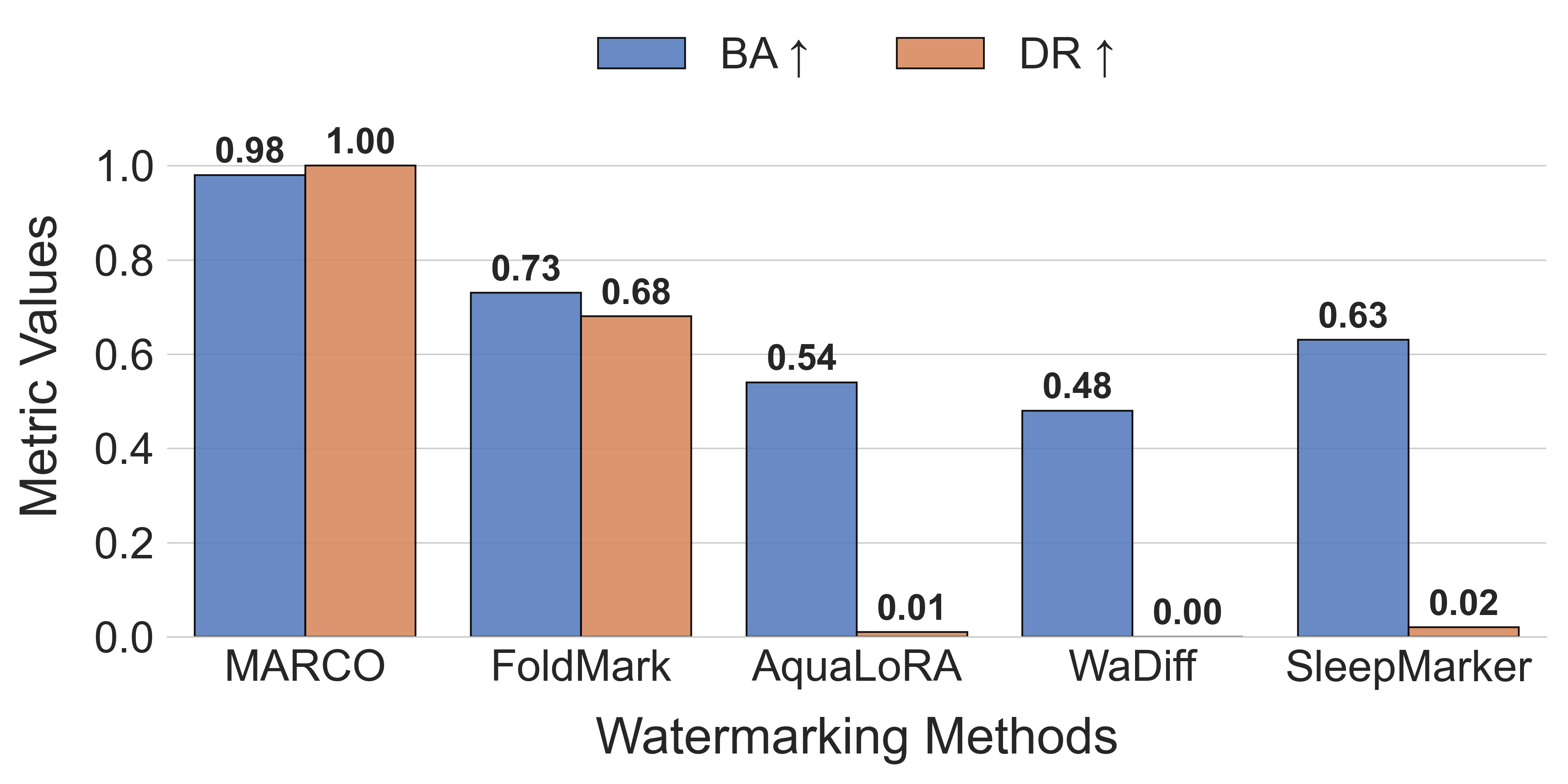}
        \caption{$\Theta:$ FrameDiff, Rotation}
    \end{subfigure}
    \hfill
    \begin{subfigure}[b]{0.32\textwidth}
        \centering
        \includegraphics[width=\linewidth]{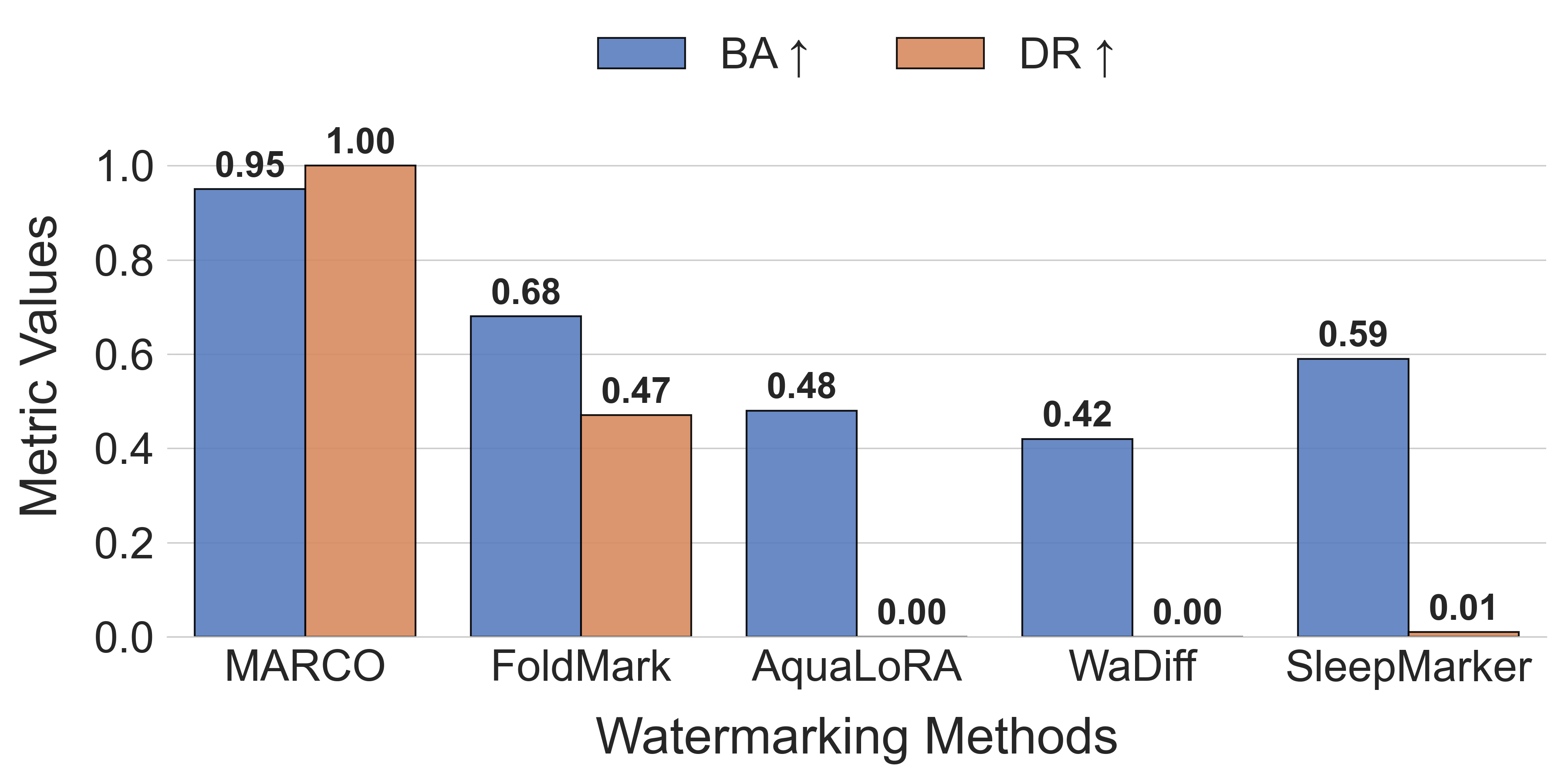}
        \caption{$\Theta:$ FrameDiff, Cropping}
    \end{subfigure}
    \hfill
    \begin{subfigure}[b]{0.32\textwidth}
        \centering
        \includegraphics[width=\linewidth]{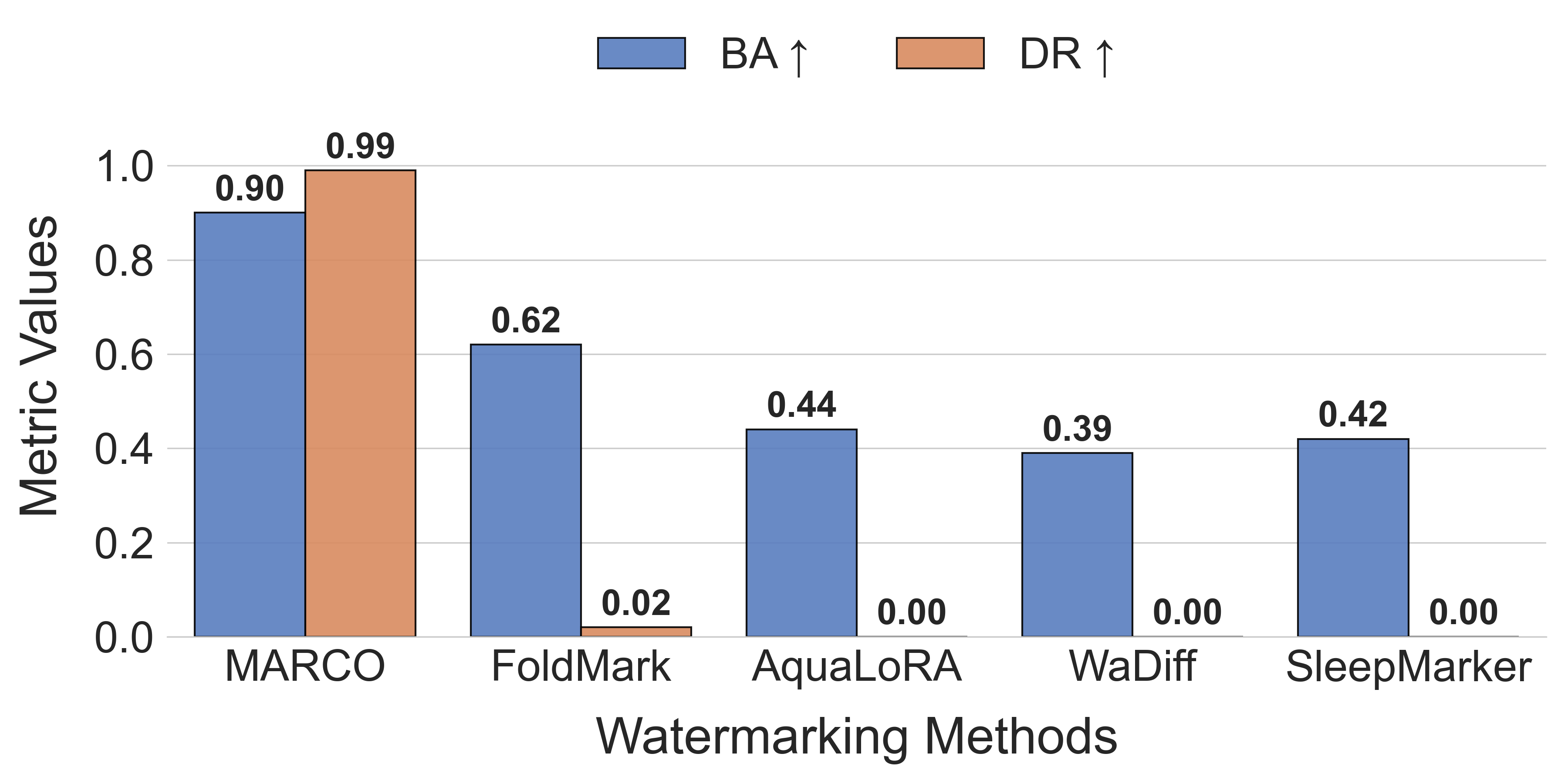}
        \caption{$\Theta:$ FrameDiff, Noise Addition}
    \end{subfigure}
    
    \par\medskip % New row
    
    % --- Row 3 ---
    \begin{subfigure}[b]{0.32\textwidth}
        \centering
        \includegraphics[width=\linewidth]{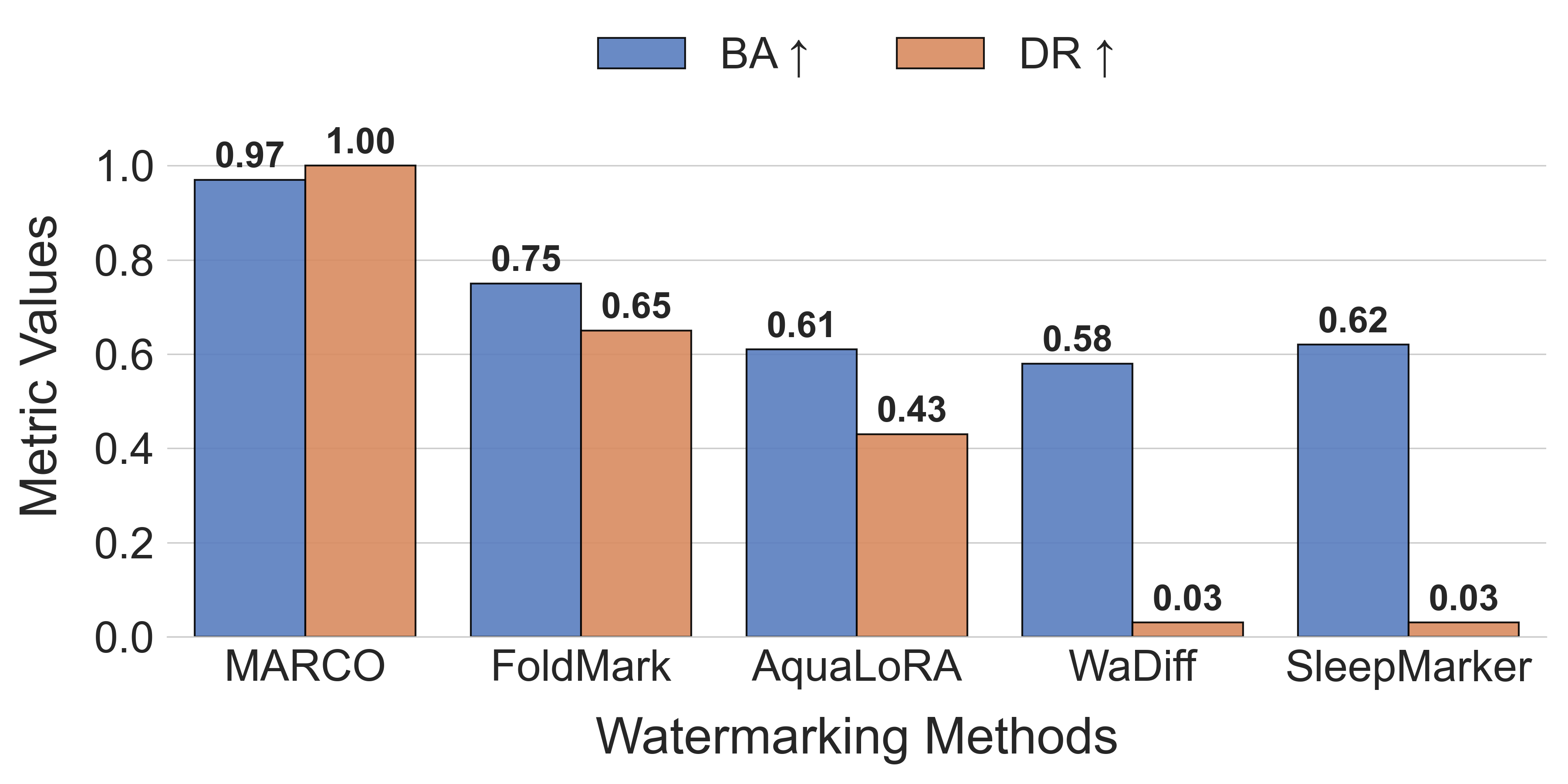}
        \caption{$\Theta:$ FrameFlow, Rotation}
    \end{subfigure}
    \hfill
    \begin{subfigure}[b]{0.32\textwidth}
        \centering
        \includegraphics[width=\linewidth]{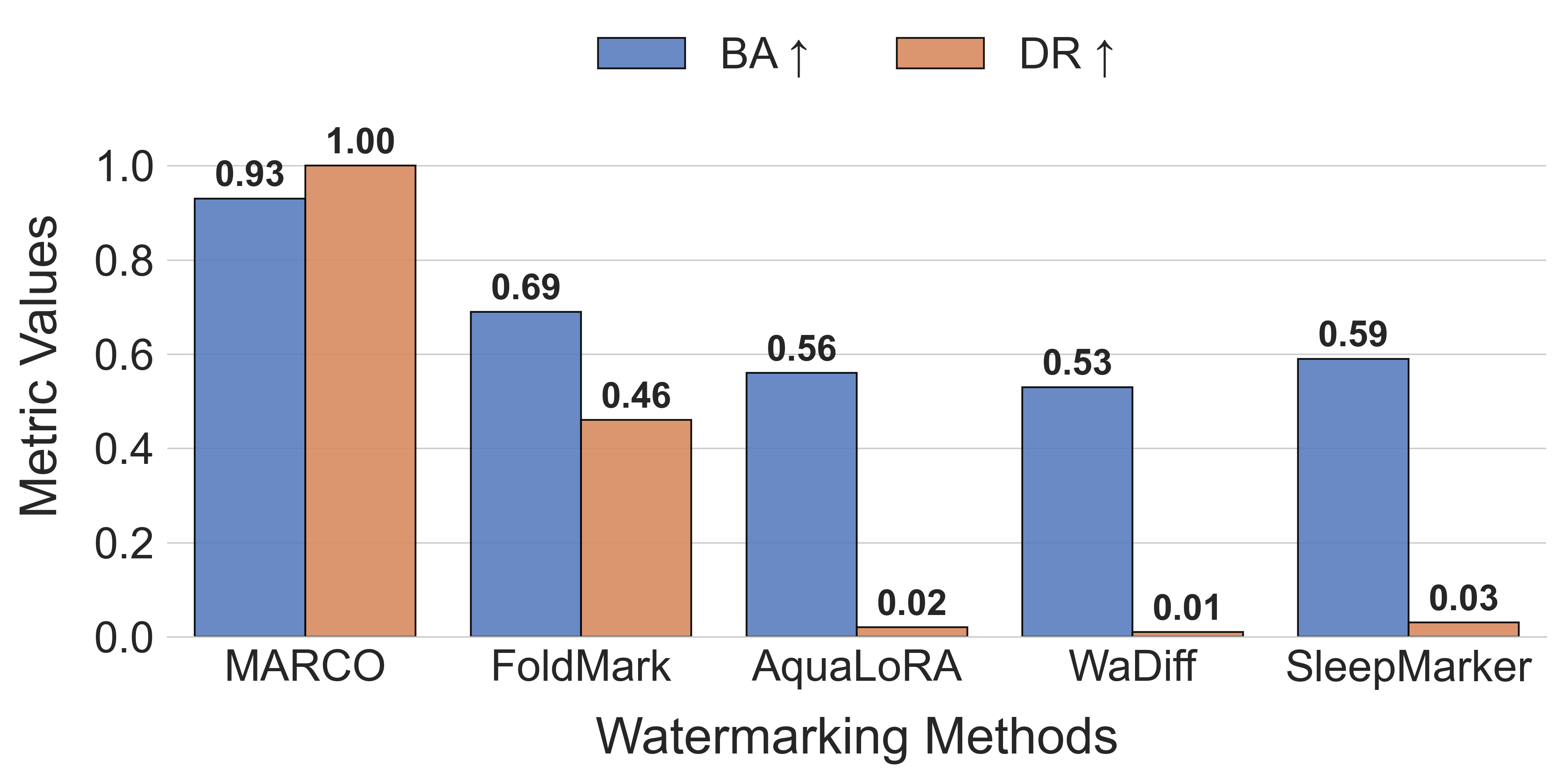}
        \caption{$\Theta:$ FrameFlow, Cropping}
    \end{subfigure}
    \hfill
    \begin{subfigure}[b]{0.32\textwidth}
        \centering
        \includegraphics[width=\linewidth]{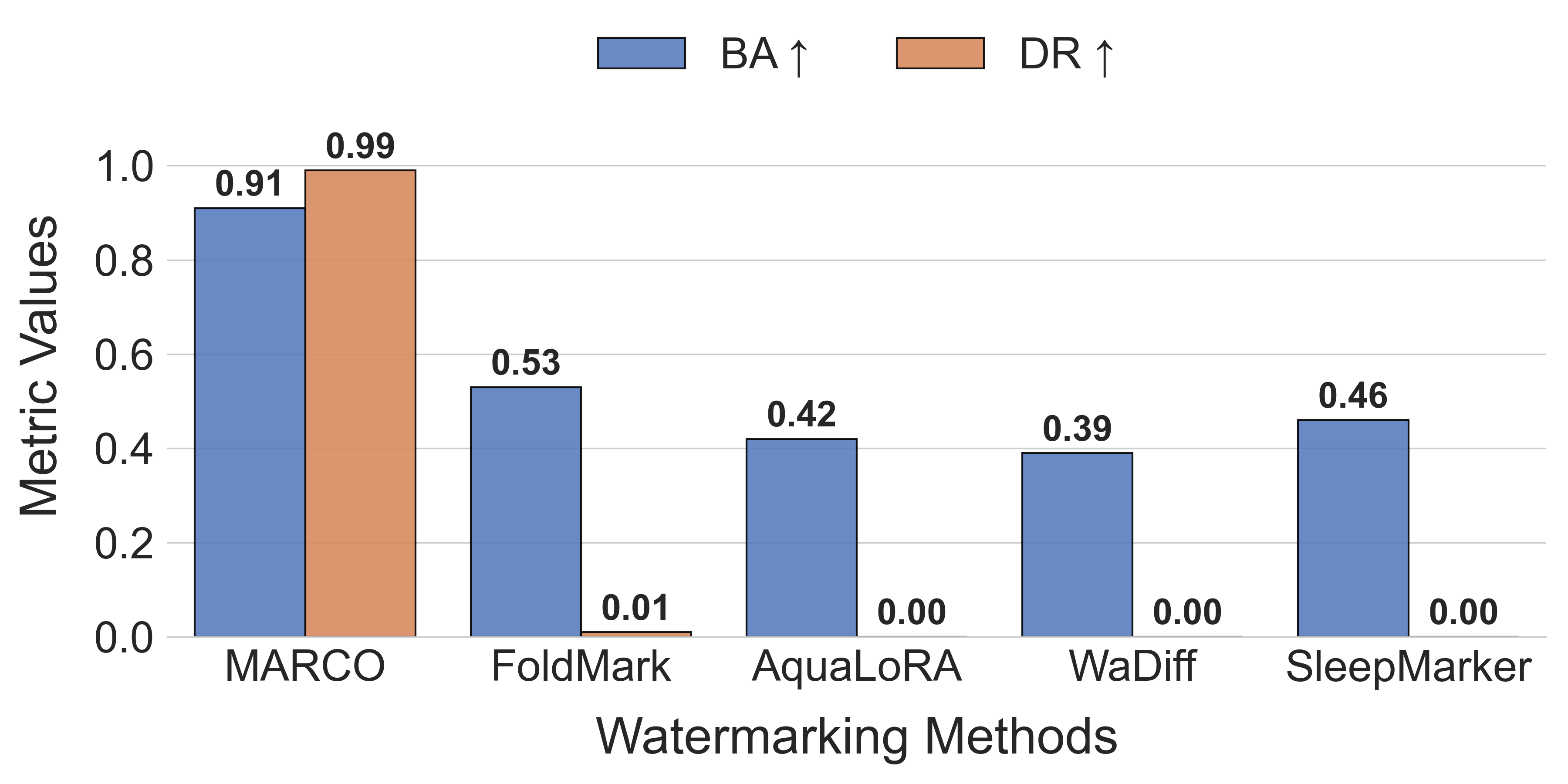}
        \caption{$\Theta:$ FrameFlow, Noise Addition}
    \end{subfigure}
    
    \par\medskip % New row
    
    % --- Row 4 ---
    \begin{subfigure}[b]{0.32\textwidth}
        \centering
        \includegraphics[width=\linewidth]{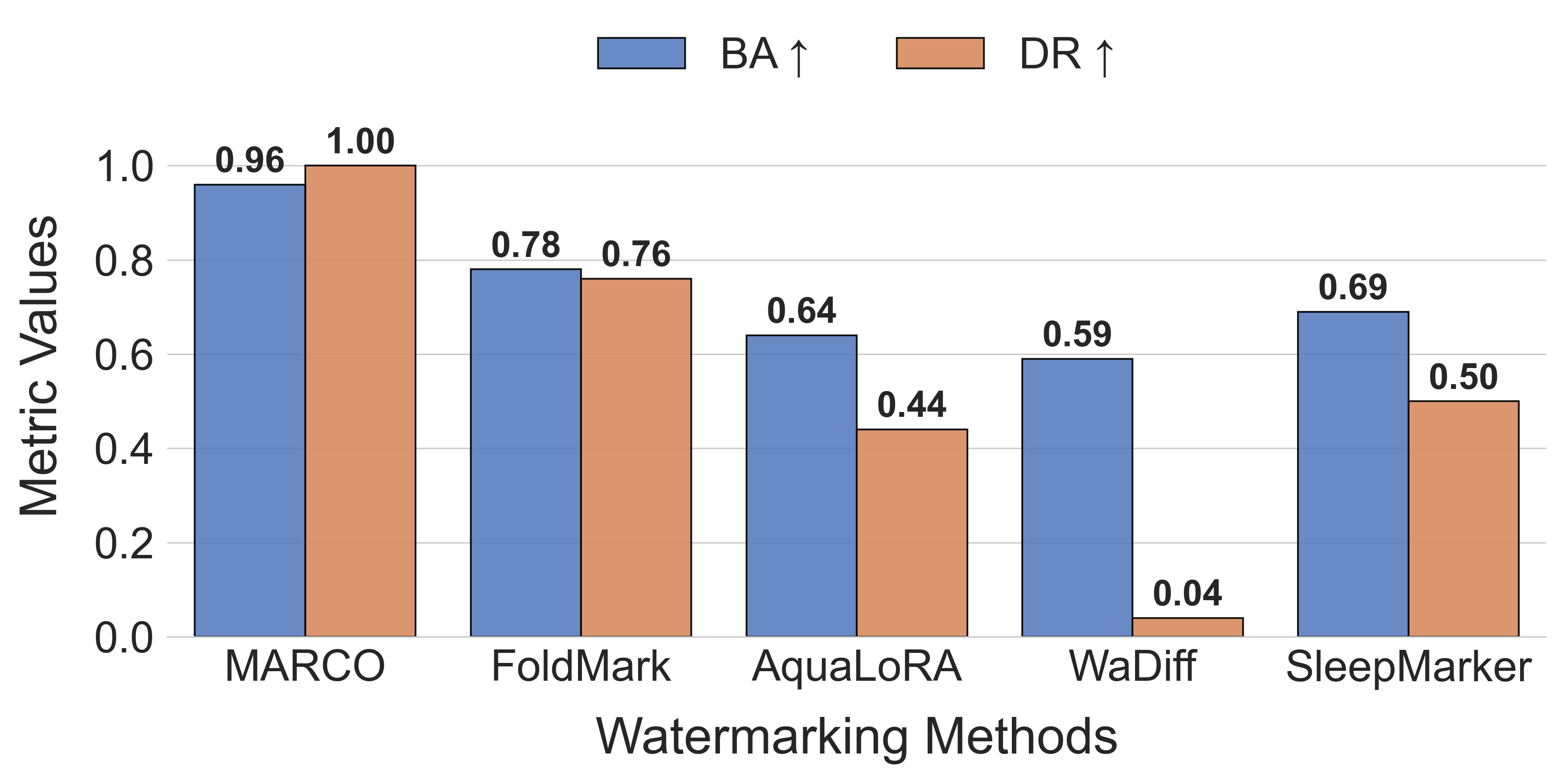}
        \caption{$\Theta:$ FoldFlow2, Rotation}
    \end{subfigure}
    \hfill
    \begin{subfigure}[b]{0.32\textwidth}
        \centering
        \includegraphics[width=\linewidth]{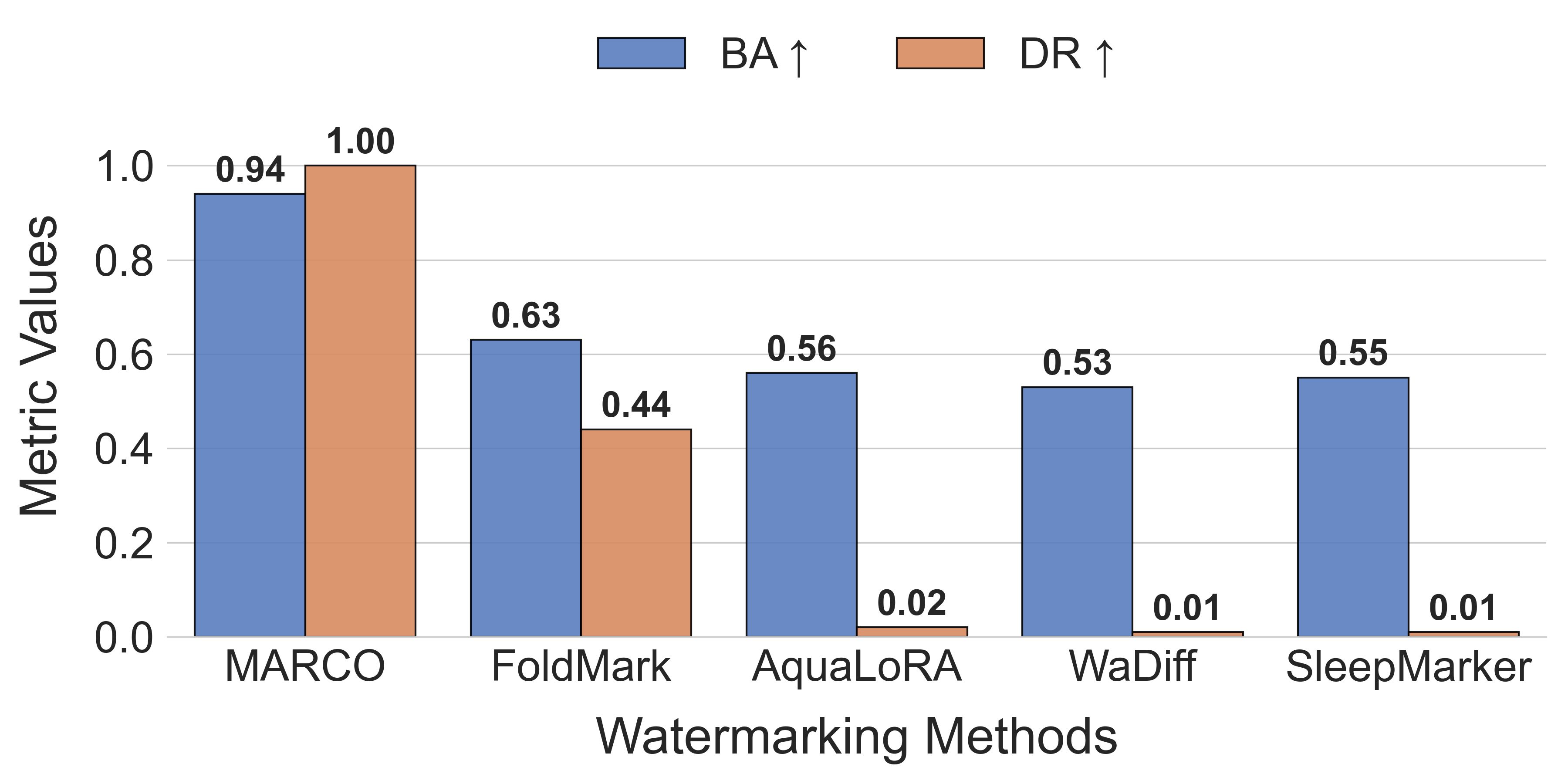}
        \caption{$\Theta:$ FoldFlow2, Cropping}
    \end{subfigure}
    \hfill
    \begin{subfigure}[b]{0.32\textwidth}
        \centering
        \includegraphics[width=\linewidth]{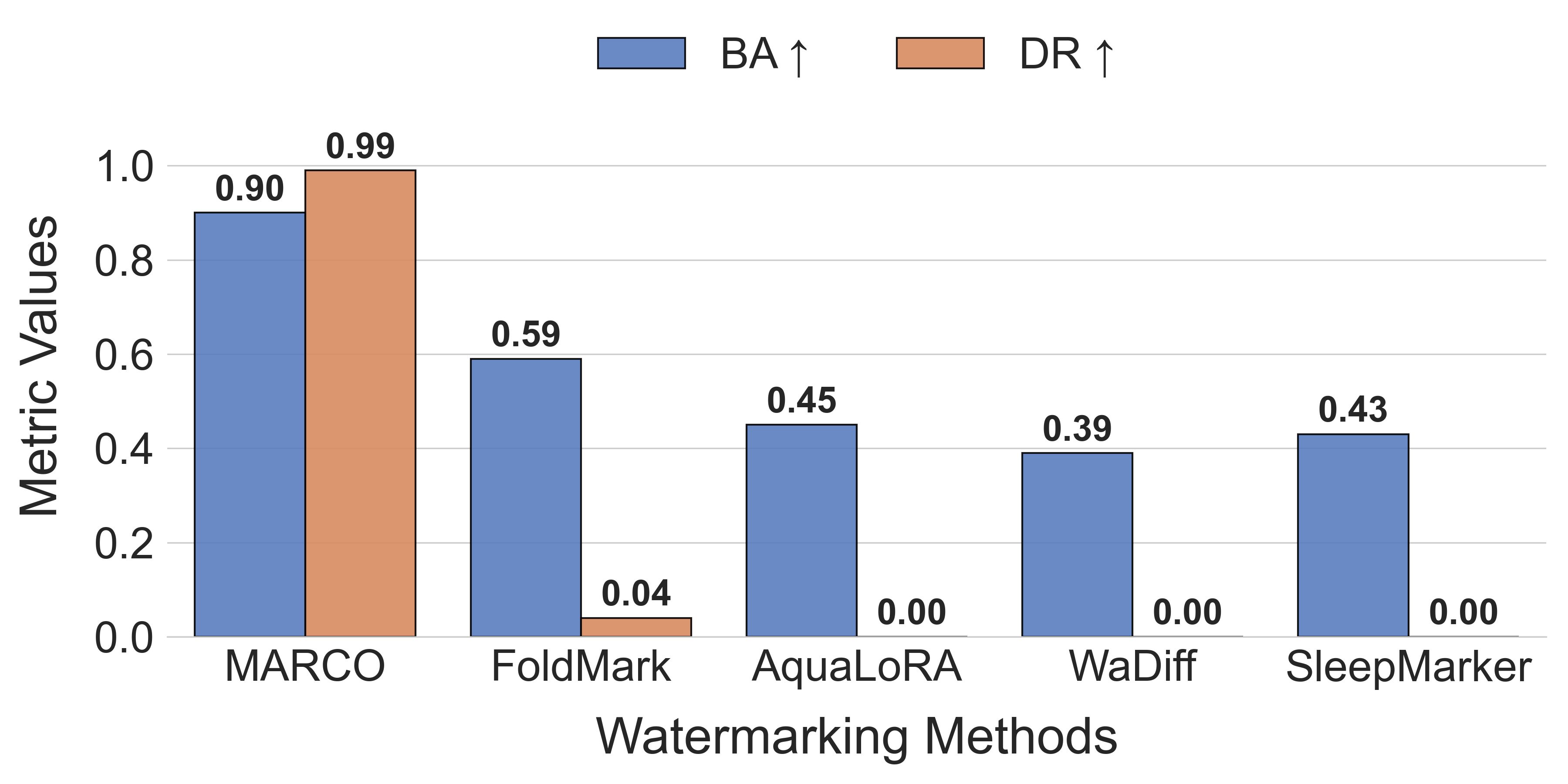}
        \caption{$\Theta:$ FoldFlow2, Noise Addition}
    \end{subfigure}
    
    \caption{Part B -- \textbf{Evaluation of Robustness of PGM Watermarking Methods against Conformation Manipulations}. After conformation manipulations, MARCO survives and outperforms the other four baselines with impressive robustness.}
    \label{fig:exp_rbn_atk_B}
\end{figure*}

\subsection{Robustness}
\begin{figure*}[t!]
    \centering
    \includegraphics[width=\linewidth]{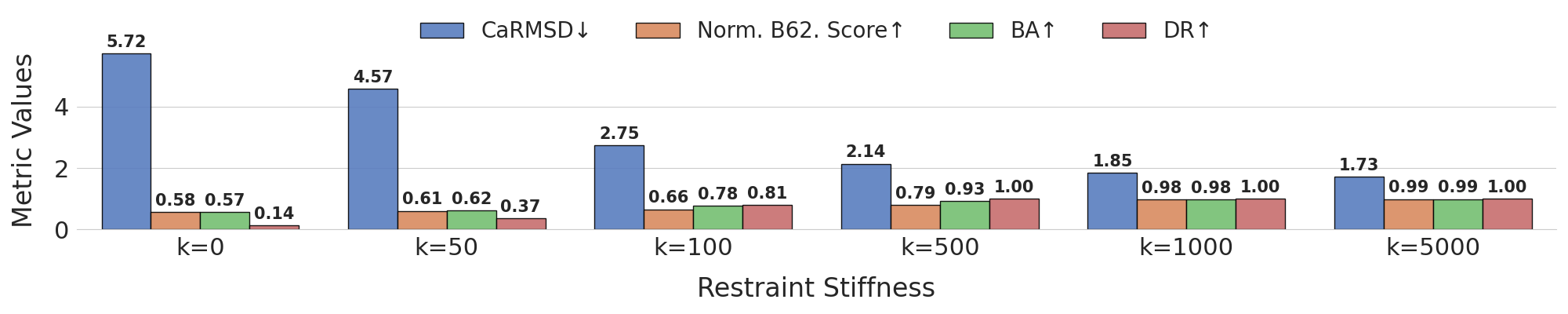}
    \caption{\textbf{MARCO vs. Energy-Based Scrubbing}. A sufficiently low $k$ results in a compromised watermark but at a non-negligible cost of fidelity.}
    \label{fig:exp_energy_based_scrubbing}
\end{figure*}

\prg{MARCO vs. Energy-Based Scrubbing}
To simulate an adaptive attacker, we employ an energy-based scrubbing attack implemented via OpenMM \cite{eastman2013OpenMM}. 
We apply a harmonic positional restraint $E_{\mathrm{rest}}=\frac{k}{2}\sum_i\|\mathbf x_i-\mathbf x_i^0\|_2^2$, where \(\mathbf x_i^0\) denotes the initial watermarked coordinate and \(k\) is the restraint stiffness in \(\mathrm{kJ\,mol^{-1}\,nm^{-2}}\). 
We evaluate MARCO's robustness by generating a watermarked dataset containing $1000$ watermarked structures using RFdiffusion. These structures are subjected to scrubbing across a stiffness range of $k \in \{0, 50, 100, 500, 1000, 5000\}$ and subsequently used to fine-tune a FrameDiff pirate model.
As illustrated in \autoref{fig:exp_energy_based_scrubbing}, MARCO demonstrates impressive robustness to this adaptive threat. 
Crucially, although extremely low $k$ values can compromise watermark recovery, they simultaneously destroy structural fidelity, rendering the resulting pirate model functionally unusable.

\begin{figure}[!t]
    % \captionsetup{labelfont={color=blue}, textfont={color=blue}}
    \centering
    \includegraphics[width=.8\linewidth]{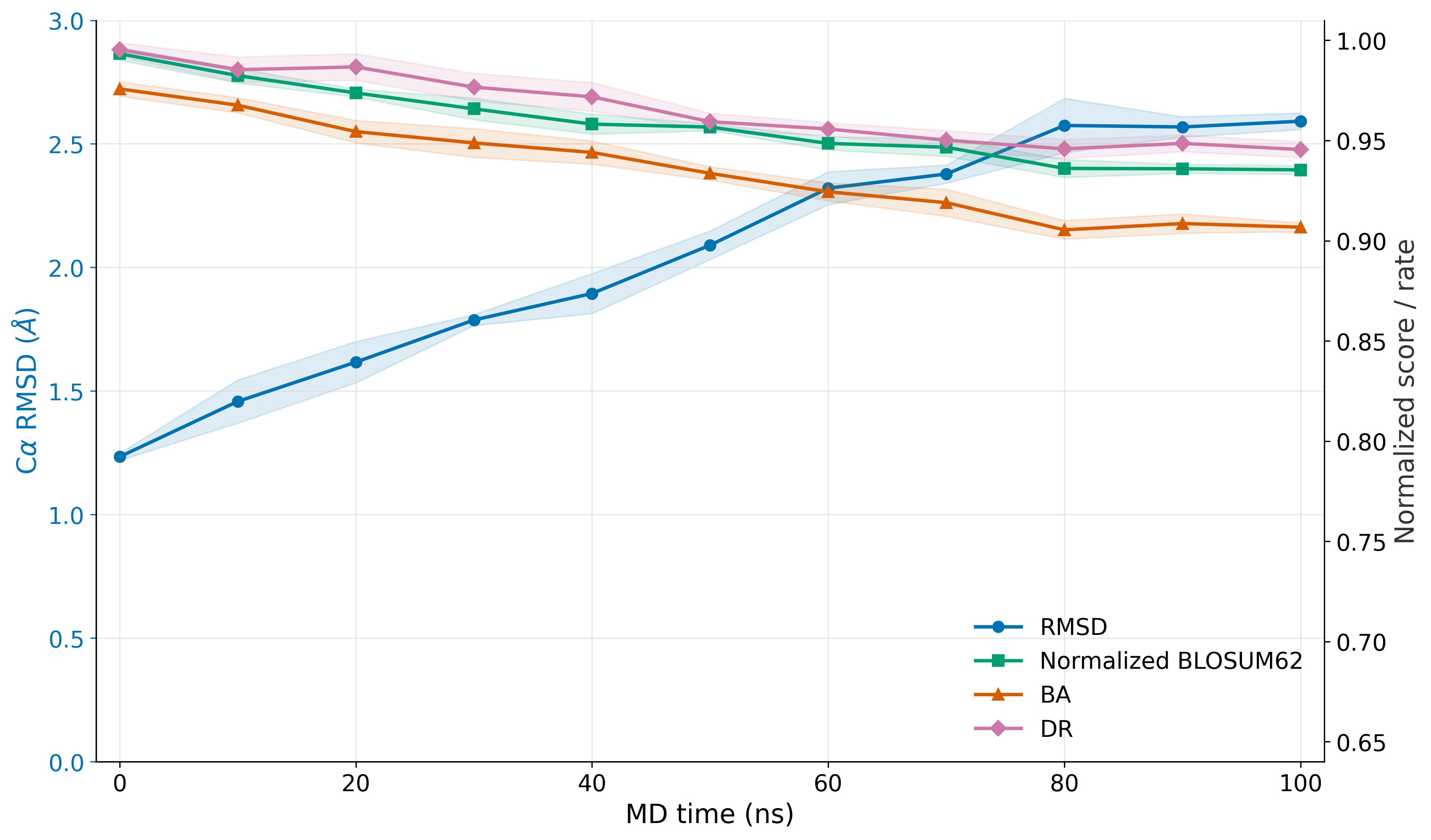}
    \caption{\textbf{MARCO vs. Molecular Dynamic Simulation}}
    \label{fig:rfdiffusion_md}
\end{figure}

Additionally, we evaluate MARCO under finite-temperature molecular dynamics using 100-ns OpenMM simulations of the same watermarked structures. Each protein is parameterized with the AMBER ff14SB force field, solvated in a TIP3P water box with a 1.0-nm buffer and 0.15-M NaCl, energy-minimized, equilibrated for 100 ps under NVT and 1 ns under NPT, and simulated without positional restraints at 300 K and 1 bar using a Langevin thermostat, Monte Carlo barostat, PME electrostatics, a 1.0-nm nonbonded cutoff, constrained hydrogen bonds, and a 2-fs timestep. Snapshots are collected every 10 ns and evaluated using $C_\alpha$-RMSD, normalized BLOSUM62, BA and DR. As depicted in \autoref{fig:rfdiffusion_md}, as time flows, both the fidelity and robustness of the watermarked structures decrease, because the shape of the structures slightly changes to reach a lower-energy state. However, the performance reaches a balance from 80ns, when the structures become stable. This suggests that MARCO is robust against molecular dynamic simulations.

\begin{figure*}[t!]
    \centering
    \includegraphics[width=\linewidth]{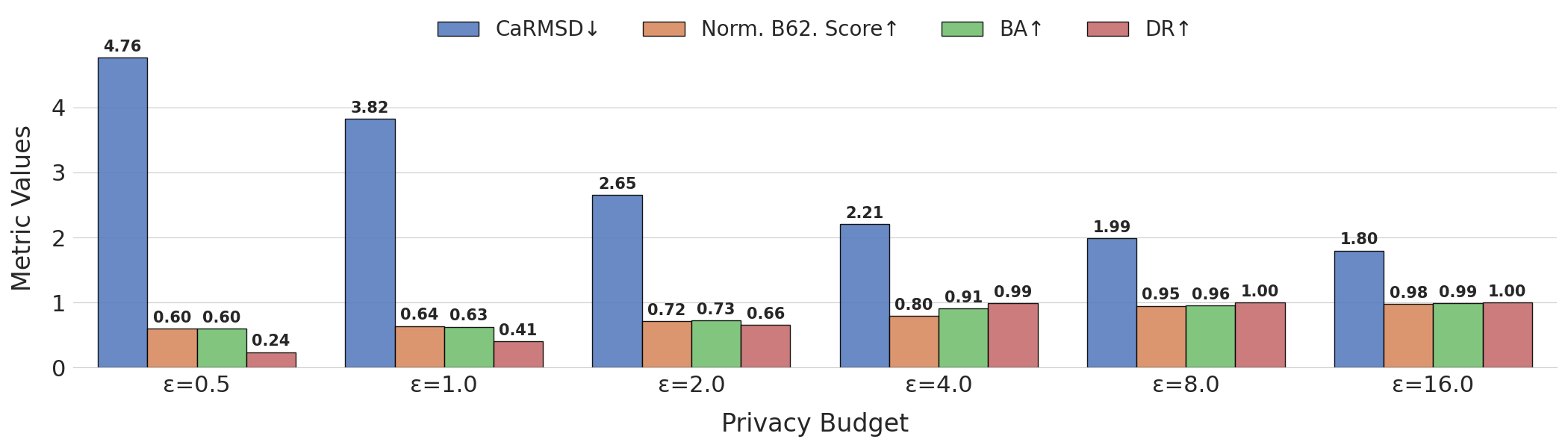}
    \caption{\textbf{MARCO vs. DPSGD}. Even if the watermark is removed with a low privacy budget, the pirate model greatly sacrifices its fidelity.}
    \label{fig:exp_dpsgd}
\end{figure*}

\prg{MARCO vs. DPSGD}
% \textit{Train a pirate model on watermarked data using DPSGD and test different $\epsilon$.}
% Furthermore, the adaptive attacker may use local regularization techniques to purge the watermark in the pirate model.
% We thus use DPSGD \cite{abadi2016DPSGD} to simulate this scenario to further evaluate the robustness of MARCO faced with local regularization.
% We set the hyperparameters $\delta$ to $0.00001$ and vary $\epsilon$ by selecting values in the set $\{0.5, 1.0, 2.0, 4.0, 8.0, 16.0\}$.
% The aforementioned watermarked dataset is once again used to fine-tune FrameDiff to derive the metric values.
We further simulate an adaptive attacker utilizing local regularization techniques to remove the watermark. Specifically, we employ DPSGD \cite{abadi2016DPSGD} to fine-tune FrameDiff on the watermarked dataset. We fixed $\delta = 10^{-5}$ and varied the privacy budget across $\epsilon \in \{0.5, 1.0, 2.0, 4.0, 8.0, 16.0\}$.
% As depicted in \autoref{fig:exp_dpsgd}, we notice that when $\epsilon \geq 4.0$, MARCO remains robust with BA $\geq 0.91$ and DR $\geq 0.99$, ensuring a promising watermark verification.
% Similarly to energy-based scrubbing, a small privacy budget can also compromise the watermark by sacrificing fidelity at a great cost, leading to an inferior pirate model.
As illustrated in \autoref{fig:exp_dpsgd}, MARCO maintains its robustness with BA $\geq 0.91$ and DR $\geq 0.99$ when the privacy budget $\epsilon \geq 4.0$. Similar to the energy-based scrubbing results, low $\epsilon$ may degrade the watermark signal. However, this comes at the cost of severe fidelity loss, rendering the resulting pirate model functionally unusable.

\prg{MARCO vs. Refolding}
In this scenario, an adaptive attacker queries the PGM protected by MARCO to unconditionally generate de novo protein structures.
The structures are then mapped back to sequences using some mapping functions.
The sequences are fed into a clean PGM to remove the watermark inside.
To simulate this attack, we employ RFDiffusion as the original PGM protected by MARCO to generate $1000$ de novo watermarked structures.
The structures are then mapped back to sequences using ProteinMPNN \cite{dauparas2022ProteinMPNN} and refolded into protein structures using the other five clean pre-trained PGMs.
We derive the metrics values by extracting the recovered watermark from the refolded structures.

\begin{figure}[t!]
    \centering
    \includegraphics[width=.6\linewidth]{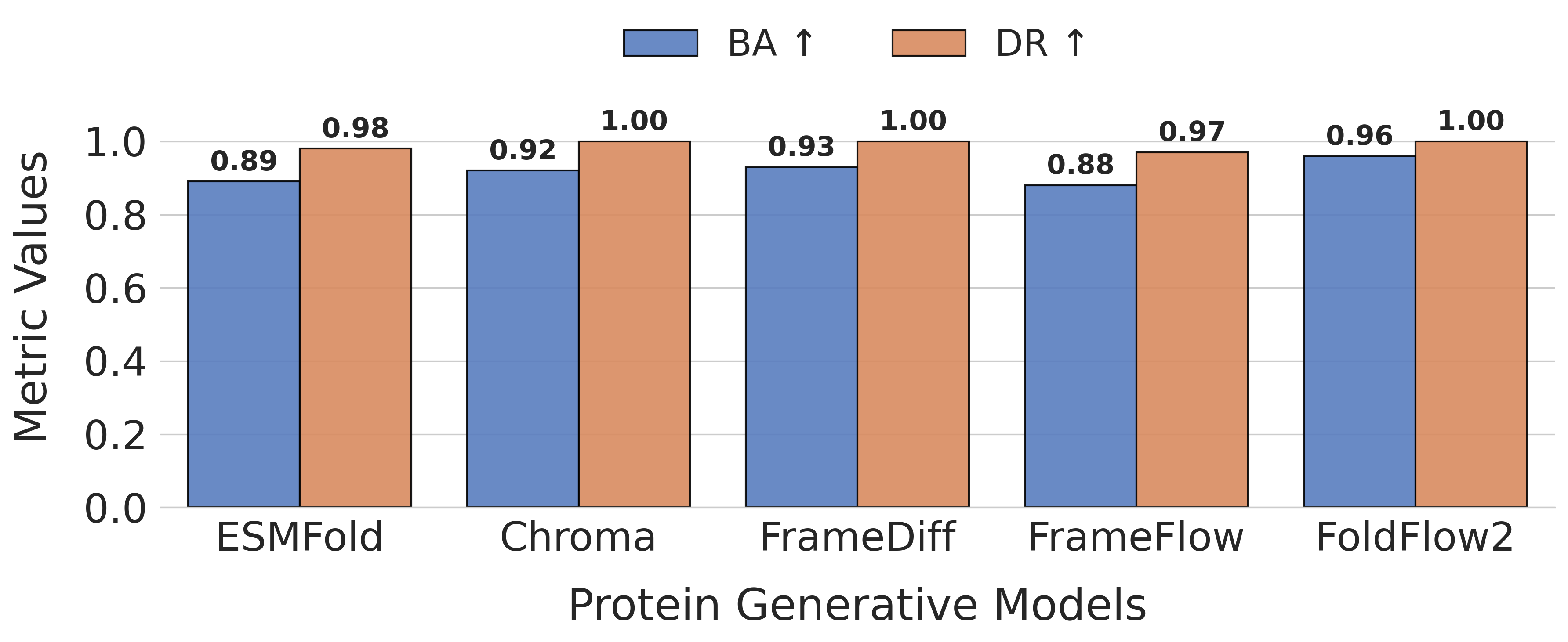}
    \caption{\textbf{MARCO vs. Refolding}. By selectively altering the translated sequences, MARCO gains sufficient robustness against refolding attack.}
    \label{fig:exp_refolding}
\end{figure}

The experimental results are depicted in \autoref{fig:exp_refolding}, where MARCO shows sufficient robustness against refolding attack.
As demonstrated in the previous experiments, MARCO does not present perfect sequential fidelity, because MARCO trades off this attributes to gain resistance against refolding attack.
The watermark signal affects the sequences translated from the watermarked structure such that structures generated by other clean PGMs given these sequences inherit the watermark.

\prg{Certified Robustness}
To prove the certified robustness of MARCO, we conduct an experiment where we use different MARCO-guarded PGMs to generate $1000$ watermarked structures.
The watermarked structures are then perturbed with random noises multiplied by different weights.
Next, we extract the watermark and compute the RMSD between the ground truth and the perturbed structures and the BA to derive a curve of the correlation between RMSD and BA.

\begin{figure}[t!]
    \centering
    \includegraphics[width=.7\linewidth]{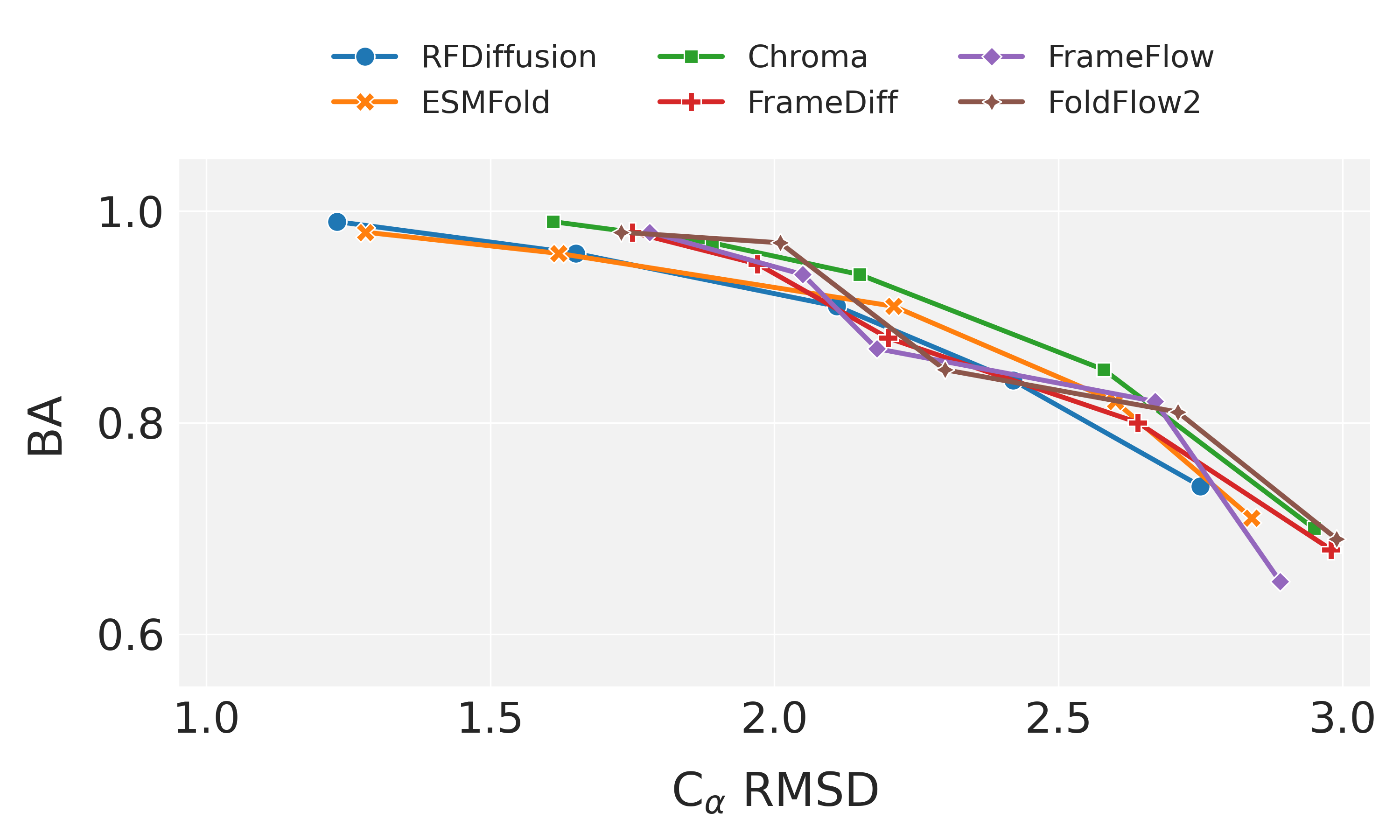}
    \caption{\textbf{Certifiable Robustness}. The robustness of MARCO is an intrinsic noise-aware geometric property.}
    \label{fig:exp_cert_rbn}
\end{figure}

In \autoref{fig:exp_cert_rbn}, we observe that the robustness of MARCO is not just an artifact of attack simulation, but an intrinsic noise-aware property.
The average BA of all PGMs remains above $0.8$ when RMSD $< 2.5$, which suffices for watermark verification.
Moreover, this relation between RMSD and BA shows that the adversary cannot guarantee watermark removal without significantly downgrade the fidelity.
This indicates that the manifold of the watermarked structures reflects the watermark itself.

% \todo{Mixed-Data Fine-Tuning Experiment}
\prg{MARCO vs. Dilution}
We evaluate MARCO when faced with fine-tuning attacks that attempt to remove the watermark by fine-tuning the pirate model with watermarked structures mixed with clean structures. We fine-tuned a FrameDiff model using varying ratios of MARCO-watermarked structures to clean structures. As shown in \autoref{tab:finetune}, the ratio does affect the effectiveness of MARCO. However, with only $25\%$ watermarked data, MARCO achieves $0.78$ BA and $0.85$ DR, demonstrating its strong robustness.
Additionally, we conduct another dilution experiment where we replace the clean structures with watermarked structures generated via FoldMark. The results in \autoref{tab:finetune_foldmark} demonstrate that the watermarked structures do not differ from the clean structures in terms of removing MARCO's watermark.

\begin{table}[t!]
    \centering
    % Increase row height slightly for a fancier look
    \renewcommand{\arraystretch}{1.5}
    \caption{\textbf{MARCO vs. Dilution.}}
    \resizebox{.4\linewidth}{!}{
    \begin{tabular}{
        >{\centering\columncolor{gray!10}}p{4cm}  % First column (Left aligned, no color)
        >{\columncolor{blue!10}}c   % 2nd Col: Light Blue
        >{\columncolor{red!10}}c    % 3rd Col: Light Red
    }
        \toprule
        % Top-left split cell: \diagbox{BottomLeft text}{TopRight text}
        \textbf{Watermarked Data Ratio} & \textbf{BA$\uparrow$} & \textbf{DR$\uparrow$}\\ 
        \midrule
        1.00 & 0.94 & 1.00 \\
        0.75 & 0.88 & 0.99 \\
        0.50 & 0.81 & 0.92 \\
        0.25 & 0.78 & 0.85 \\
        0.10 & 0.73 & 0.68 \\ 
        \bottomrule
    \end{tabular}
    }
    \label{tab:finetune}
\end{table}

\begin{table}[t!]
    % \captionsetup{labelfont={color=blue}, textfont={color=blue}}
    \centering
    % Increase row height slightly for a fancier look
    \renewcommand{\arraystretch}{1.5}
    \caption{\textbf{MARCO vs. FoldMark Dilution.}}
    \resizebox{.4\linewidth}{!}{
    \begin{tabular}{
        >{\centering\columncolor{gray!10}}p{4cm}  % First column (Left aligned, no color)
        >{\columncolor{blue!10}}c   % 2nd Col: Light Blue
        >{\columncolor{red!10}}c    % 3rd Col: Light Red
    }
        \toprule
        % Top-left split cell: \diagbox{BottomLeft text}{TopRight text}
        \textbf{MARCO Data Ratio} & \textbf{BA$\uparrow$} & \textbf{DR$\uparrow$}\\ 
        \midrule
        1.00 & 0.92 & 0.99 \\
        0.75 & 0.85 & 0.97 \\
        0.50 & 0.82 & 0.92 \\
        0.25 & 0.75 & 0.73 \\
        0.10 & 0.71 & 0.65 \\
        \bottomrule
    \end{tabular}
    }
    \label{tab:finetune_foldmark}
\end{table}

% \todo{Geometry-Aware Baselines}
\prg{MARCO vs. Geometry-Aware Baselines}
We elevate the level of robustness evaluation of MARCO by comparing it with a geometry-aware baseline. We compare MARCO with MarkSplatter \cite{huang2025marksplatter}, a SOTA 3D point-cloud watermark. We adapted MarkSplatter by treating the 3D atomic coordinates of 1,000 RFdiffusion-generated proteins as unstructured point clouds to fit into its workflow. The results are listed in \autoref{tab:geo-aware}, from which we conclude that MARCO outperforms MarkSplatter in watermarking PGMs due to its original design aiming at protein structure prediction.

\begin{table*}[t!]
    \centering
    % Increase row height slightly for a fancier look
    \renewcommand{\arraystretch}{1.5}
    \caption{\textbf{MARCO vs. Geometry-Aware Baselines.}}
    \resizebox{\textwidth}{!}{
    \begin{tabular}{
        >{\columncolor{gray!10}}c  % First column (Left aligned, no color)
        >{\columncolor{blue!10}}c   % 2nd Col: Light Blue
        >{\columncolor{red!10}}c    % 3rd Col: Light Red
        >{\columncolor{purple!10}}c  % 4th Col: Light Green
        >{\columncolor{orange!10}}c % 5th Col: Light Orange
        >{\columncolor{blue!10}}c   % 2nd Col: Light Blue
        >{\columncolor{red!10}}c    % 3rd Col: Light Red
        >{\columncolor{purple!10}}c  % 4th Col: Light Green
        >{\columncolor{orange!10}}c % 5th Col: Light Orange
    }
        \toprule
        % Top-left split cell: \diagbox{BottomLeft text}{TopRight text}
        \diagbox{\textbf{Methods}}{\textbf{Metrics}} & \textbf{$C_\alpha$RMSD$\downarrow$} & \textbf{Norm. B62$\uparrow$} & \textbf{BA$\uparrow$} & \textbf{DR$\uparrow$} & \textbf{BA(Rotation)$\uparrow$} & \textbf{DR(Rotation)$\uparrow$} & \textbf{BA(Crop)$\uparrow$} & \textbf{DR(Crop)$\uparrow$}\\ 
        \midrule
        MARCO & 1.23 & 0.99 & 0.98 & 1.00 & 0.97 & 1.00 & 0.93 & 0.99 \\
        MarkSplatter & 2.71 & 0.74 & 0.95 & 1.00 & 0.73 & 0.68 & 0.63 & 0.35 \\
        \bottomrule
    \end{tabular}
    }
    \label{tab:geo-aware}
\end{table*}

\subsection{Radioactivity}
\prginit{MARCO vs. Various Architectures}
% To further evaluate the radioactivity of MARCO, we first randomly create a subset that contains $1000$ protein sequences from the training dataset.
% Then, we use the four watermarking methods to generate watermarked structures with the subset using RFDiffusion, ESMFold, and Chroma as $\Theta$.
% Next, we finetune the three pretrained models to be the pirate models $\Upsilon$, namely FrameDiff, Frameflow, and FoldFlow2 with the paired sequences and watermarked structures.
% This is because these three models are relatively simple to finetune due to limited resources.
% Eventually, we extract the recovered watermark from the output of $\Upsilon$ and compute the metric values.
To evaluate radioactivity, we randomly create a subset of 1,000 sequences from the training data. We then generate watermarked datasets using RFdiffusion, ESMFold, and Chroma as the source PGMs $\Theta$ with a fixed watermark. 
Subsequently, we fine-tune three computationally efficient pirate models $\Upsilon$ including FrameDiff, FrameFlow, and FoldFlow2 using these paired sequences and watermarked structures. Finally, we quantify radioactivity by extracting the recovered watermark from the outputs of $\Upsilon$s.

\begin{figure*}[t!] % 'p' encourages placing this tall figure on its own page
    % \captionsetup{labelfont={color=blue}, textfont={color=blue}}
    \centering
    
    % --- Row 1 ---
    \begin{subfigure}[b]{0.32\textwidth}
        \centering
        \includegraphics[width=\linewidth]{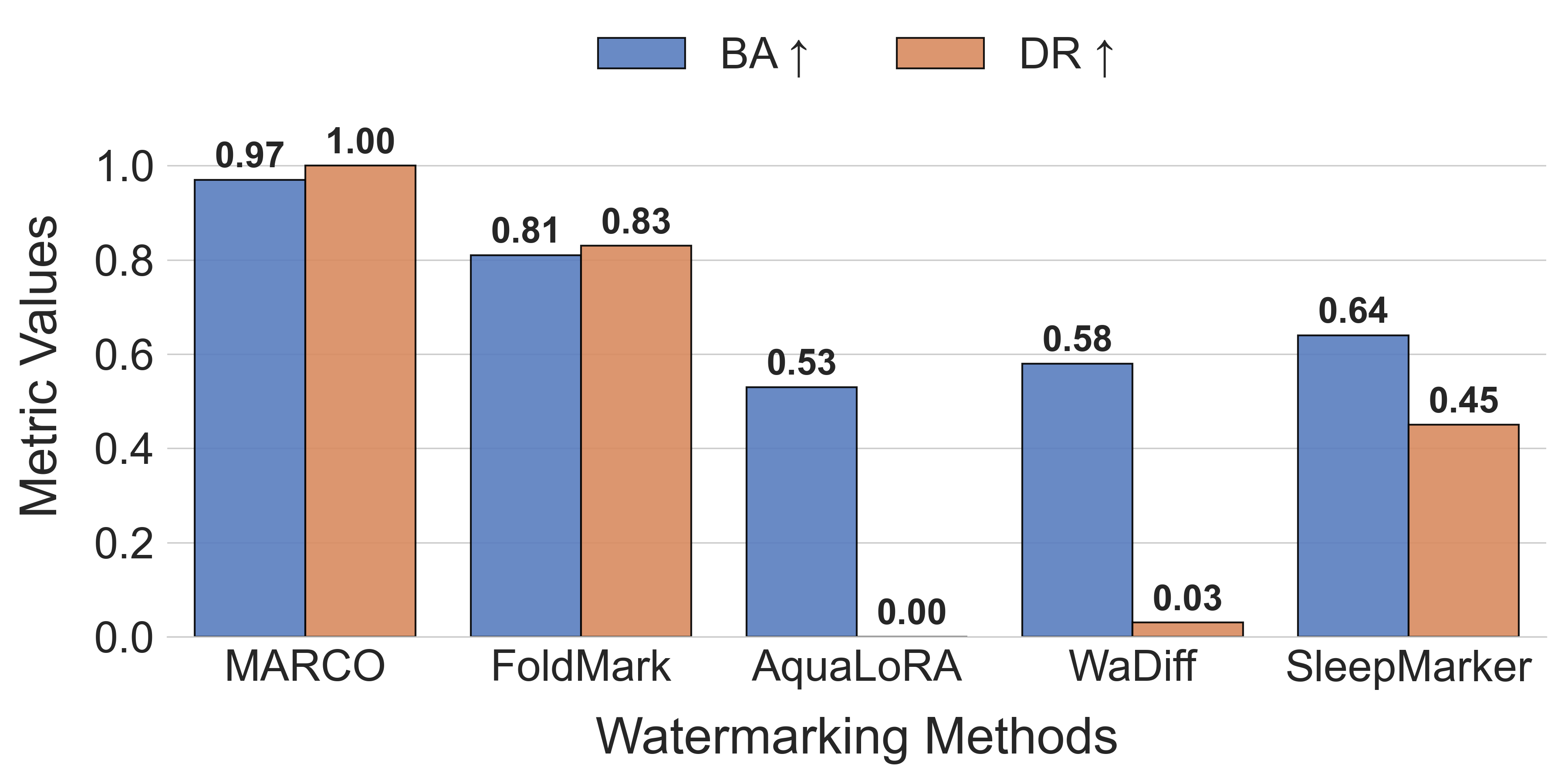}
        \caption{$\Theta$: RFDiffusion, $\Upsilon$: FrameDiff}
    \end{subfigure}
    \hfill
    \begin{subfigure}[b]{0.32\textwidth}
        \centering
        \includegraphics[width=\linewidth]{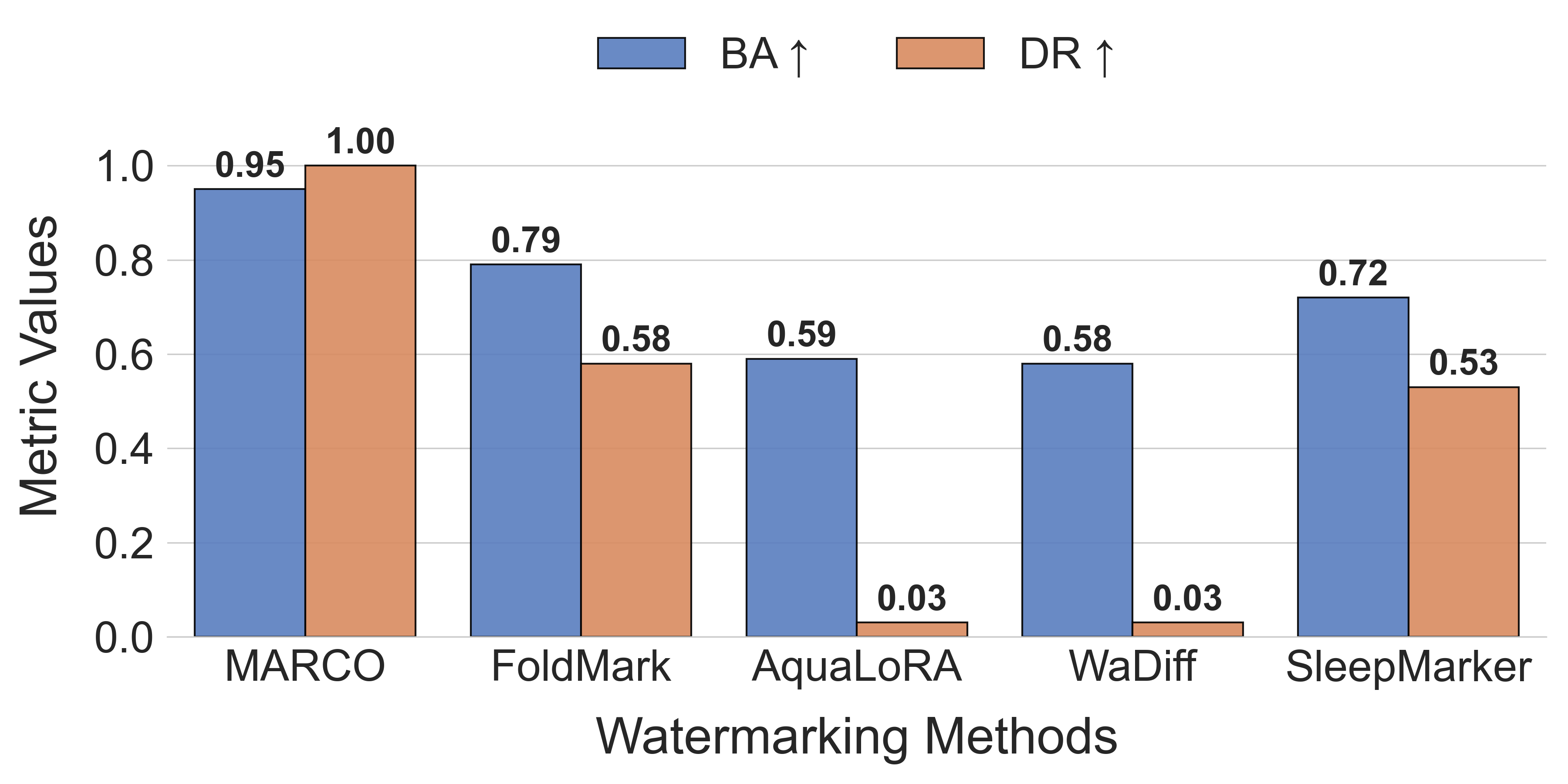}
        \caption{$\Theta$: RFDiffusion, $\Upsilon$: FrameFlow}
    \end{subfigure}
    \hfill
    \begin{subfigure}[b]{0.32\textwidth}
        \centering
        \includegraphics[width=\linewidth]{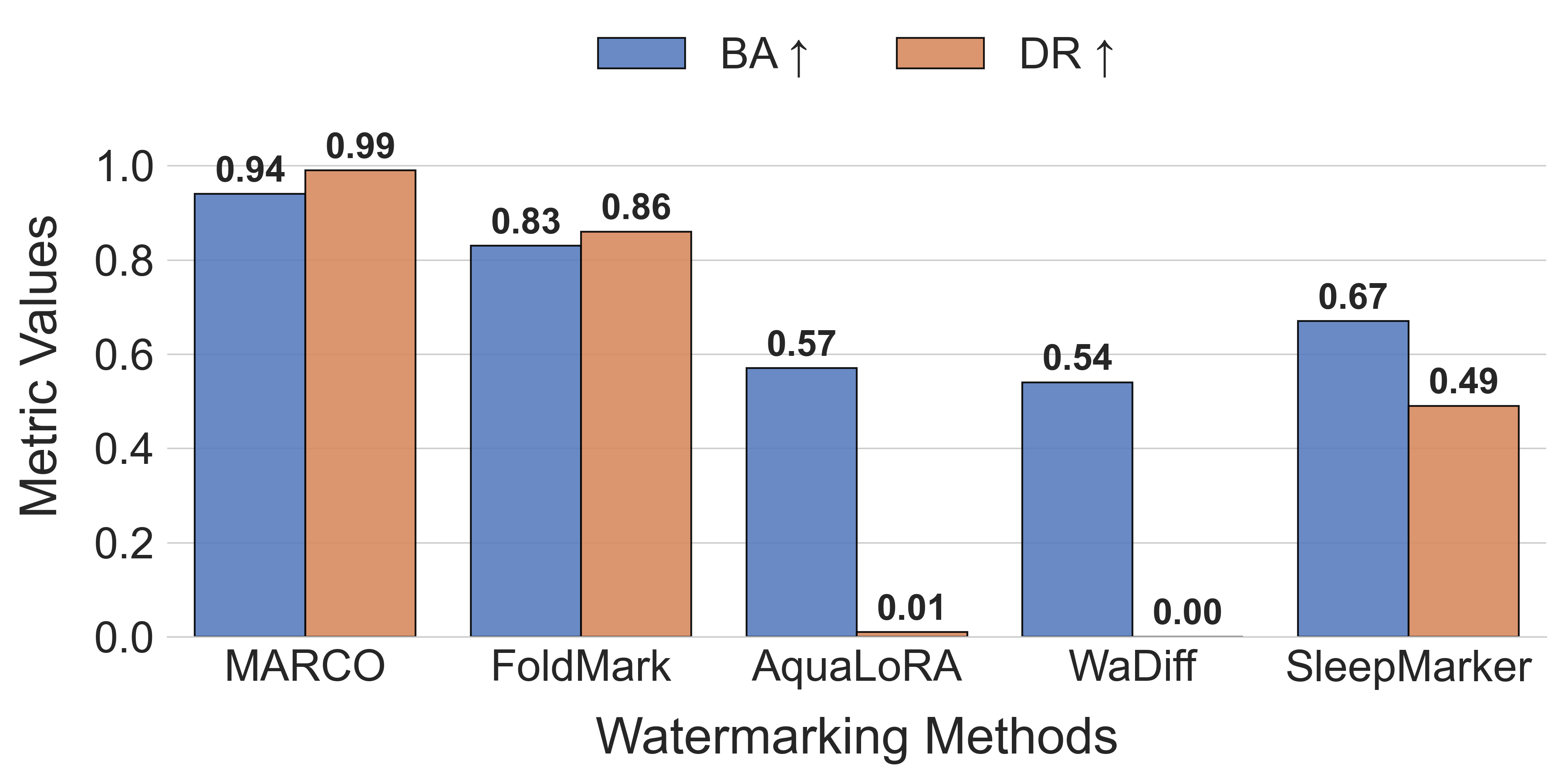}
        \caption{$\Theta$: RFDiffusion, $\Upsilon$: FoldFlow2}
    \end{subfigure}
    
    \par\medskip % New row
    
    % --- Row 2 ---
    \begin{subfigure}[b]{0.32\textwidth}
        \centering
        \includegraphics[width=\linewidth]{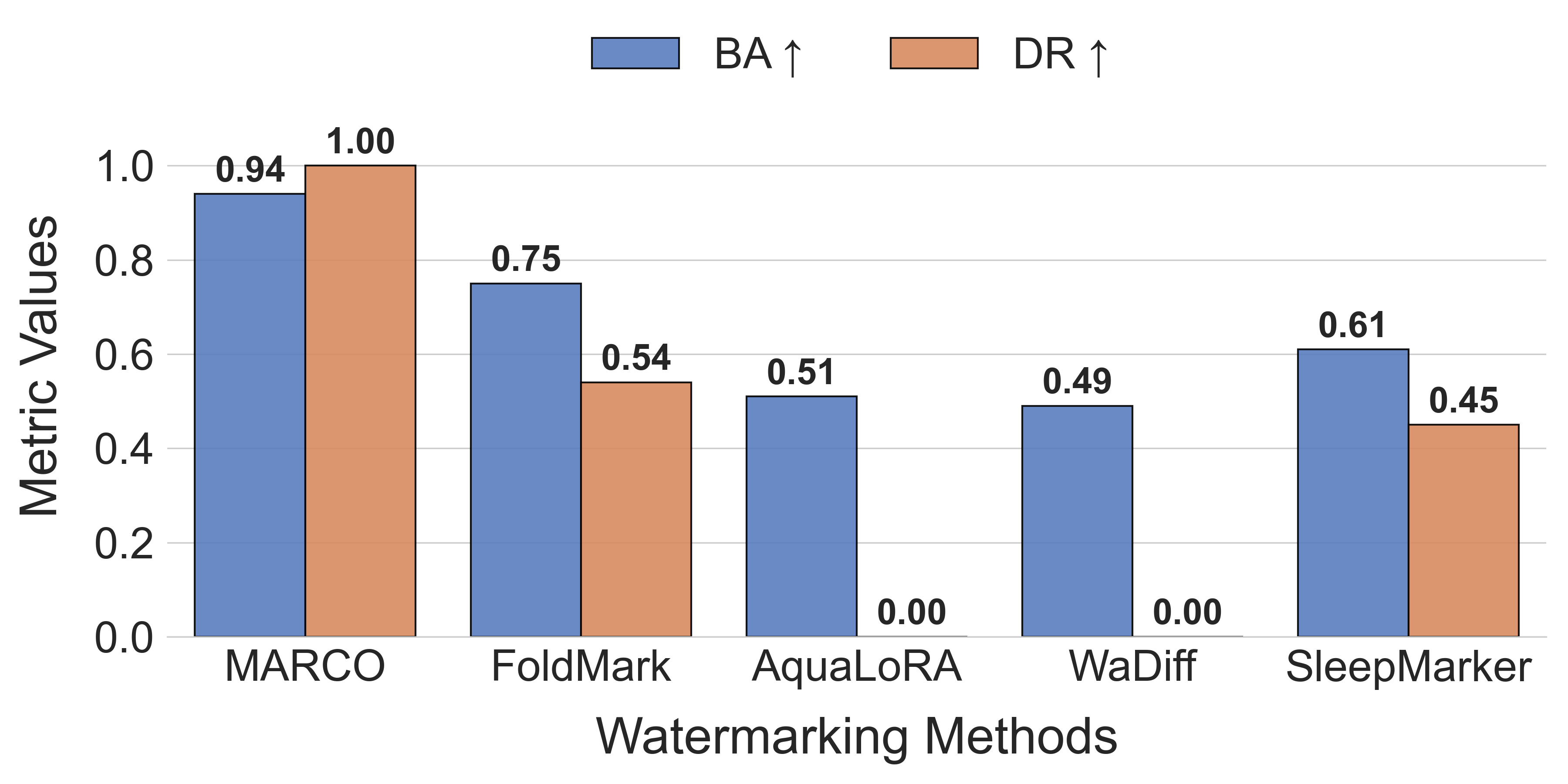}
        \caption{$\Theta$: ESMFold, $\Upsilon$: FrameDiff}
    \end{subfigure}
    \hfill
    \begin{subfigure}[b]{0.32\textwidth}
        \centering
        \includegraphics[width=\linewidth]{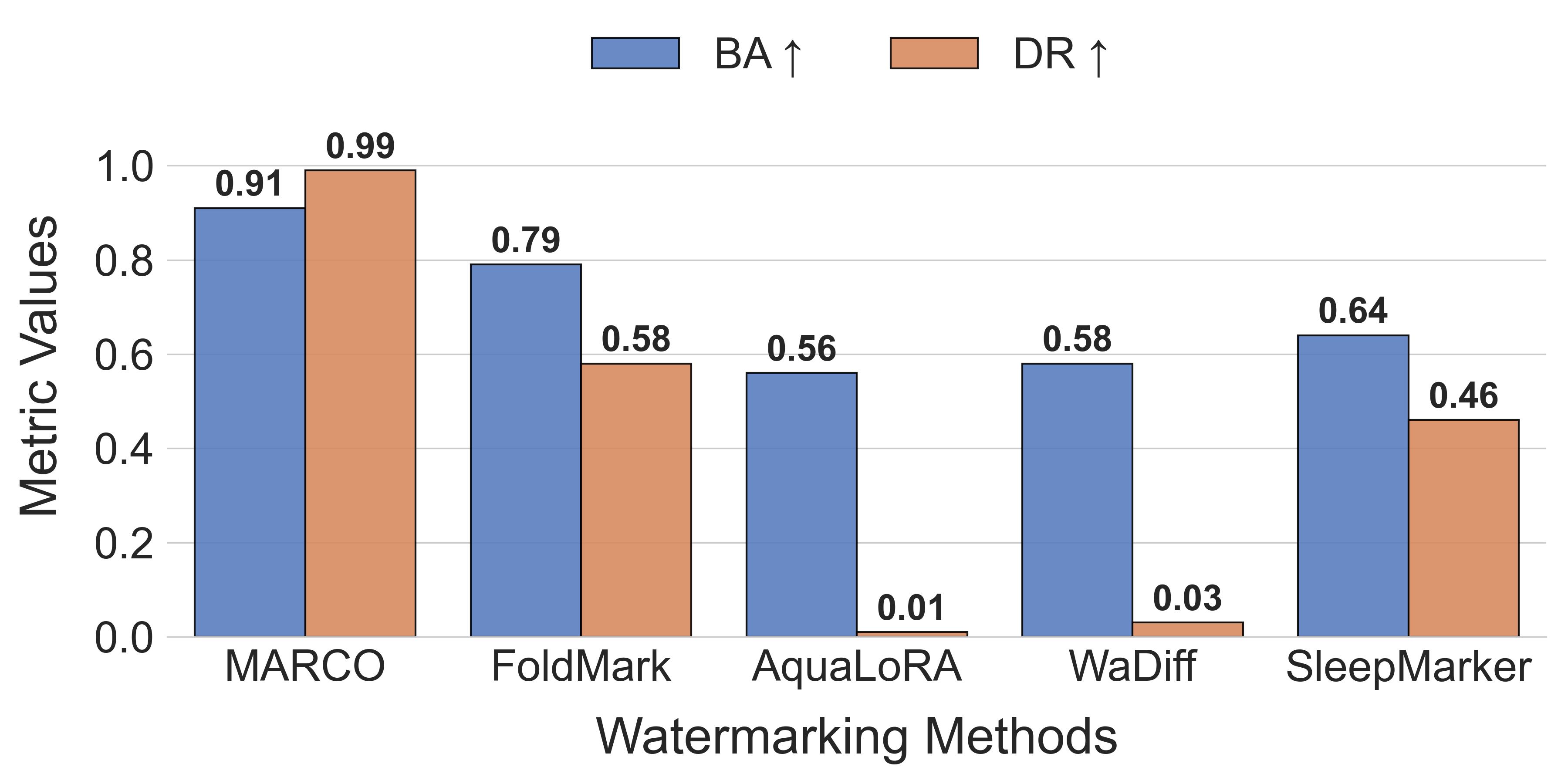}
        \caption{$\Theta$: ESMFold, $\Upsilon$: FrameFlow}
    \end{subfigure}
    \hfill
    \begin{subfigure}[b]{0.32\textwidth}
        \centering
        \includegraphics[width=\linewidth]{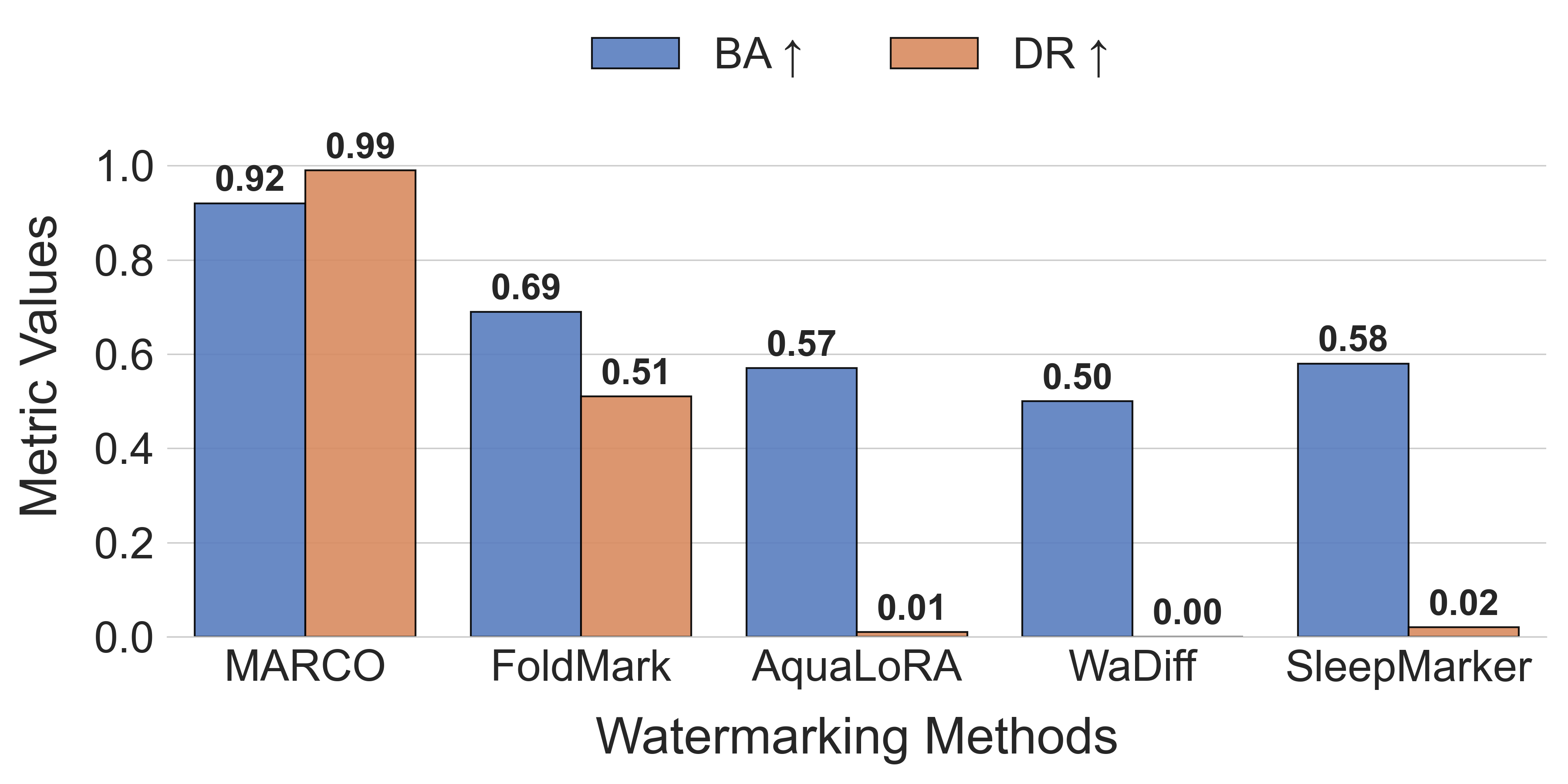}
        \caption{$\Theta$: ESMFold, $\Upsilon$: FoldFlow2}
    \end{subfigure}

    \par\medskip % New row

    % --- Row 3 ---
    \begin{subfigure}[b]{0.32\textwidth}
        \centering
        \includegraphics[width=\linewidth]{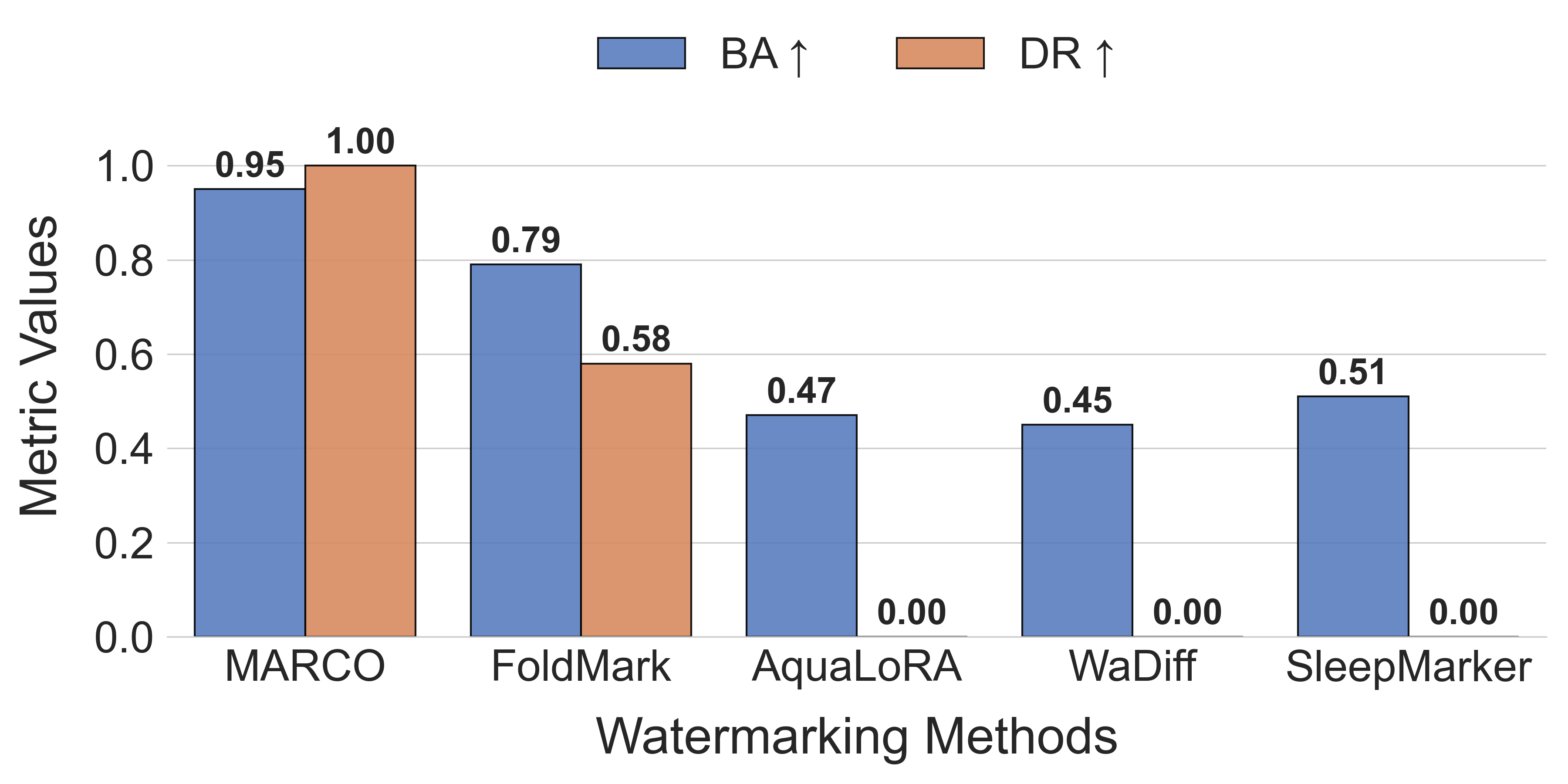}
        \caption{$\Theta$: Chroma, $\Upsilon$: FrameDiff }
    \end{subfigure}
    \hfill
    \begin{subfigure}[b]{0.32\textwidth}
        \centering
        \includegraphics[width=\linewidth]{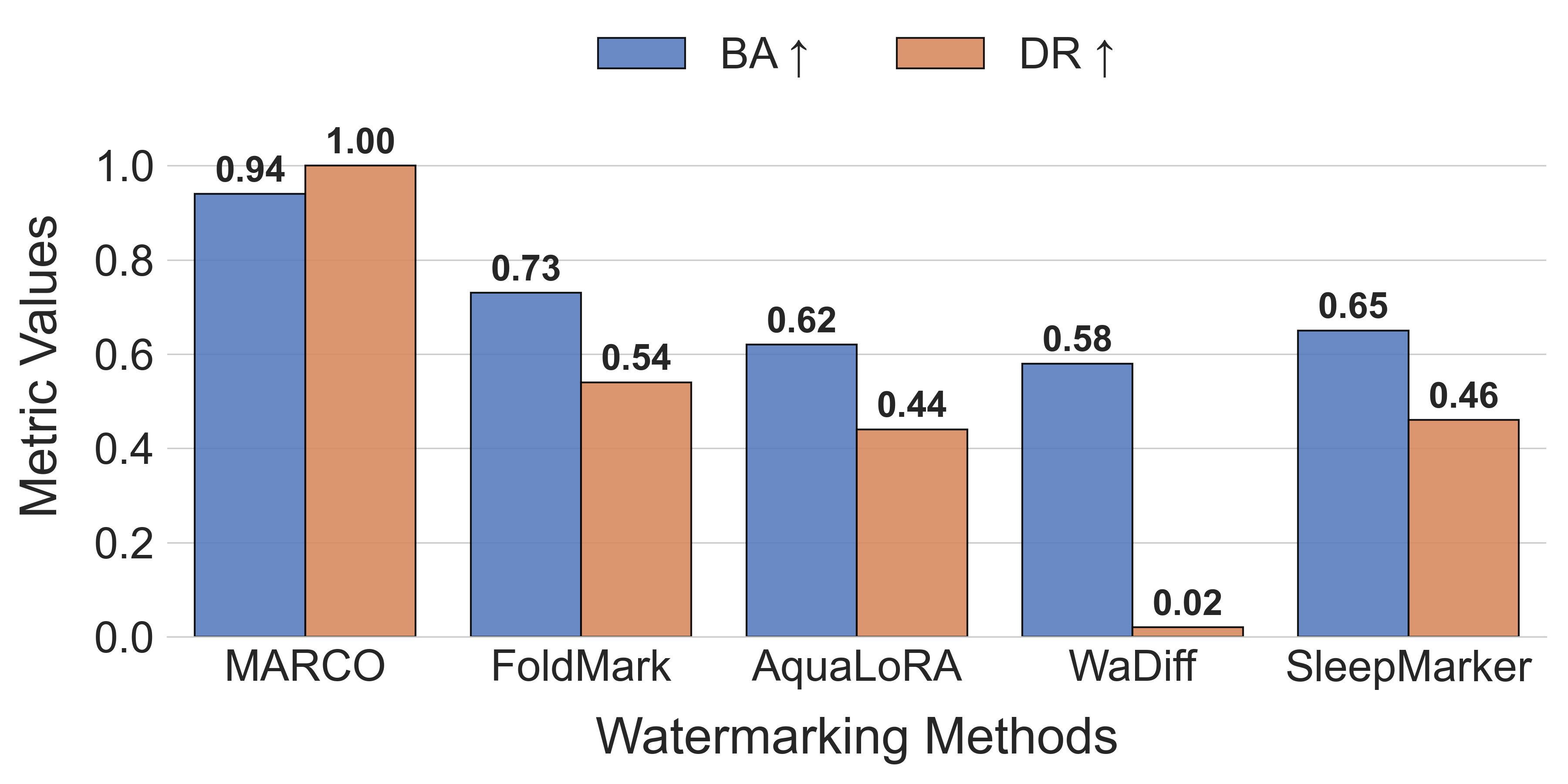}
        \caption{$\Theta$: Chroma, $\Upsilon$: FrameFlow}
    \end{subfigure}
    \hfill
    \begin{subfigure}[b]{0.32\textwidth}
        \centering
        \includegraphics[width=\linewidth]{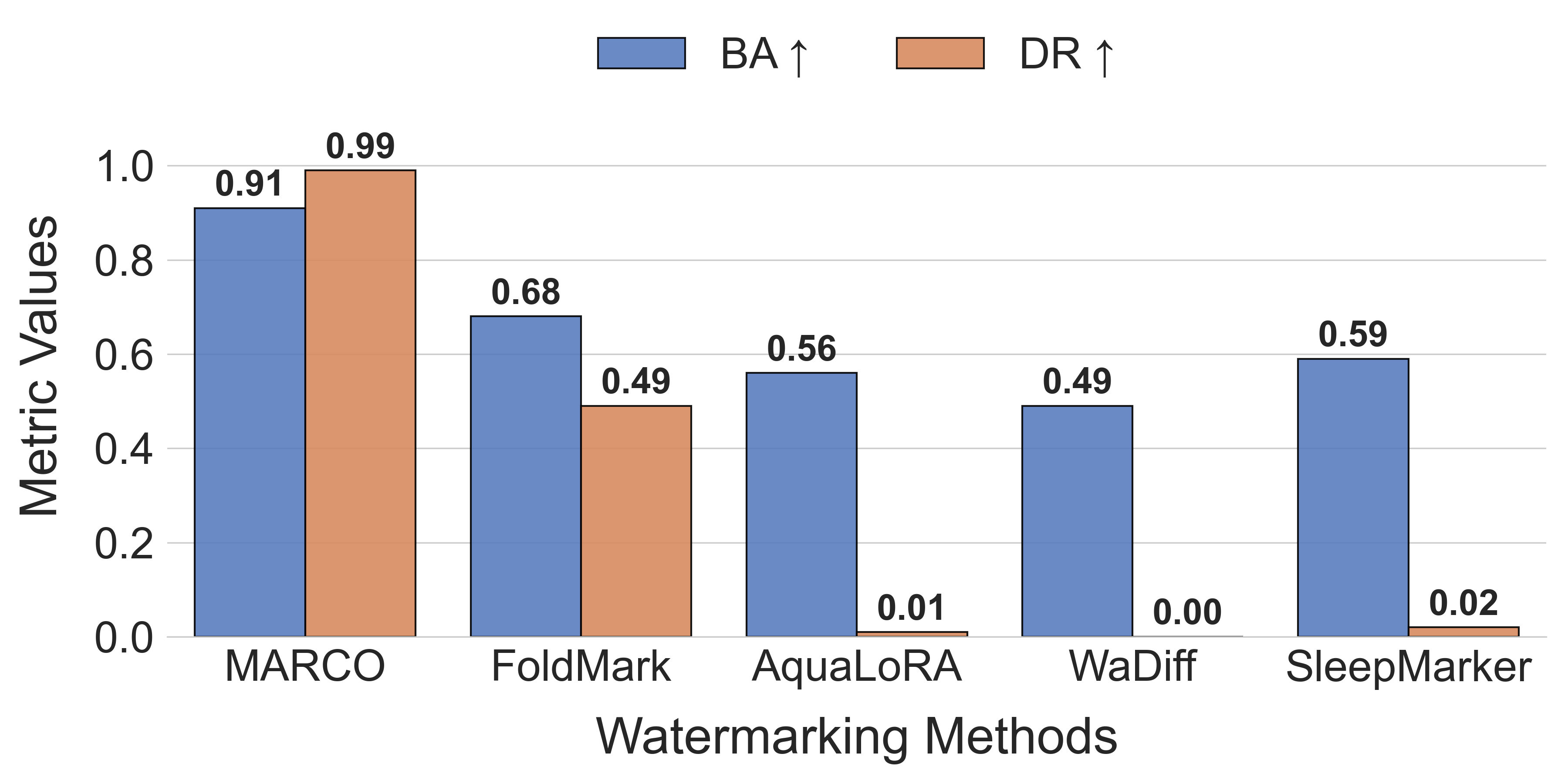}
        \caption{$\Theta$: Chroma, $\Upsilon$: FoldFlow2}
    \end{subfigure}
    
    \caption{\textbf{Evaluation of Radioactivity of PGM Watermarking Methods with Various Model Architectures}. MARCO shows superior radioactivity by transferring its watermark signal to models trained on its watermarked data regardless of model architectures.}
    \label{fig:exp_radioactivity}
\end{figure*}

% We show the results in \autoref{fig:exp_radioactivity}, where MARCO presents impressive radioactivity.
% By reaching over $0.90$ BA in watermark extraction among all cases, MARCO demonstrates its superiority in transfering its watermark signal to the pirate models trained on the watermarked data, regardless of the model architecture of either the original PGM or the pirate model.
% The baselines, in contrast, fail to demonstrate such strong radioactivity.
% Notably, some attempts to correctly extract the recovered watermark are successful for the three baselines.
% This may be due to the fine-tuning process affects the output distribution of $\Theta$, and $\Upsilon$ learns this biased distribution in the model extraction process.
% Based on these facts, we may conclude that the bias created in the reverse process step by step of MARCO could be much more radioactive.
As illustrated in \autoref{fig:exp_radioactivity}, MARCO demonstrates superior radioactivity, consistently achieving BA $\geq 0.90$ and DR $\geq 0.99$. This confirms that MARCO effectively propagates the watermark signal to the pirate model, regardless of the architectural disparities between the source and pirate models. In contrast, the baselines fail to exhibit comparable radioactivity. Their sporadic success likely stems from the pirate models learning simple distributional shifts introduced during the baselines' fine-tuning process. However, MARCO's robust performance indicates that the bias injected iteratively during the reverse diffusion process establishes a more persistent and transferable signature.

% \todo
% \textit{Prove that MARCO can infect pirate models of different architectures.}

\subsection{Capacity}
\begin{figure*}[t]
    \centering
    
    % --- Row 1 ---
    \begin{subfigure}[b]{0.45\textwidth}
        \centering
        \includegraphics[width=\linewidth]{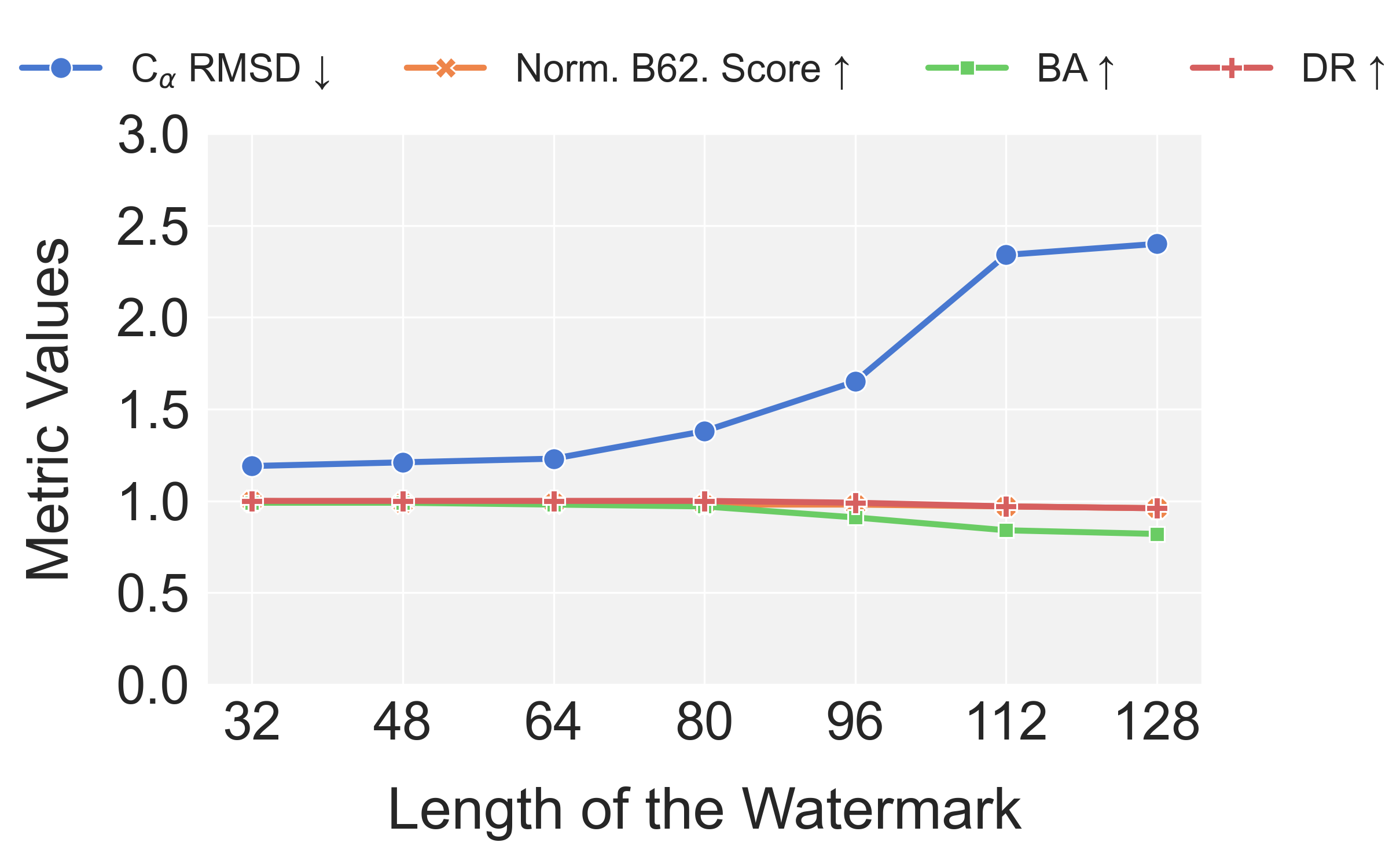}
        \caption{$\Theta$: RFDiffusion}
        \label{fig:exp_capacity_rfdiffusion}
    \end{subfigure}
    % \hspace{5pt}
    \begin{subfigure}[b]{0.45\textwidth}
        \centering
        \includegraphics[width=\linewidth]{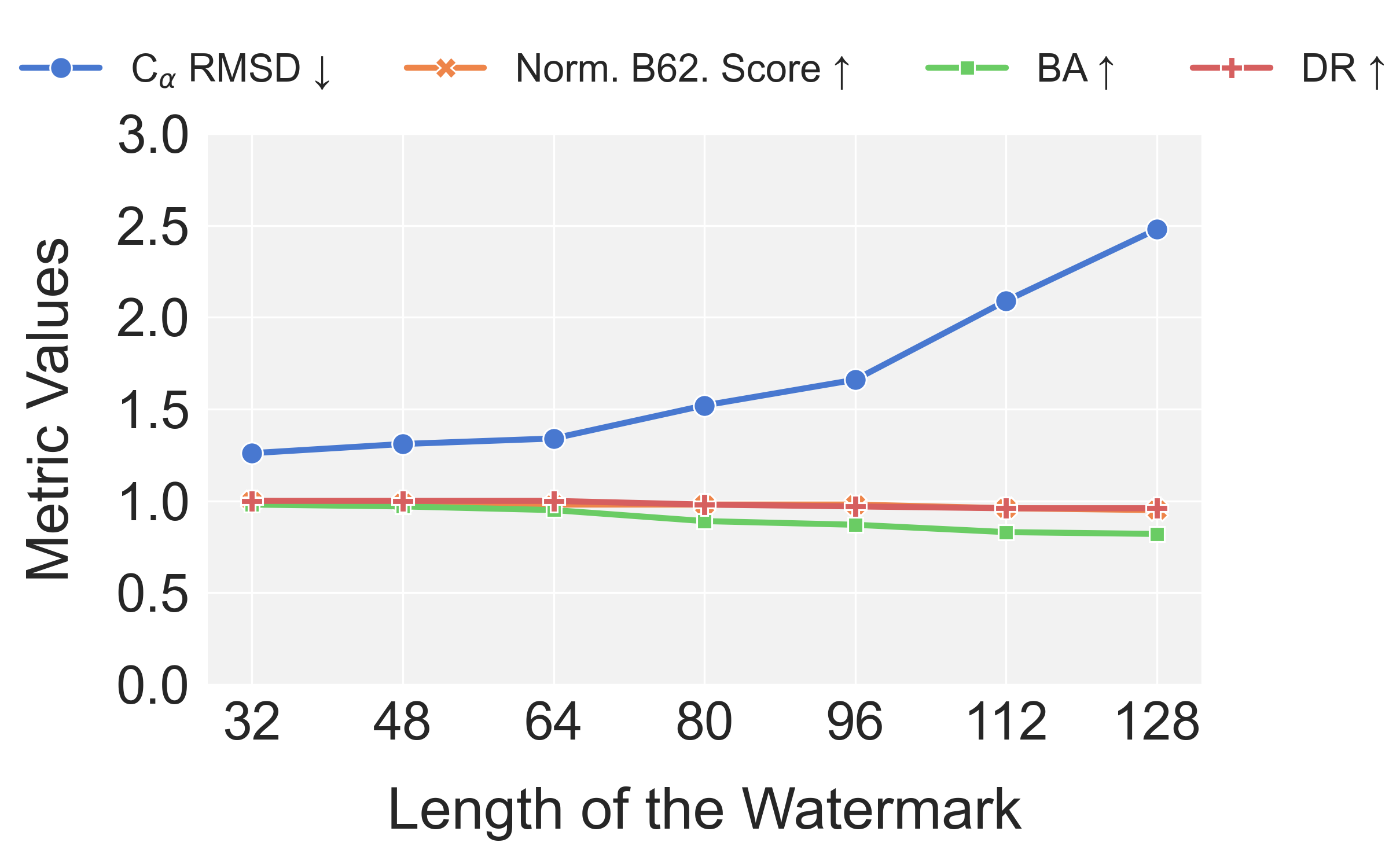}
        \caption{$\Theta$: ESMFold}
        \label{fig:exp_capacity_esmfold}
    \end{subfigure}
    % \hspace{5pt}

    % --- Row 2 ---

    \par\bigskip % Force a new row with some vertical spacing
    \begin{subfigure}[b]{0.45\textwidth}
        \centering
        \includegraphics[width=\linewidth]{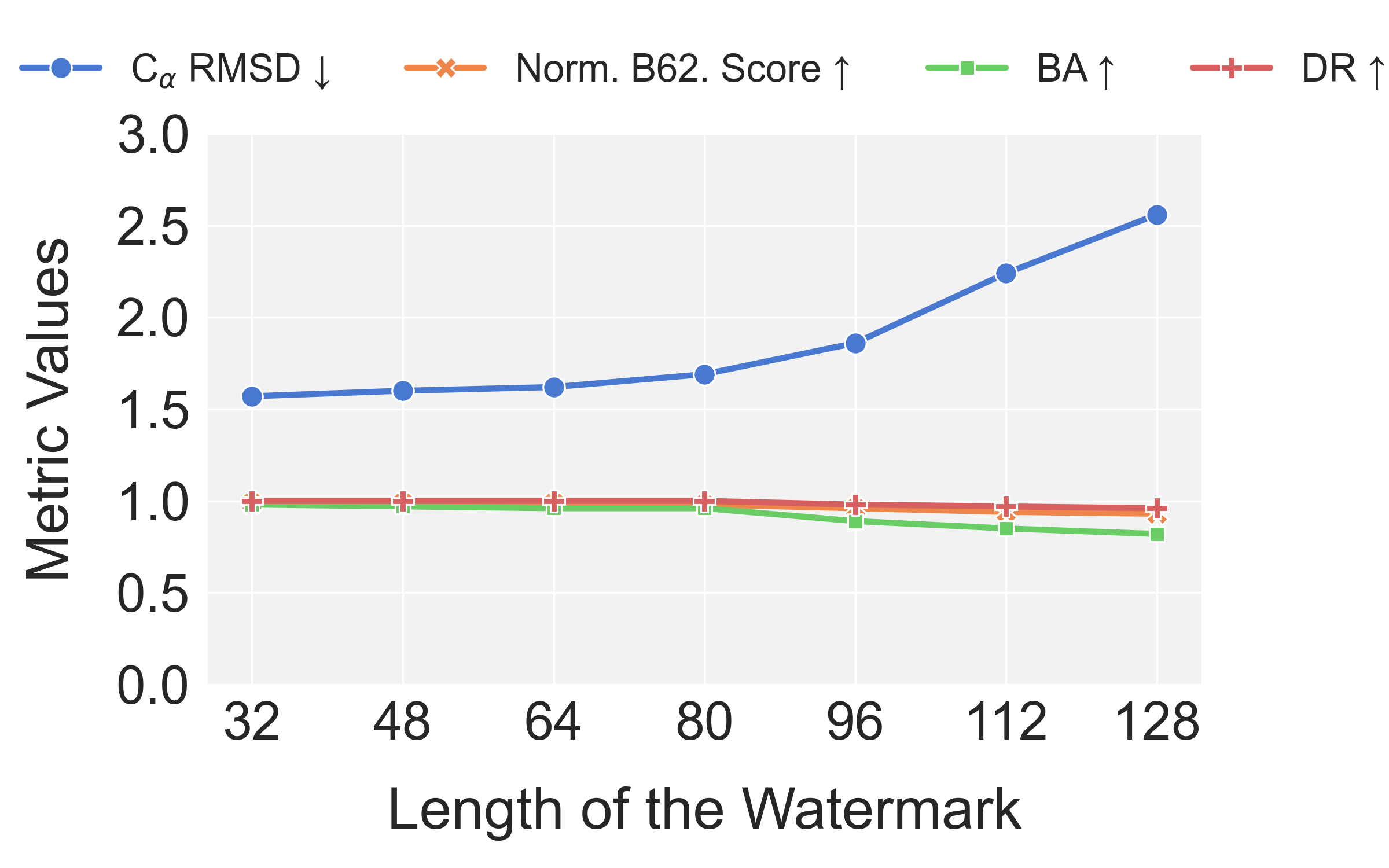}
        \caption{$\Theta$: Chroma}
        \label{fig:exp_capacity_chroma}
    \end{subfigure}
    % \par\bigskip % Force a new row with some vertical spacing
    \begin{subfigure}[b]{0.45\textwidth}
        \centering
        \includegraphics[width=\linewidth]{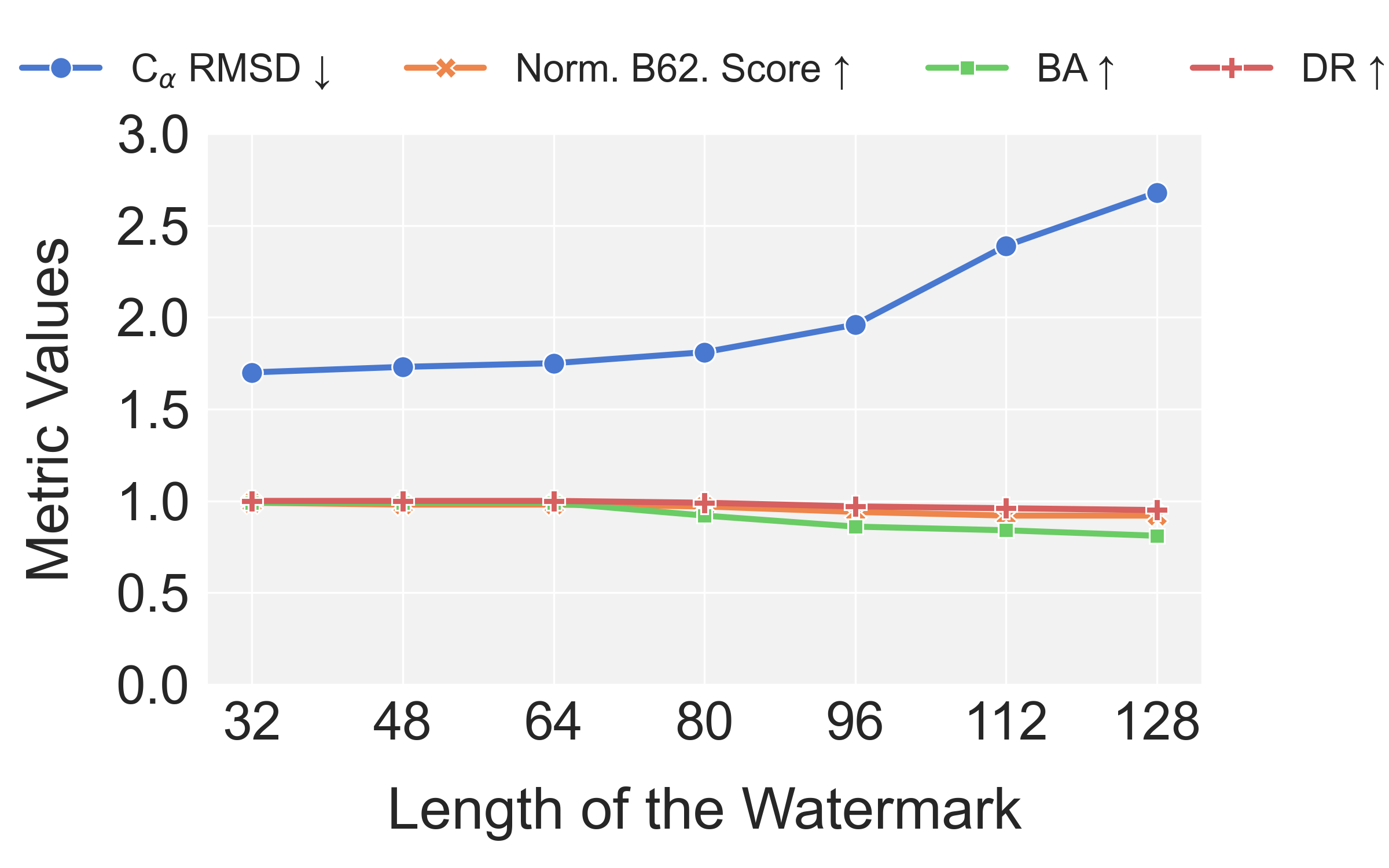}
        \caption{$\Theta$: FrameDiff}
        \label{fig:exp_capacity_FrameDiff}
    \end{subfigure}
    % \hspace{5pt}

    % --- Row 3 ---

    \par\bigskip % Force a new row with some vertical spacing
    \begin{subfigure}[b]{0.45\textwidth}
        \centering
        \includegraphics[width=\linewidth]{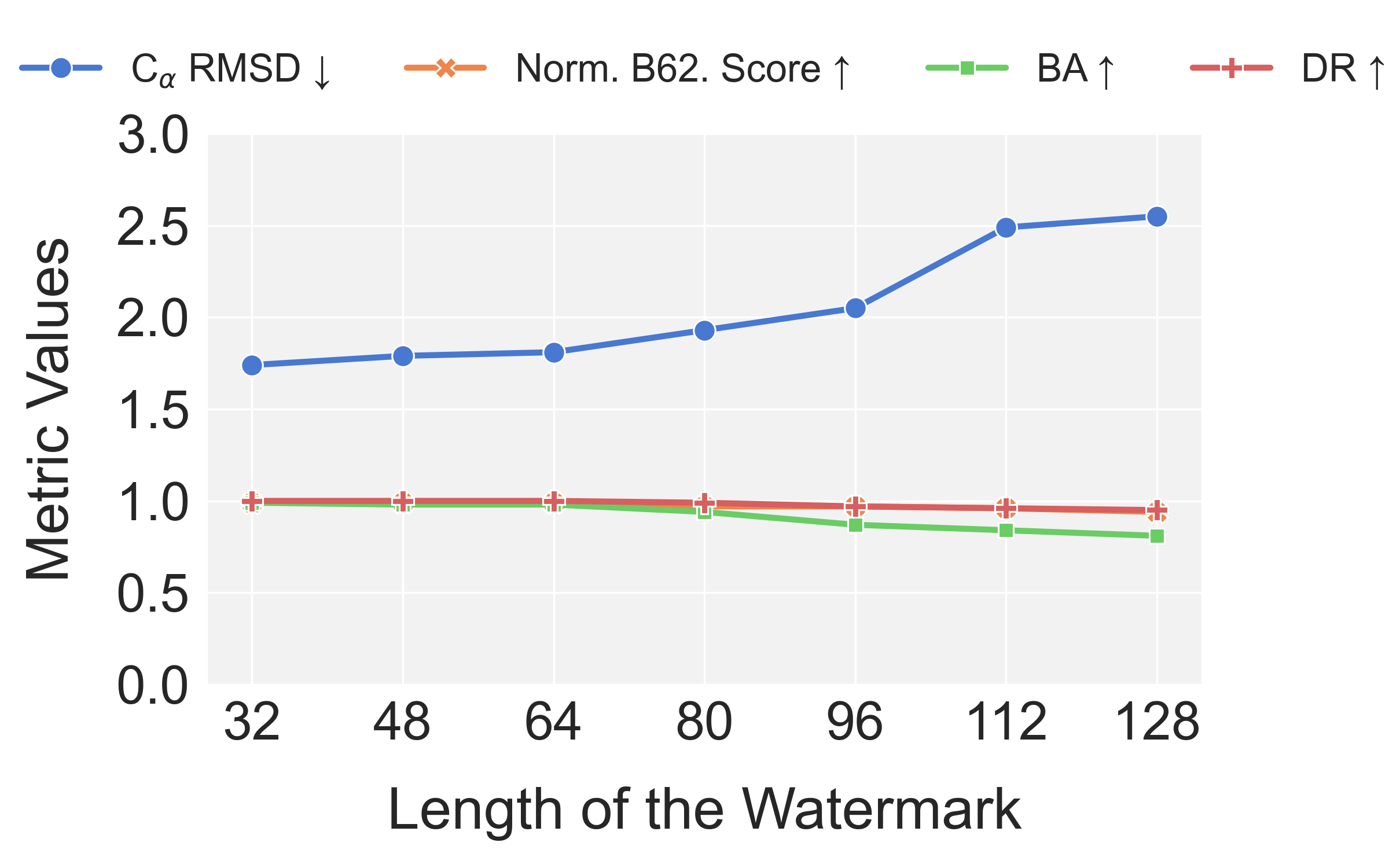}
        \caption{$\Theta$: FrameFlow}
        \label{fig:exp_capacity_frameflow}
    \end{subfigure}
    % \hspace{5pt}
    \begin{subfigure}[b]{0.45\textwidth}
        \centering
        \includegraphics[width=\linewidth]{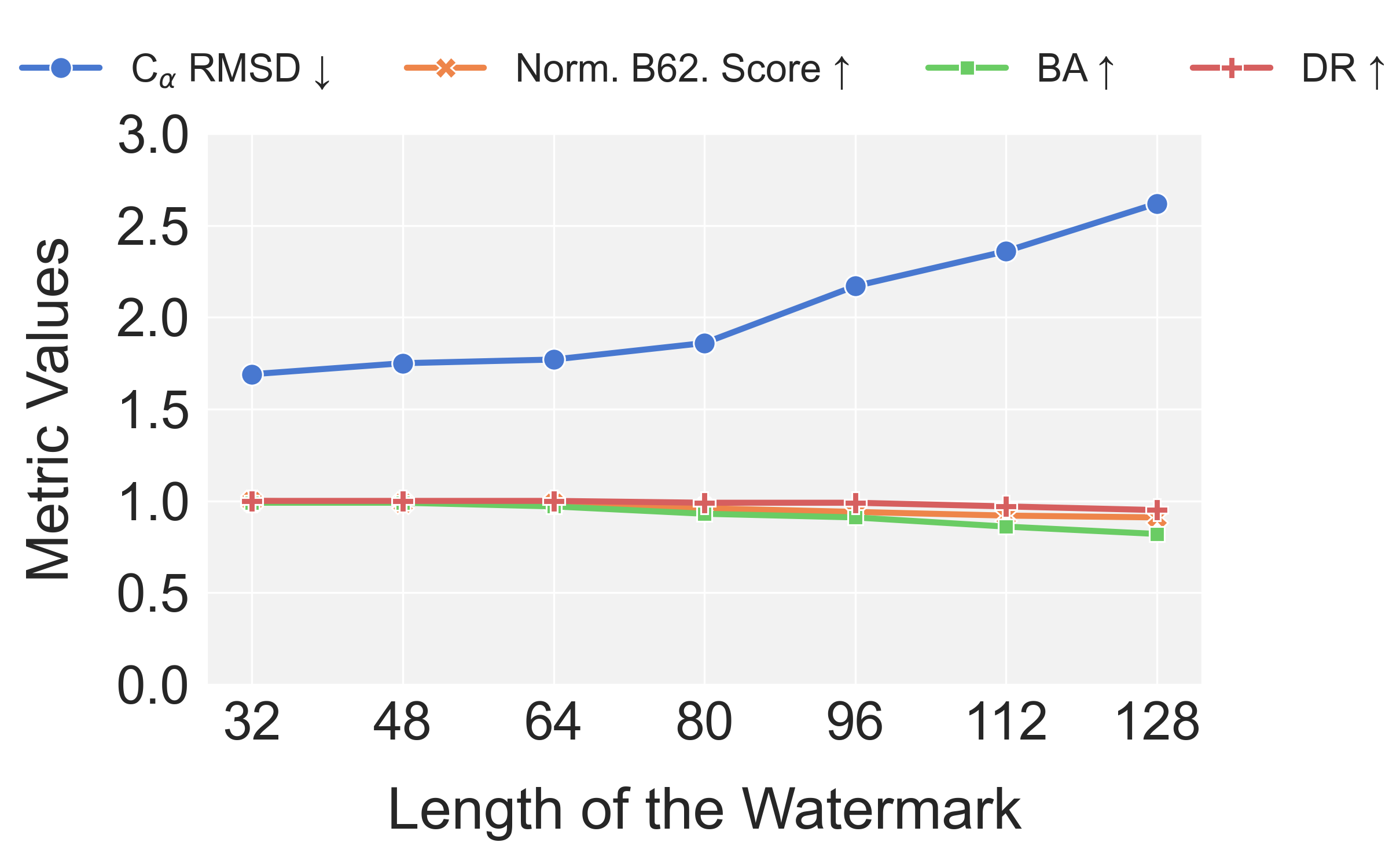}
        \caption{$\Theta$: FoldFlow2}
        \label{fig:exp_capacity_FoldFlow2}
    \end{subfigure}
    
    % \par\bigskip % Force a new row
    
    % --- Row 3 ---

    \caption{\textbf{Evaluation of Capacity of MARCO}. As the length of the watermark increases, both the fidelity and robustness drops. }
    \label{fig:exp_capacity}
\end{figure*}

% To demonstrate the capacity of MARCO, we vary the length of the watermark $L$ in a set $\{32, 48, 64, 80, 96, 112, 128\}$ and generate watermarked structures using the PGMs along with the watermark methods.
% We then evaluate how varying $L$ affects the fidelity and the robustness of the watermarking methods by computing the metric values.
To assess the capacity of MARCO, we varied the watermark length $l$ across the set $\{32, 48, \dots, 128\}$ and evaluated the impact on fidelity and robustness.
% The experimental results are illustrated in \autoref{fig:exp_capacity}, where MARCO demonstrates a sufficient capacity to store identity information in the watermark without significantly decreasing the fidelity and robustness.
% Among all cases, we notice that the normalized BLOSUM62 score and detection rate remain stable as the length of the watermark $L$ increases.
% This indicates that even though the perturbation can affect the molecular conformation to some extent, it is still redundant to maintain sequential similarity and reach a sufficiently high bit accuracy to support the watermark detection.
% In contrast, the RMSD and BA values fluctuate more.
% We find that when $L \geq 96$, the RMSD and BA vary greatly, indicating that there may be a threshold of $L$ that draws a line for trading off capacity with fidelity and robustness.
% Although the decreased BA can still guarantee valid watermark detection, it may not suffice to verify a user's identification.
% Nonetheless, $L = 64$ already provides considerable fidelity and robustness with enough coding space for identification verification.
As shown in \autoref{fig:exp_capacity}, MARCO effectively encodes identity information without compromising performance. Both the normalized BLOSUM62 score and DR remain stable as $l$ increases, indicating that the induced perturbations are sufficiently redundant to preserve sequential similarity and detection reliability.  However, RMSD and Bit Accuracy (BA) exhibit greater sensitivity; specifically, for $l \geq 96$, we observe significant fluctuations, suggesting a saturation threshold where capacity begins to degrade fidelity. While the reduced BA at high $l$ values still supports basic detection, it may compromise precise user identification. Nonetheless, the default length of $l=64$ offers an optimal trade-off, providing a vast coding space with robust fidelity.

\subsection{Ablation Study}
\prginit{Impact of the Loss Terms and the Attack Layer}
% To show what impact the loss terms have on MARCO, we conduct an ablation study, where we remove the loss terms one by another by setting the corresponding weight parameter $\lambda$ to $0$.
% We also evaluate the efficacy of the attack layer $\mcl A$ by canceling it during the training process.
% In this evaluation, we derive the BA and DR after launching the noise addition manipulation on the watermarked structures.
% This is because noise addition decreases the BA and DR mostly in the previous experiments.
% We select and fix the pre-trained $\Theta$ to be RFDiffusion.
To quantify the contribution of individual components, we conduct an ablation study by sequentially setting the weights $\lambda=0$ and disabling the Attack Layer $\mathcal{A}$. We utilize RFdiffusion as the fixed PGM and evaluate robustness specifically under noise addition, the most disruptive attack observed in prior experiments.

% As shown in \autoref{tab:ablation_losses}, the experimental results demonstrate that all components listed in the table are crucial to the performance of MARCO.
% The watermark loss $\mcl L_{\text{wm}}$ minimizes the $L^2$ norm of the rotation matrix and atom coordinate perturbations.
% Removing $\mcl L_{\text{wm}}$ results in a significantly decreased fidelity, but the BA and DR remain high.
% The distance matrix loss $\mcl L_{\text{dm}}$ controls the inter-atom distances within the same molecule.
% Without its presence, the fidelity drops significantly.
% Similarly, the performance drop also happens without the torsion angle loss $\mcl L_{\text{tor}}$ that controls the rotation of the atoms.
% As expected, the robustness vanishes when the extraction loss $\mcl L_{\text{ext}}$ that provides the core functionality is removed.
% The adversarial loss $\mcl L_{\text{adv}}$ contributes to both fidelity and robustness by helping the encoder hide the watermark signal more covertly.
% Lastly, without the attack layer, conformational manipulation can easily penetrate the watermark defense.
% Thus, the experimental results support our design by showing the necessity of all components.
As presented in \autoref{tab:ablation_losses}, every component is critical to MARCO's performance. The watermark loss $\mathcal{L}_{\text{wm}}$, which constrains perturbation magnitudes, is essential for fidelity. Its removal causes significant structural degradation, even though robustness remains high. Similarly, omitting the distance matrix Loss $\mathcal{L}_{\text{dm}}$ and torsion angle Loss $\mathcal{L}_{\text{tor}}$ results in substantial fidelity drops. The extraction loss $\mathcal{L}_{\text{ext}}$ is foundational; without it, robustness vanishes entirely. The adversarial loss $\mathcal{L}_{\text{adv}}$ serves a dual role, enhancing both fidelity and robustness. Finally, the attack layer proves vital for generalization. Its exclusion leaves the watermark highly vulnerable to conformational manipulations.

% \todo{Distributional Indistinguishability}

\prg{Distributional Indistinguishability}
We further measure the distributional indistinguishability between clean predictions and watermarked structures using the Fréchet Distance (FD). We trained two MARCOs with and without $\mcl L_{\text{adv}}$. We generated 1,000 triplets of unwatermarked and watermarked proteins via RFdiffusion using the two MARCO variants to compute the FD. With $\mcl L_{\text{adv}}$, the FD is $2.27$, whereas the FD increases to $16.31$ without it. This suggests that $\mcl L_{\text{adv}}$ significantly improves covertness.

\prg{Impact of the Training Epochs}
% Additionally, we vary the number of training epochs to evaluate its impact on the performance of MARCO by following the settings above.
% The results are listed in \autoref{tab:ablation_epochs}, where we observe that when the number of epochs exceeds $25$, the performance increment of MARCO becomes less significant.
% The performance of MARCO with the number of epochs $100$ does not differ from that of MARCO trained for $50$ epochs.
% We thus conclude that $50$ epochs are sufficient to make MARCO have considerable performance.
We further investigate the impact of training duration on MARCO's performance by varying the epoch count under identical experimental conditions. As detailed in \autoref{tab:ablation_epochs}, performance gains diminish significantly beyond 25 epochs. Notably, the model exhibits negligible improvement when scaling from 50 to 100 epochs, indicating convergence. Consequently, we determine that 50 epochs are sufficient to achieve optimal performance stability.

\begin{table}[t!]
    \centering
    % Increase row height slightly for a fancier look
    \renewcommand{\arraystretch}{1.5}
    \caption{\textbf{Impact of Training Epochs}.}
    \resizebox{.6\linewidth}{!}{
    \begin{tabular}{
        >{\columncolor{gray!10}}c  % First column (Left aligned, no color)
        >{\columncolor{blue!10}}c   % 2nd Col: Light Blue
        >{\columncolor{red!10}}c    % 3rd Col: Light Red
        >{\columncolor{purple!10}}c  % 4th Col: Light Green
        >{\columncolor{orange!10}}c % 5th Col: Light Orange
    }
        \toprule
        % Top-left split cell: \diagbox{BottomLeft text}{TopRight text}
        \diagbox{\textbf{Epochs}}{\textbf{Metrics}} & \textbf{$C_\alpha$RMSD$\downarrow$} & \textbf{Norm. B62$\uparrow$} & \textbf{BA$\uparrow$} & \textbf{DR$\uparrow$} \\ 
        \midrule
        $5$ & 2.44 & 0.71 & 0.62 & 0.17 \\ 
        $10$ & 1.62 & 0.92 & 0.84 & 0.96 \\
        $25$ & 1.41 & 0.97 & 0.96 & 1.00 \\
        $50$ & 1.23 & 0.98 & 0.98 & 1.00 \\ 
        $100$ & 1.23 & 0.98 & 0.99 & 1.00 \\ 
        \bottomrule
    \end{tabular}
    }
    \label{tab:ablation_epochs}
\end{table}

\begin{table}[t!]
    \centering
    % Increase row height slightly for a fancier look
    \renewcommand{\arraystretch}{1.5}
    \caption{\textbf{Impact of Loss Terms and the Attack Layer}.}
    \resizebox{.5\linewidth}{!}{
    \begin{tabular}{
        >{\columncolor{gray!10}}c  % First column (Left aligned, no color)
        >{\columncolor{blue!10}}c   % 2nd Col: Light Blue
        >{\columncolor{red!10}}c    % 3rd Col: Light Red
        >{\columncolor{purple!10}}c  % 4th Col: Light Green
        >{\columncolor{orange!10}}c % 5th Col: Light Orange
    }
        \toprule
        % Top-left split cell: \diagbox{BottomLeft text}{TopRight text}
        \diagbox{\textbf{Terms}}{\textbf{Metrics}} & \textbf{$C_\alpha$RMSD$\downarrow$} & \textbf{Norm. B62$\uparrow$} & \textbf{BA$\uparrow$} & \textbf{DR$\uparrow$} \\ 
        \midrule
        \textbf{With All} & 1.23 & 0.99 & 0.98 & 1.00 \\ 
        \textbf{W/O} $\mcl L_{\text{wm}}$ & 2.13 & 0.79 & 0.97 & 1.00 \\ 
        \textbf{W/O} $\mcl L_{\text{dm}}$ & 1.63 & 0.92 & 0.98 & 1.00 \\ 
        \textbf{W/O} $\mcl L_{\text{tor}}$ & 1.56 & 0.87 & 0.97 & 1.00 \\ 
        \textbf{W/O} $\mcl L_{\text{ext}}$ & 1.25 & 0.99 & 0.02 & 0.00 \\ 
        \textbf{W/O} $\mcl L_{\text{adv}}$ & 1.34 & 0.97 & 0.92 & 0.99 \\ 
        \textbf{W/O} $\mcl A$ & 1.22 & 0.99 & 0.05 & 0.00 \\ 
        \bottomrule
    \end{tabular}
    }
    \label{tab:ablation_losses}
\end{table}

% \prg{Impact of the Denoising Steps}
% We also test the impact of the number of denoising steps in the reverse process on the performance of MARCO using the same settings.
% As illustrated in \autoref{tab:ablation_rev_steps}, the increase in MARCO's performance stops when the number of steps exceeds $100$.
% However, the number of steps drastically affects the performance of MARCO when it increases from $10$ to $100$.
% We thus believe that $100$ steps for the reverse process suffice for MARCO.
% We further assess the impact of the denoising step count on MARCO's efficacy. As detailed in \autoref{tab:ablation_rev_steps}, performance improves drastically as steps increase from 10 to 100 but plateaus thereafter. Consequently, we identify 100 steps as the optimal threshold for the reverse process.

\prg{Inference Overhead}
% It takes in total 16.71 hours to finish training MARCO for $50$ epochs using one NVIDIA A100 GPU.
% Converted into the exact amount of money, it is equal to $57.32$\$ for renting a single A100 GPU on Amazon Web Services.
% This shows that MARCO can be very efficiently trained and deployed.
Regarding computational overhead, training MARCO for 50 epochs requires approximately 16.71 hours on a single NVIDIA A100 GPU. This translates to a cloud computing cost of \$57.32 (based on standard AWS pricing), underscoring the framework's training efficiency and economic feasibility for deployment.
% \todo{Inference Overhead}
Moreover, we measure the inference overhead of MARCO and present the results in \autoref{tab:ifr_ovh}.
We discover that the ratio of the increased time to the clean reference time is low and is therefore trivial to the protein structure generation.

\begin{table}[t!]
    \centering
    % Increase row height slightly for a fancier look
    \renewcommand{\arraystretch}{1.5}
    \caption{\textbf{Inference Overhead (s/structure)}.}
    \resizebox{.6\linewidth}{!}{
    \begin{tabular}{
        >{\columncolor{gray!10}}c  % First column (Left aligned, no color)
        >{\columncolor{blue!10}}c   % 2nd Col: Light Blue
        >{\columncolor{red!10}}c    % 3rd Col: Light Red
    }
        \toprule
        % Top-left split cell: \diagbox{BottomLeft text}{TopRight text}
        \textbf{Model} & \textbf{Clean Inference Time} & \textbf{MARCO Inference Time}\\ 
        \midrule
        RFDiffusion & 28.8 & 32.2 \\
        ESMFold & 8.7 & 8.9 \\
        \bottomrule
    \end{tabular}
    }
    \label{tab:ifr_ovh}
\end{table}

\prg{Negative Control}
We evaluate MARCO's specificity against negative structures. We make the trained watermark decoder extract a pre-defined 64-bit watermark from $100$ natural structures and clean predicted structures to compute the BA and DR. The results in \autoref{tab:fpr_detection} show that MARCO cannot falsely retrieve the pre-defined watermark from the natural or the clean predicted structures, demonstrating strong specificity.

\begin{table}[t!]
    % \captionsetup{labelfont={color=blue}, textfont={color=blue}}
    \centering
    \renewcommand{\arraystretch}{1.5}
    \caption{\textbf{Negative Control Evaluation.}}
    \resizebox{.5\linewidth}{!}{
    \begin{tabular}{
        >{\centering\columncolor{gray!10}}p{4cm}  
        >{\columncolor{blue!10}}c   
        >{\columncolor{red!10}}c   
    }
        \toprule
        \textbf{Structure Type} & \textbf{BA $\downarrow$} & \textbf{DR $\downarrow$}\\ 
        \midrule
        Natural Structures & 0.48 & 0.00 \\
        Clean Predicted Structures & 0.49 & 0.00 \\ 
        \bottomrule
    \end{tabular}
    }
    \label{tab:fpr_detection}
\end{table}

\section{Limitations \& Future Work}\label{sect:future_work}
Currently, MARCO is specialized for diffusion-based architectures. Auto-regressive Protein Language Models (PLMs) such as ESM3 \cite{hayes2025ESM3}, SaProt \cite{suSaprot}, and BioMistral \cite{labrak2024Biomistral} remain outside our current scope. However, inspired by Kirchenbauer et al. \cite{kirchenbauer2023watermark}, we can design a new paradigm for PLM watermarking by dynamically dividing the amino acid dictionary into a red and a green set with some biological constraints. Additionally, we have not yet conducted wet-lab experiments to validate the recovery of watermarks from physically synthesized and re-digitized proteins.
Future research will focus on generalizing MARCO to discrete modalities such as PLMs and the latest genome language model AlphaGenome \cite{AlphaGenome2026}, and adapting to emerging architectures like BoltzGen \cite{stark2025boltzgen}. 

For wet-lab verification of a synthesized protein, we envision two complementary approaches. First, after confirming the sequence by DNA sequencing and LC–MS/MS, the protein structure could be resolved using X-ray crystallography, cryo-EM, or NMR and processed by MARCO’s decoder to recover the registered watermark~\cite{durbin1996proteincrystallization}. Second, a lower-cost antibody-based assay could be enabled by concentrating part of the watermark into a biologically tolerant, surface-exposed conformational epitope. A specific monoclonal antibody could then detect the watermarked protein through native ELISA, immunoprecipitation, or surface plasmon resonance~\cite{davies1990antibody}.

\end{document}